\documentclass[12pt,notitlepage,a4paper]{article}

\pdfoutput=1

\usepackage{a4wide}
\usepackage{epsfig}
\usepackage{color,graphicx}
\usepackage{cite}
\usepackage{amssymb,amsmath}

\newcommand{\be}{\begin{equation}}
\newcommand{\ee}{\end{equation}}
\newcommand{\bea}{\begin{eqnarray}}
\newcommand{\eea}{\end{eqnarray}}

\begin{document}

\begin{center}  

\vskip 2cm 

\centerline{\Large {\bf An $\mathcal{N}=1$ Lagrangian for an $\mathcal{N}=3$ SCFT}}

\vskip 1cm

\renewcommand{\thefootnote}{\fnsymbol{footnote}}

   \centerline{
    {\large \bf Gabi Zafrir${}^{a}$} \footnote{gabi.zafrir@unimib.it}}

\vspace{1cm}
\centerline{{\it ${}^a$ Dipartimento di Fisica, Universit\`a di Milano-Bicocca \& INFN, Sezione di Milano-Bicocca,}}
\centerline{{\it I-20126 Milano, Italy}}
\vspace{1cm}

\end{center}

\vskip 0.3 cm

\setcounter{footnote}{0}
\renewcommand{\thefootnote}{\arabic{footnote}}   
   
      \begin{abstract}
     
		We propose that a certain $4d$ $\mathcal{N}=1$ $SU(2)\times SU(2)$ gauge theory flows in the IR to an $\mathcal{N}=3$ SCFT plus a single free chiral field. The specific $\mathcal{N}=3$ SCFT has rank $1$ and a dimension three Coulomb branch operator. The flow is generically expected to land at the $\mathcal{N}=3$ SCFT deformed by the marginal deformation associated with said Coulomb branch operator. We also present a discussion about the properties expected of various RG invariant quantities from $\mathcal{N}=3$ superconformal symmetry, and use these to test our proposal. Finally, we discuss a generalization to another $\mathcal{N}=1$ model that we propose is related to a certain rank $3$ $\mathcal{N}=3$ SCFT through the turning of certain marginal deformations.
		
      \end{abstract}

\tableofcontents

\section{Introduction}

The study of conformal field theories plays an important role in the study of quantum field theory in general. These type of theories usually appear at the end points of RG flows and have a larger group of spacetime symmetries, the conformal group rather than the Poincare group. One can also enlarge the symmetry further by adding supersymmetry, leading to superconformal field theories (SCFTs). The largest amount of supersymmetry one can have without introducing gravity is sixteen supercharges, corresponding to $\mathcal{N}=4$ supersymmetry in four dimensions. This class of theories appears to posses the largest amount of spacetime symmetry making it an interesting and active research ground.

While the constraints on the dynamics imposed by the large amount of supersymmetry help in the study of such theories, they also make the behavior of these theories less rich. This in turn motivates the study of theories with less supersymmetry. In that spirit, $\mathcal{N}=2$ theories and SCFTs were, and still are, studied extensively. These present richer dynamics than that present in $\mathcal{N}=4$ theories, yet the supersymmetry is still sufficient to make significant progress.

This present the question of the place of $\mathcal{N}=3$ theories. Despite being an intermediate case, with potentially richer dynamics than $\mathcal{N}=4$ theories, but with more supersymmetry than $\mathcal{N}=2$ theories, these theories are not as widely studied as their counterparts. A large part of this is due to the fact that these were only recently discovered, as it was suspected for a long time that these may not exist. This follows from the observation that any Lagrangian theory possessing $\mathcal{N}=3$ supersymmetry, actually possesses $\mathcal{N}=4$ supersymmetry, and so there are no purely $\mathcal{N}=3$ supersymmetric Lagrangian theories. The curious thing is that while no purely $\mathcal{N}=3$ supersymmetric Lagrangian theory exists, ones with no Lagrangian manifesting the $\mathcal{N}=3$ supersymmetry exist. Notably, $\mathcal{N}=3$ SCFTs can be constructed in string theory as the theories living on D$3$-branes in the presence of S-folds, which can be thought of as generalizations of orientifolds to also include an action of the $SL(2,\mathbb{Z})$ symmetry of type IIB string theory. These were first considered in \cite{GER}, with the construction being later generalized in \cite{ATsf,GER1}.     

The non-Lagrangian\footnote{Here, we use the nomenclature 'non-Lagrangian' for theories with no known Lagrangian manifesting all their supersymmetry.} nature of these theories seemingly adds to the complexity of dealing with these types of theories, though that is not the whole story. After all, there are numerous known non-Lagrangian $\mathcal{N}=2$ SCFTs, see for instance \cite{MN,AW,Gai,CD,CD1}, and many of them have been well studied. However, most of these non-Lagrangian $\mathcal{N}=2$ SCFTs can be related to Lagrangian theories, either by gauging part of their global symmetries \cite{ArS}, or by going to special points on the Coulomb branch \cite{ArD,APSW}. The $\mathcal{N}=3$ SCFTs appear not to posses such relations, at least if one insists on maintaining $\mathcal{N}=2$ supersymmetry. This leads them to be somewhat isolated compared to many $\mathcal{N}=2$ non-Lagrangian theories, which adds to their inaccessibility. Note though, that $\mathcal{N}=3$ SCFTs can be reached, for instance, by mass deformations of non-Lagrangian $\mathcal{N}=2$ SCFTs \cite{ALLM,ZafTW}, so they are not completely isolated.  

While it seems difficult to access these theories using $\mathcal{N}=2$ Lagrangian theories, it might be possible to access them via $\mathcal{N}=1$ Lagrangian theories. Here the idea is to consider an $\mathcal{N}=1$ Lagrangian theories, and deform it by an operator such that it goes to the $\mathcal{N}=3$ SCFT. The deformation may be either marginal or relevant. The former case corresponds to the situation where the $\mathcal{N}=3$ SCFT and a certain $\mathcal{N}=1$ Lagrangian theory share the same conformal manifold. That is there is an underlying $\mathcal{N}=1$ SCFT with a conformal manifold, that at one point becomes the $\mathcal{N}=1$ Lagrangian theory at zero coupling and at another one becomes the $\mathcal{N}=3$ SCFT. This is similar to the relation between the $USp(2N)$ and $SO(2N+1)$ $\mathcal{N}=4$ super Yang-Mills theories, but now applied to a case where only $\mathcal{N}=1$ supersymmetry is preserved generally. The existence of such $\mathcal{N}=1$ SCFTs, that are built from Lagrangian theories with a conformal manifold containing the free point, was first considered in \cite{LS}, and further also in \cite{GKSTW}. The classification of such $\mathcal{N}=1$ SCFTs, where the gauge content of the Lagrangian theory at the free point consists of a simple Lie group, was carried out in \cite{RSZcl}. These types of relations but between these types of $\mathcal{N}=1$ SCFTs and $\mathcal{N}=2$ non-Lagrangian theories were studied in \cite{RZCM}. 

An alternative case is when the deformation is relevant. In these cases, we consider an $\mathcal{N}=1$ Lagrangian theory that flows in the IR to the $\mathcal{N}=3$ SCFT, potentially with the addition of decouped free fields. Various examples of such flows, involving an $\mathcal{N}=2$ non-Lagrangian theory as the end point, are known \cite{GRW,MS,MS1,AMSad,ASS,BGad,MNS,AMS,ZafE6}. The purpose of this article is to study these types of relations for $\mathcal{N}=3$ SCFTs. Specifically, we shall employ the strategy of \cite{ZafE6}, and use the anomalies of an $\mathcal{N}=3$ SCFT to conjecture an $\mathcal{N}=1$ asymptotically free $SU(2)\times SU(2)$ gauge theory that we postulate flows in the IR to the $\mathcal{N}=3$ SCFT plus a free decoupled chiral field. To be precise, the $\mathcal{N}=3$ SCFT possesses a conformal manifold on which only $\mathcal{N}=1$ SUSY is preserved, and we expect the $\mathcal{N}=1$ Lagrangian theory to flow to a generic point on this conformal manifold.

The specific $\mathcal{N}=3$ SCFT in question is the one with the moduli space $\mathbb{C}^3/\mathbb{Z}_3$, considered in \cite{NT}. This theory seems to be the simplest non-trivial purely $\mathcal{N}=3$ SCFT, on account of having the smallest central charges. The structure of the moduli space suggests that it has both a dimension three Coulomb and Higgs branch operators, and so we expect from the reasoning of \cite{RZCM}, that it also has an $\mathcal{N}=1$ only preserving conformal manifold. This can indeed be confirmed as we shall show in this article.  

 An interesting question then is how can we test this conjecture, as it involves strong coupling dynamics. A standard way to test such relations is using various RG invariant quantities, like anomalies or the superconformal index, that can be evaluated from the Lagrangian. However, these will be needed to be compared against those of the $\mathcal{N}=3$ SCFT. Unfortunately, the superconformal index for the $\mathcal{N}=3$ SCFT in question is not known\footnote{The superconformal index for $\mathcal{N}=3$ SCFTs engineered using S-folds was considered in \cite{ISin,AIin} using their gravity dual. However, the cases we consider here all have low rank so it is not clear how helpful the gravity dual is for these cases.}. To tackle this, we shall use the power of $\mathcal{N}=3$ supersymmetry. Specifically, $\mathcal{N}=3$ supersymmetry places some restrictions on the form of the superconformal index \cite{Evt1}, and we can check that the resulting index is consistent with $\mathcal{N}=3$ superconformal symmetry.

In fact, for the case we consider, we can do better. This comes about as the moduli space of this $\mathcal{N}=3$ SCFT is known, and imply the presence of various operator that span it. These are analogues of Higgs and Coulomb branch operators in $\mathcal{N}=2$ SCFTs. We can then form a guess for the superconformal index of this SCFT by taking the contributions of these multiplets. We shall see that the resulting index matches remarkably well with the index of the gauge theory, after the removal of the decoupled chiral. This not only gives evidence that the index can be that of an $\mathcal{N}=3$ SCFT, but also that it can be the index of this specific $\mathcal{N}=3$ SCFT.

Finally, we consider generalizations to other models. These considerations shall lead us to an additional $\mathcal{N}=1$ model with a conformal manifold, which we conjecture it shares with an $\mathcal{N}=3$ SCFT. The $\mathcal{N}=3$ SCFT in question is the one with moduli space $\mathbb{C}^9/G(3,3,3)$ introduced in \cite{ATsf}, where we use the notation used in the reference. This proposal can again be checked by comparing various RG invariant quantities, like anomalies and the superconformal index. For the latter we again compare against the index expected based on the spectrum of operators spanning the moduli space. Additionally, we also find a related $\mathcal{N}=1$ model, again with a conformal manifold, whose anomalies and superconformal index appear consistent with that of $\mathcal{N}=3$ SCFTs, but whose properties do not match any known $\mathcal{N}=3$ SCFT. It is not clear whether a relation to an $\mathcal{N}=3$ SCFT, like the ones considered here, also holds for this case.

The structure of this article is as follows. We begin in section \ref{sec:pre} with some preliminaries on some of the properties of $\mathcal{N}=3$ SCFTs, and how these can be used to help conjecture and test $\mathcal{N}=1$ models related to $\mathcal{N}=3$ SCFTs. These considerations will lead us to conjecture an $\mathcal{N}=1$ $SU(2)\times SU(2)$ gauge theory that we postulate flows in the IR to an $\mathcal{N}=3$ SCFT plus a free decoupled chiral field. We then move on to section \ref{sec:model} to study this model in detail. This allows us to uncover evidence in support of this conjecture. In section \ref{sec:gen} we consider generalizations, where we present another $\mathcal{N}=1$ model that we conjecture is dual to an $\mathcal{N}=3$ SCFT. This conjecture is then tested using similar methods. Finally, the appendix collects many properties of $4d$ superconformal theories. Specifically, the relation between their symmetries, anomalies and the decomposition of superconformal multiplets. These are then used in the course of this article.

\section{Preliminaries}
\label{sec:pre}

The first thing we would like to consider is the implications of $\mathcal{N}=3$ supersymmery on various RG invariant quantities. The purpose here is two fold. First, it can be used as a tool to help build $\mathcal{N}=1$ Lagrangian models that might be related to $\mathcal{N}=3$ theories. Second, it provides us with a method to test whether a given $\mathcal{N}=1$ Lagrangian model can be related to an $\mathcal{N}=3$ theory.

We will be particularly interested in two types of RG invariant quantities. The first is the superconformal index, while the second are the 't Hooft anomalies, particularly the central charges. We shall keep with the general strategy of \cite{RZCM,ZafE6}, and use the anomalies as a tool to help conjecture models and the index to test them. When we discuss the index, it will also be important to consider the structure of the moduli space of $\mathcal{N}=3$ theories. This is because the index of $\mathcal{N}=3$ theories is in general not known, and so we don't immediately have anything to compare with. However, the structure of the moduli space implies the existence of certain multiplets that can be used to formulate a conjecture for the superconformal index.  

This section, as well as the proceeding ones, rely on various properties of $\mathcal{N}=3$ theories. For convenience, these are summarized in the appendix.

\subsection{Index and the moduli space}
\label{N3mds}

We will want to test our proposals by matching the superconformal index. However, the superconformal index of $\mathcal{N}=3$ SCFTs is generally unknown. However, we do know the structure of the moduli space of many of these theories. From this, we can infer of the existence of various operators that are related to the moduli space. This allows us to determine some properties of these theories, and formulate a minimal conjecture for their superconformal indices by taking the contributions of only these multiplets. We shall consider the structure of the moduli space, and its implication on the spectrum of operators in this section.

We begin by considering the structure of the moduli space of $\mathcal{N}=3$ SCFTs. Here we shall concentrate only on a specific family of $\mathcal{N}=3$ SCFTs, that can be realized in string theory by D$3$-branes probing the so-called S-folds \cite{GER}. The moduli space of such theories have a special structure given by $(\mathbb{C}^3)^n/\Gamma$, where $\Gamma$ is a complex reflection group. As the name suggests, complex reflection groups are groups generated by complex reflections. These are the transformations of $\mathbb{C}^n$, spanned by the coordinates $z_i$, acting as $z_i \rightarrow \gamma z_i$, for some $i$ with $\gamma$ being a complex number obeying $|\gamma|=1$. In the special case where $\gamma$ is real, the groups are known as real reflections groups, and include the Weyl groups of Lie groups. For more information on complex reflection groups, oriented towards their application to the moduli space of $\mathcal{N}=3$ SCFTs, see \cite{ATsf,CC,BMRl,TyZ} and references therein. Here we will be interested in a specific family of complex reflection groups usually denoted as $G(k,p,n)$. These are defined by their action on $\mathbb{C}^n$, and generated by the permutations of the $z_i$ coordinates, which are real reflections, and the transformations:

\be
(z_1 , z_2 , ... , z_n) \rightarrow (e^{\frac{2\pi a_1 i}{k}} z_1 , e^{\frac{2\pi a_2 i}{k}} z_2 , ... , e^{\frac{2\pi a_n i}{k}} z_n) ,
\ee  
for all $a_i$'s obeying $a_1 + a_2 + ... + a_n = m p$, for some integer $m$. For the $\mathcal{N}=3$ SCFTs we considered here, $\Gamma$ is known to be of type $G(k,p,n)$ with some restrictions. Notably, we must have that $k=1,2,3,4$ or $6$, with $k=1,2$ leading to $\mathcal{N}=4$ SCFTs. These are necessary so that $G(k,p,n)$ is crystallographic, that is it preserves a lattice, which we can identify with the electric-magnetic charge lattice that this theory has on the Coulomb branch. Additionally, there are some restrictions on $p$, see \cite{ATsf} for the details. It should also be noted that the structure of the moduli space of $\mathcal{N}=3$ SCFTs might be more varied, see \cite{ABMm1,ABMm2} for some examples and discussions.

Let us consider the operators spanning the moduli space. For this it is convenient to introduce complex coordinates $z_i$ and $\bar{z}_i$ for $i=1,...,n$, where each pair of $z_i$ and its conjugate span a $\mathbb{C}^3$. The $z_i$ coordinates are taken to transform in the fundamental of the $SU(3)$ part of the $\mathcal{N}=3$ R-symmetry and with $U(1)$ charge $2$. The $\mathcal{N}=3$ R-symmetry then acts on the moduli space through this action on the coordinates. The group $G(k,p,n)$, which we orbifold by, acts on the $n$ $z_i$ coordinates in the same manner as previously discussed, though now each $z_i$ is a triplet of complex numbers. In the field theory, these coordinates represent scalar fields whose vevs parametrize the moduli space, and for the special case of $\mathcal{N}=4$ are just the six real scalars in the vector multiplet. 

These sit in the short multiplets $B_1 \bar{B}_1 [0,0]^{(1,0;2)}_{1}$, and $B_1 \bar{B}_1 [0,0]^{(0,1;-2)}_{1}$ for the complex conjugate \cite{ABMm2}. Here we use the notations of \cite{CID}. The two letters represent the shortening conditions obeyed by the operator with respect to $Q$ and $\bar{Q}$. The remaining terms represent the charges of the ground state under the Lorentz group, dilatation symmetry and R-symmetry, see the appendix for details. Together these form the $\mathcal{N}=3$ vector multiplet, which is the same as the $\mathcal{N}=4$ vector multiplet.

While the basic coordinates are given by $z_i$ and $\bar{z}_i$, generically these are not 'gauge' invariant under the modded group $\Gamma$ and need to be combined to form invariants. To illustrate this, it is convenient to consider the example of $\mathcal{N}=4$ super-YM. Here the scalars in $z$ and $\bar{z}$ combine to form the six scalars in the $\mathcal{N}=4$ vector multiplets, whose vevs span the moduli space. The structure of the moduli space then, for an $\mathcal{N}=4$ gauge theory with gauge group $G$, is very familiar. The six scalars are in the adjoint of $G$, but can be simultaneously diagonalized in the vacuum and so the latter is parametrized by a $rank(G)$ collection of six scalars. These span the space $(\mathbb{C}^3)^{rank(G)}$ playing the role of $z_i$ and $\bar{z}_i$. However, not all choices are distinct. This comes about as we still need to take into account the actions of the gauge symmetry on these fields that maintain the diagonal choice. The latter are known to be just the Weyl group of the gauge group $G$, $\mathcal{W}_G$. This gives the moduli space $(\mathbb{C}^3)^{rank(G)}/\mathcal{W}_G$, which is a special case of $(\mathbb{C}^3)^{n}/\Gamma$ when $\Gamma$ is a crystallographic real reflection group.

The fields spanning the moduli space are then given by combinations invariant under $\mathcal{W}_G$. This reflects the fact that the diagonal entries in the adjonit scalar fields in the vector multiplets are not gauge invariant. The operators that actually span the moduli space then are various gauge invarint combinations of the scalars in the vector multiplets. When the latter are written in terms of their diagonal entries, they precisely provide a combination of $z_i$ and $\bar{z}_i$ that are invariant under $\mathcal{W}_G$. A similar story happens also for the $\mathcal{N}=3$ SCFTs we consider here, but with the modded group not necessary a Weyl group. We shall next illustrate this with some specific examples. We shall concentrate here only on moduli spanning operators of the shortest type, $B_1 \bar{B}_1$.

\subsubsection*{$\mathcal{N}=4$ $U(N)$ theory}    

Let us begin by considering the case when $\Gamma=G(1,1,N)=S_N$, that is the symmetric group. The most well well known case featuring this space is the $\mathcal{N}=4$ $U(N)$ super Yang-Mills theory. The coordinates of the moduli space are given by $z_i$ and $\bar{z}_i$ for $i=1...,N$, and the group $G(1,1,N)$ acts by permutations of the $z_i$ coordinates. 

The function parametrizing the moduli space are then given by combinations of $z_i$ and $\bar{z}_i$ invariant under permutations. To consider this, it is convenient to look for invariants made from increasing number of fields. First, we can consider invariants made from a single coordinate. There are precisely two of these, given by:

\be
z_1 + z_2 + ... + z_N \; , \; \bar{z}_1 + \bar{z}_2 + ... + \bar{z}_N \nonumber .
\ee 

The first gives the multiplet $B_1 \bar{B}_1 [0,0]^{(1,0;2)}_{1}$, and the second the multiplet $B_1 \bar{B}_1 [0,0]^{(0,1;-2)}_{1}$. Together they form the $\mathcal{N}=4$ multiplet $B_1 \bar{B}_1 [0,0]^{(0,1,0)}_{1}$. In the Lagrangian description, this describes the field given by $Tr(\phi)$, where we use $\phi$ for the three adjoint chiral fields in the theory.   

Next, we can consider invariants made from two coordinates. There are six of these, given by:

\bea
& & z^2_1 + z^2_2 + ... + z^2_N \; , \; z_1 \bar{z}_1 + z_2 \bar{z}_2 + ... z_N \bar{z}_N \; , \; \bar{z}^2_1 + \bar{z}^2_2 + ... + \bar{z}^2_N \; , \nonumber \\ \nonumber & & z_1 z_2 + z_1 z_3 + ... + z_{N-1} z_N \; , \; z_1 \bar{z}_2 + z_1 \bar{z}_3 + ... z_N \bar{z}_{N-1} \; , \; \bar{z}_1 \bar{z}_2 + \bar{z}_1 \bar{z}_3 + ... + \bar{z}_{N-1} \bar{z}_N.
\eea

Here the first three terms exist for every $N$, while the last three exist only if $N>1$. These terms gives two copies of the multiplets\footnote{We expect the combination of coordinates made from the product of $k$ $z_i$ coordinates and $l$ $\bar{z}_i$ coordinates to be associated with the multiplet $B_1 \bar{B}_1 [0,0]^{(k,l;2(k-l))}_{k+l}$. This comes as we expect this combination to be the ground state of a superconformal multiplet obeying the same shortening conditions. As it is a product of scalar fields, it should itself be a scalar with dimension and $U(1)$ R-charge given by the sum of those of its constitutes. Furthermore, since the basic coordinates are scalars, the $SU(3)$ representations needs to be multiplied symmetrically, and we do not contact $SU(3)$ indices between $z_i$ and $\bar{z}_i$. These leads to the $SU(3)$ representation of the product being $[k,l]$. However, we do note that it might be possible to achieve the symmetric product by using a mixed product in both the $i$ and $SU(3)$ indices of the $z_i$ coordinates. These should then give multiplets of type longer than $B_1 \bar{B}_1$, and we will not consider these here.} $B_1 \bar{B}_1 [0,0]^{(2,0;4)}_{2}$, $B_1 \bar{B}_1 [0,0]^{(1,1;0)}_{2}$ , $B_1 \bar{B}_1 [0,0]^{(0,2;-4)}_{2}$ and together form two copies of the $\mathcal{N}=4$ multiplets $B_1 \bar{B}_1 [0,0]^{(0,2,0)}_{2}$. In the Lagrangian description these describe the fields given by $Tr(\phi)^2$ and $Tr(\phi^2)$.

Next, we can consider invariants made from three coordinates. There are fourteen of these, given by:

\bea
& & z^3_1 + z^3_2 + ... + z^3_N \; , \; z^2_1 \bar{z}_1 + z^2_2 \bar{z}_2 + ... z^2_N \bar{z}_N \; , \; z_1 \bar{z}^2_1 + z_2 \bar{z}^2_2 + ... z_N \bar{z}^2_N \; , \; \bar{z}^3_1 + \bar{z}^3_2 + ... + \bar{z}^3_N \; , \nonumber \\ \nonumber & & z^2_1 z_2 + z^2_1 z_3 + ... + z^2_N z_{N-1} \; , \; z^2_1 \bar{z}_2 + z^2_1 \bar{z}_3 + ... z^2_N \bar{z}_{N-1} \; , \; z_1 \bar{z}_1 z_2 + z_1 \bar{z}_1 z_3 + ... z_N \bar{z}_N z_{N-1} \; , \\ \nonumber & & z_1 \bar{z}^2_2 + z_1 \bar{z}^2_3 + ... z_N \bar{z}^2_{N-1} \; , \; \bar{z}_1 \bar{z}_2 z_2 + \bar{z}_1 \bar{z}_3 z_3 + ... \bar{z}_N \bar{z}_{N-1} z_{N-1} \; , \; \bar{z}^2_1 \bar{z}_2 + \bar{z}^2_1 \bar{z}_3 + ... + \bar{z}^2_N \bar{z}_{N-1} \; , \\ \nonumber & & z_1 z_2 z_3 + z_1 z_2 z_4  + ... + z_{N-2} z_{N-1} z_N \; , \; z_1 z_2 \bar{z}_3 + z_1 z_2 \bar{z}_4 + ... z_{N-1} z_N \bar{z}_{N-2} \; , \\ \nonumber & & z_1 \bar{z}_2 \bar{z}_3 + z_1 \bar{z}_2 \bar{z}_4 + ... + z_N \bar{z}_{N-1} \bar{z}_{N-2} \; , \; \bar{z}_1 \bar{z}_2 \bar{z}_3 + \bar{z}_1 \bar{z}_2 \bar{z}_3 + ... + \bar{z}_{N-2} \bar{z}_{N-1} \bar{z}_N.
\eea

Here the first four terms exist for every $N$, the next six exist only if $N>1$, and the last four exist only if $N>2$. These terms give three copies of the multiplets $B_1 \bar{B}_1 [0,0]^{(3,0;6)}_{3}$, $B_1 \bar{B}_1 [0,0]^{(0,3;-6)}_{3}$ and four copies of the multiplets $B_1 \bar{B}_1 [0,0]^{(2,1;2)}_{3}$, $B_1 \bar{B}_1 [0,0]^{(1,2;-2)}_{3}$. Together these form three copies of the $\mathcal{N}=4$ multiplets $B_1 \bar{B}_1 [0,0]^{(0,3,0)}_{3}$, and one copy of the $\mathcal{N}=4$ multiplet $B_1 \bar{B}_1 [0,0]^{(1,1,1)}_{3}$. The last multiplet also requires other $\mathcal{N}=3$ multiplets longer than of type $B_1 \bar{B}_1$, and so would not be considered here. In the Lagrangian description these describe the fields given by $Tr(\phi)^3$, $Tr(\phi^2)Tr(\phi)$ and $Tr(\phi^3)$. For the second case we have that $(0,2,0)_{SU(4)} \otimes (0,1,0)_{SU(4)}\rightarrow (0,3,0)_{SU(4)} \oplus (1,1,1)_{SU(4)}\oplus ...$, with the second term in the decomposition leading to the presence of the $B_1 \bar{B}_1 [0,0]^{(1,1,1)}_{3}$ multiplet.

The general structure now becomes apparent. The $\mathcal{N}=3$ multiplets are given by a choice of a combination of $k$ $z_i$ coordinates and $l$ $\bar{z}_i$ coordinates, and give the $\mathcal{N}=3$ multiplet $B_1 \bar{B}_1 [0,0]^{(k,l;2(k-l))}_{k+l}$. Combinations differing by changing some of the $z_i$ coordinates to their conjugates combine to form the $\mathcal{N}=4$ multiplet $B_1 \bar{B}_1 [0,0]^{(0,k+l,0)}_{k+l}$, with the remaining ones forming more general $\mathcal{N}=4$ $B_1 \bar{B}_1$ multiplets. Multipets are differentiated by the number of coordinates sharing the same index. In the Lagrangian theory, we can order the invariants by the the number of traces they use, and there are precisely $N$ basic single trace combinations. This is mapped to the fact that there are only $N$ combinations of fields one can build without repeating the index of the coordinates.  

\subsubsection*{$\mathcal{N}=4$ $USp(2N)$ theory}

As the next example, we consider the case when $\Gamma=G(2,1,N)$, which is the Weyl group of $USp(2N)$ and $SO(2N+1)$. Naturally, this case is featured by the corresponding $\mathcal{N}=4$ super Yang-Mills theories. This group is generated by permutations of the $N$ coordinates, as well as by a reflection on any of them. The latter imply that only the combinations $z^2_i$, $z_i \bar{z}_i$ and $\bar{z}^2_i$ can be used to build invariants. The invariants then follow the same pattern as in the $U(N)$ case, as permutation invariant combinations of these. For example, with two fields we have the following combinations:

\be \nonumber
z^2_1 + z^2_2 + ... + z^2_N \; , \; z_1 \bar{z}_1 + z_2 \bar{z}_2 + ... + z_N \bar{z}_N \; , \; \bar{z}^2_1 + \bar{z}^2_2 + ... + \bar{z}^2_N .
\ee

These give the multiplets $B_1 \bar{B}_1 [0,0]^{(2,0;4)}_{2}$, $B_1 \bar{B}_1 [0,0]^{(1,1;0)}_{2}$ , $B_1 \bar{B}_1 [0,0]^{(0,2;-4)}_{2}$ and together form the $\mathcal{N}=4$ multiplet $B_1 \bar{B}_1 [0,0]^{(0,2,0)}_{2}$. In the Lagrangian description this describes the field given by $Tr(\phi^2)$.

Likewise, with four fields we have the combinations:

\bea \nonumber
& & z^4_1 + z^4_2 + ... + z^4_N \; , \; z^3_1 \bar{z}_1 + z^3_2 \bar{z}_2 + ... + z^3_N \bar{z}_N \; , \; z^2_1 \bar{z}^2_1 + z^2_2 \bar{z}^2_2 + ... + z^2_N \bar{z}^2_N \; , \\ \nonumber & & z_1 \bar{z}^3_1 + z_2 \bar{z}^3_2 + ... + z_N \bar{z}^3_N \; , \; \bar{z}^4_1 + \bar{z}^4_2 + ... + \bar{z}^4_N \; , \; z^2_1 z^2_2 + z^2_1 z^2_3 + ... + z^2_{N-1} z^2_N \; , \\ \nonumber & & z^2_1 z_2 \bar{z}_2 + z^2_1 z_3 \bar{z}_3 + ... + z^2_N z_{N-1} \bar{z}_{N-1} \; , \; z_1 z_2 \bar{z}_1 \bar{z}_2 + z_1 z_3 \bar{z}_1 \bar{z}_3 + ... + z_{N-1} z_N \bar{z}_{N-1} \bar{z}_N \; , \\ \nonumber & & z^2_1 \bar{z}^2_2 + z^2_1 \bar{z}^2_3 + ... + z^2_N \bar{z}^2_{N-1} \; , \; z_1 \bar{z}_1 \bar{z}^2_2 + z_1 \bar{z}_1 \bar{z}^2_3 + ... + z_N \bar{z}_N \bar{z}^2_{N-1} \; , \; \bar{z}^2_1 \bar{z}^2_2 + \bar{z}^2_1 \bar{z}^2_3 + ... + \bar{z}^2_{N-1} \bar{z}^2_N .
\eea

Here the last six only exist if $N>1$. These give two copies of the multiplets $B_1 \bar{B}_1 [0,0]^{(4,0;8)}_{4}$, $B_1 \bar{B}_1 [0,0]^{(3,1;4)}_{4}$ , $B_1 \bar{B}_1 [0,0]^{(1,3;-4)}_{4}$ , $B_1 \bar{B}_1 [0,0]^{(0,4;-8)}_{4}$, and three copies of the multiplet $B_1 \bar{B}_1 [0,0]^{(2,2;0)}_{4}$. Together these form two copies of the $\mathcal{N}=4$ multiplet $B_1 \bar{B}_1 [0,0]^{(0,4,0)}_{4}$, and one copy of the $\mathcal{N}=4$ multiplet $B_1 \bar{B}_1 [0,0]^{(2,0,2)}_{4}$. The last multiplet also requires other $\mathcal{N}=3$ multiplets longer than of type $B_1 \bar{B}_1$, and so would not be considered here. In the Lagrangian description these describe the fields given by $Tr(\phi^2)^2$ and $Tr(\phi^4)$. For the first case we have that $(0,2,0)_{SU(4)} \otimes_{symmetric} (0,2,0)_{SU(4)}\rightarrow (0,4,0)_{SU(4)} \oplus (2,0,2)_{SU(4)}\oplus ...$, with the second term in the decomposition leading to the presence of the $B_1 \bar{B}_1 [0,0]^{(2,0,2)}_{4}$ multiplet.

We can again continue to go to higher orders, though the structure should now be apparent. 

\subsubsection*{$\mathcal{N}=3$ SCFT with $\Gamma=G(3,1,1)$}

As the next example, we consider the case when $\Gamma=G(3,1,1)=\mathbb{Z}_3$. This case corresponds to a genuine $\mathcal{N}=3$ SCFT. Here the moduli space has a single coordinate $z_1$ on which $\Gamma$ acts as $z_1 \rightarrow e^{\frac{2\pi i}{3}} z_1$. It is straightforward to see that there are three basic invariants we can build:

\be \nonumber
z^3_1 \; , \; z_1 \bar{z}_1 \; , \; \bar{z}^3_1 .
\ee

These give the $\mathcal{N}=3$ multiplets $B_1 \bar{B}_1 [0,0]^{(3,0;6)}_{3}$, $B_1 \bar{B}_1 [0,0]^{(1,1;0)}_{2}$, and $B_1 \bar{B}_1 [0,0]^{(0,3;-6)}_{3}$. All other invariant can be written as products of these basic combinations and so correspond to products of the three basic multiplets. It should be noted, though, that each combination appears only once, which implies relations between the various multiplets. For instance, there is only one $z^3_1 \bar{z}^3_1$ operators implying that the product of the $B_1 \bar{B}_1 [0,0]^{(3,0;6)}_{3}$ and $B_1 \bar{B}_1 [0,0]^{(0,3;-6)}_{3}$ multiplets should be equal to the cubic product of the $B_1 \bar{B}_1 [0,0]^{(1,1;0)}_{2}$ multiplet \cite{NT,LLMM,BPP}.

\subsubsection*{$\mathcal{N}=3$ SCFT with $\Gamma=G(3,3,3)$}

As our final example, we consider the case of $\Gamma=G(3,3,3)$. This case corresponds to a genuine $\mathcal{N}=3$ SCFT. The group structure here is more complicated, consisting of permutations of the three coordinates, $(z_1,z_2,z_3)$, as well as the transformation $(z_1,z_2,z_3)\rightarrow (e^{\frac{2\pi i k}{3}} z_1, e^{\frac{2\pi i l}{3}} z_2, e^{\frac{2\pi i m}{3}} z_3)$, where $l+k+m=3n$ for some integer $n$. In this case the spectrum of invariants is much richer. The first invariant is made of two fields:

\be \nonumber
z_1 \bar{z}_1 + z_2 \bar{z}_2 + z_3 \bar{z}_3 .
\ee

This gives the multiplet $B_1 \bar{B}_1 [0,0]^{(1,1;0)}_{2}$, which corresponds to the energy-momentum tensor of the $\mathcal{N}=3$ SCFT.  

At the three field level we have four different invariants:

\be \nonumber
z^3_1 + z^3_2 + z^3_3 \; , \; z_1 z_2 z_3 \; , \; \bar{z}^3_1 + \bar{z}^3_2 + \bar{z}^3_3 \; , \; \bar{z}_1 \bar{z}_2 \bar{z}_3.
\ee

These give two copies of the multiplets $B_1 \bar{B}_1 [0,0]^{(3,0;6)}_{3}$ and $B_1 \bar{B}_1 [0,0]^{(0,3;-6)}_{3}$. 

At the four fields level we again have four different invariants:

\be \nonumber
z^2_1 \bar{z}^2_1 + z^2_2 \bar{z}^2_2 + z^2_3 \bar{z}^2_3 \; , \; z_1 z_2 \bar{z}_1 \bar{z}_2 + z_1 z_3 \bar{z}_1 \bar{z}_3 + z_2 z_3 \bar{z}_2 \bar{z}_3 \; , \; z^2_1 \bar{z}_2 \bar{z}_3 + z^2_2 \bar{z}_1 \bar{z}_3 + z^2_3 \bar{z}_1 \bar{z}_2 \; , \; \bar{z}^2_1 z_2 z_3 + \bar{z}^2_2 z_1 z_3 + \bar{z}^2_3 z_1 z_2 \; .
\ee

These give four copies of the multiplet $B_1 \bar{B}_1 [0,0]^{(2,2;0)}_{4}$. Note that one copy can be identified with the square of the energy-momentum tensor multiplet, but the rest are additional multiplets. 

We can continue and consider multiplets made from more fields, but we will not need that in this article.

\subsection{Anomalies}

Next, we consider the 't Hooft anomalies of $\mathcal{N}=3$ SCFTs. First, there are the a and c central charges, which for $\mathcal{N}=3$ SCFTs must be equal \cite{AoMe}. It is convenient to write these as:

\be
a = c = \frac{n_v}{4} .
\ee 

For Lagrangian theories, like $\mathcal{N}=4$ theories, $n_v$ is the number of vector multiplets. For non-Lagrangian $\mathcal{N}=2$ theories, $n_v$ can in many cases be obtained from the dimensions of the independent Coulomb branch operators, $\Delta_i$, using \cite{ShTa}:

\be \label{env}
n_v = 4(2a-c)=\sum_i (2\Delta_i - 1) ,
\ee 
where the sums runs over all the independent Coulomb branch operators. This relation is known to hold for many $\mathcal{N}=2$ SCFTs, though it is not clear whether the assumption used in \cite{ShTa} to derive it are satisfied for the $\mathcal{N}=3$ SCFTs we consider here, see the discussion in \cite{NT}. It should be noted that this relation is known to fail when the gauge group has disconnected components. For instance, it is possible to engineer $\mathcal{N}=3$ SCFTs by gauging a discrete symmetry of $\mathcal{N}=4$ SCFTs, see \cite{AMdc,BPP,Evt2}. In these cases, \eqref{env} will not be obeyed, and instead a and c will be equal to those of the underlying $\mathcal{N}=4$ SCFT. We shall for the most part not consider these types of $\mathcal{N}=3$ SCFTs here.

Instead, we will mostly concentrate on the $\mathcal{N}=3$ SCFTs that can be build using S-folds. These were originally introduced in \cite{GER}, and further studied in \cite{ATsf}. For the rank $1$ case, the central charges were originally evaluated using \eqref{env} in\cite{NT}, and the results obtained there match the results obtained from other methods, notably the detailed study of the Coulomb branch geometry of rank $1$ $\mathcal{N}=2$ theories performed in \cite{ALLM00,ALLM0,ALLM,ALLM1}. As the higher rank cases can be thought of as generalizations of the rank $1$ theories, we shall assume that this relation indeed holds also for them.

The structure of the moduli space, and therefore the dimensions of the independent Coulomb branch operators, are known for this class of $\mathcal{N}=3$ SCFTs, allowing us to determine the central charges for these cases using \eqref{env}. We can then use the anomalies as a tool to look for models that can flow to $\mathcal{N}=3$ SCFTs. The idea is that we seek an $\mathcal{N}=1$ Lagrangian model made from $n^{(1)}_v$ vector multiplets and $n^{(\frac{2}{3})}_c + n^{(\frac{1}{3})}_c$ chiral fields. We further demand that there is a non-anomalous $U(1)_R$ symmetry such that $n^{(\frac{2}{3})}_c$ of the chirals have R-charge of $\frac{2}{3}$ and $n^{(\frac{1}{3})}_c$ of them have R-charge of $\frac{1}{3}$. We take this specific choice as we wish to maintain the possibility that the R-charges of all gauge invariant BPS operators be a multiplet of $\frac{2}{3}$, which can be achieved if the gauge symmetry is such that invariants can only be made from a pair of fields with R-charge of $\frac{1}{3}$. We could take $n^{(\frac{1}{3})}_c=0$ in which case it is guaranteed that the R-charges of all gauge invariant BPS operators be a multiplet of $\frac{2}{3}$, but in that case the theory must be conformal at weak coupling\footnote{This follows as in that case the condition for the $U(1)$ R-symmetry to be non-anomalous is equal to the condition that $1$-loop beta function vanishes at zero coupling.}. By allowing fields with R-charge $\frac{1}{3}$ we are extending our possible models to include RG flows. 

It should be noted that naively it is possible to have BPS operators with non-integer dimension in $\mathcal{N}=3$ SCFTs, so these restrictions may not be necessary. This can be seen, for instance, in the structure of short representations of the $\mathcal{N}=3$ superconformal summarized in the appendix. However, we are not aware of any example of an $\mathcal{N}=3$ SCFT containing a BPS operator with non-integer dimension. As was discussed in the previous subsection, the dimension of the BPS operators spanning the moduli space of $\mathcal{N}=3$ SCFTs appear to be mostly integer, see \cite{AoMe,ABMm1,ABMm2} for some discussions on these issues. As a result, it is sensible to try to limit our search space by insisting on this.

We can constrain the values of $n^{(1)}_v$, $n^{(\frac{2}{3})}_c$ and $n^{(\frac{1}{3})}_c$ by demanding that the resulting anomalies of the $\mathcal{N}=1$ R-symmetry match the anomalies of the target $\mathcal{N}=3$ SCFT. In general there might be some gauge invariants with R-charge of $\frac{2}{3}$. These give free chiral fields that are expected to decouple. Say that the number of these fields is $n_{free}$. Then we must demand that the anomalies are consistent with those of the $\mathcal{N}=3$ SCFT plus the free fields. This gives the constraints: 

\be \label{anomrel}
\frac{n_v}{4} + \frac{n_{free}}{48} = \frac{1}{48} (n^{(\frac{2}{3})}_c - n^{(\frac{1}{3})}_c + 9 n^{(1)}_v) \; , \; \frac{n_v}{4} + \frac{n_{free}}{24} = \frac{1}{48} (2n^{(\frac{2}{3})}_c + n^{(\frac{1}{3})}_c + 6 n^{(1)}_v) .
\ee

We can then search for a solution to these constraints obeying several restrictions, notably, that the theory is non-anomalous and further that the R-symmetry giving these charges is anomaly free. Once such a solution is found, it can be subjected to more intricate tests. First, we must actually verify that the R-symmetry we found can be the superconformal R-symmetry. This may necessitates the introduction of superpotential terms so as to break other possible R-symmetries, essentially forcing the found R-symmetry to be the superconformal one. If this is indeed possible, we shall want to subject the proposed model to further tests, usually by comparing other RG invariant quantities.  

Before giving an example, we note that if we consider the $\mathcal{N}=3$ SCFT as an $\mathcal{N}=1$ SCFT, then from the $\mathcal{N}=1$ viewpoint, there is an $SU(2)\times U(1)$ global symmetry which is the commutant of the $\mathcal{N}=1$ $U(1)_R$ R-symmetry in the $\mathcal{N}=3$ $U(1)_R\times SU(3)$ R-symmetry. The $\mathcal{N}=3$ supersymmetry also constrains the anomalies involving this $SU(2)\times U(1)$ global symmetry, which in turn are also expressible in terms of $n_v$. In theories where these symmetries are manifest then we can also use these anomalies to test potential flows. However, in the cases we consider here, these will not be manifest in the Lagrangian theory so we will not consider these anomalies further here. 

Finally, we want to illustrate the search strategy, and the discussion done so far, with an example. This in fact is the main case that is studied in this article. For this, we consider one of the simplest non-trivial $\mathcal{N}=3$ SCFT, the one with rank $1$ and dimension three Coulomb branch operator. This $\mathcal{N}=3$ SCFT is relatively well studied, having been considered in \cite{NT,ALLM}. Since it has a one dimensional Coulomb branch spanned by an operator with dimension three, we see from \eqref{env} that $n_v = 2 \Delta - 1 = 5$ for this $\mathcal{N}=3$ SCFT. The moduli space in turn is known to be $\mathbb{C}^3/\mathbb{Z}_3$, and is one of the examples we did in the previous subsection.

We want to search for an $\mathcal{N}=1$ Lagrangian model that can flow to this $\mathcal{N}=3$ SCFT, employing the search strategy discussed here. For this we use \eqref{anomrel} to constrain the values of $n^{(\frac{2}{3})}_c$, $n^{(\frac{1}{3})}_c$ and $n^{(1)}_v$, where we note that as $n_{free}$ is the number of gauge invariants with R-charge $\frac{2}{3}$, it is not an independent parameter. These lead to the two constrains:

\be \label{mconst}
n^{(\frac{2}{3})}_c - n_{free} = 5(8 - n^{(1)}_v) \; , \; n^{(\frac{1}{3})}_c = 4(n^{(1)}_v - 5) .
\ee

All that is left now is to go over all possibilities, and see if we can find a consistent solution to \eqref{mconst}. Indeed, going over the possible cases, we find the following solution:

\be
n^{(1)}_v = 6 \; , \; n^{(\frac{1}{3})}_c = 4 \; , \; n^{(\frac{2}{3})}_c = 11 \; , \; n_{free} = 1 .
\ee  

The specific realization of this solution involves an $SU(2)\times SU(2)$ gauge theory, which gives six vector multiplets. The matter content includes a bifundamental chiral of R-charge $\frac{1}{3}$, which gives the four chiral fields with that R-charge. We also note that there is a single quadratic gauge invariant one can build from it, giving the correct $n_{free}$. Finally, we also have an adjoint chiral under one of the $SU(2)$ groups, let us call it $SU(2)_1$, a chiral in the doublet of $SU(2)_1$ and a chiral field in the $(\bold{2},\bold{3})$ of $SU(2)_1\times SU(2)_2$. We note that the resulting theory is non-anomalous and that the R-symmetry providing the assigned charges is anomaly free. This then gives a consistent solution to \eqref{anomrel}, which ensures that the central charges of the theory be equal to those of the $\mathcal{N}=3$ SCFT plus a free chiral field, if the R-symmetry we used is the superconformal R-symmetry. We next study this model in greather detail to see whether this can be enforced. 

Before that, we want to note that the bifundamental is the only field sensitive to the $\mathbb{Z}_2$ center of $SU(2)_2$. As a result, all gauge invariants must be made from an even number of bifundamentals. As this is the only field with R-charge of $\frac{1}{3}$, this ensures that the R-charges of all gauge invariants be multiples of $\frac{2}{3}$. 

\section{The model}
\label{sec:model}

\begin{figure}
\center
\includegraphics[width=0.50\textwidth]{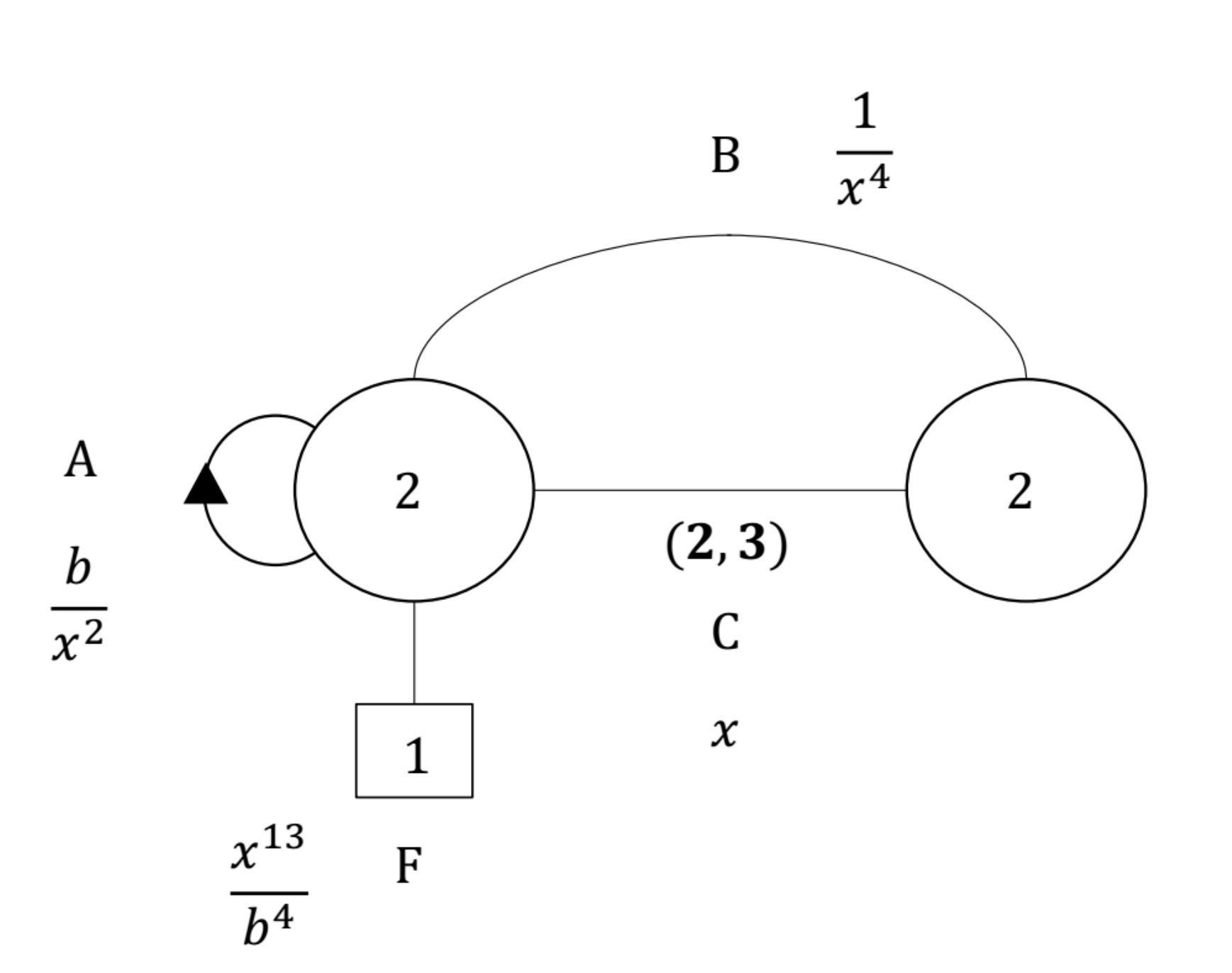} 
\caption{The quiver diagram of the proposed $\mathcal{N}=1$ model. As usual in quiver diagrams, circles represent gauge symmetries while boxes represent flavor ones. Next to each matter field is written its symbol as well as its charges under the global non-R symmetries. Finally we note that the $(\bold{2},\bold{3})$ below the line associated with the field $C$ gives it representations under the two $SU(2)$ groups.}
\label{quiverN3}
\end{figure}

Let us analyze in detail the proposed dual of the rank $1$ $\mathcal{N}=3$ SCFT with dimension three Coulomb branch operator. The matter content consists of an $SU(2)\times SU(2)$ gauge theory with chiral fields in various representations. The list of chiral fields, with their charges under the gauge and flavor symmetries can be seen in table \ref{N33MC}. This is also summarized in the quiver description in figure \ref{quiverN3}. This theory has non-anomalous global symmetry given by $U(1)_b \times U(1)_x \times U(1)_R$. Particularly, it has a non-anomalous R-symmetry under which the bifundamental chiral field, $B$, has R-charge $\frac{1}{3}$, while the rest have free R-charge $\frac{2}{3}$.

\begin{table}[htbp]
\begin{center}
	\begin{tabular}{|c||c|c|c|c|c|}
		\hline
		Field & $SU(2)_1$ & $SU(2)_2$ & $U(1)_b$ & $U(1)_x$ & $U(1)_{R}$\\
		\hline
		$A$ & $\boldsymbol{3}$ & $\boldsymbol{1}$ & $1$ & $-2$ & $\frac{2}{3}$ \\
		
		$C$ & $\boldsymbol{2}$ & $\boldsymbol{3}$ & $0$ & $1$ & $\frac{2}{3}$ \\
		
		$F$ & $\boldsymbol{2}$ & $\boldsymbol{1}$ & $-4$ & $13$ & $\frac{2}{3}$ \\
		
		$B$ & $\boldsymbol{2}$ & $\boldsymbol{2}$ & $0$ & $-4$ & $\frac{1}{3}$ \\
		\hline
	\end{tabular}
\end{center}
\caption{The gauge and matter content of the proposed dual of the rank $1$ $\mathcal{N}=3$ SCFT with dimension three Coulomb branch operator. The field entry stands for the symbol we used for that chiral field, where the rest of the entries give the charges of that chiral fields under the gauge and flavor symmetries.} 
\label{N33MC}
\end{table}

Without a superpotential this R-charge is not the superconformal one. This occurs due to mixing with $U(1)_b$ and $U(1)_x$. We can determine the mixing using the technique of a maximization \cite{Amax}. Specifically, we define the trial R-symmetry $U(1)^{trial}_R = U(1)_R + \gamma_b U(1)_b + \gamma_x U(1)_x$ and compute:

\bea
Tr \left((U(1)^{trial}_R)^3 \right) & = & 6 + 3 (-\frac{1}{3} + \gamma_b - 2\gamma_x)^3 + 6 (-\frac{1}{3} + \gamma_x)^3 \nonumber \\ & + & 2 (-\frac{1}{3} -4 \gamma_b + 13\gamma_x)^3 + 4 (-\frac{2}{3} - 4\gamma_x)^3, \nonumber \\ Tr \left(U(1)^{trial}_R \right) & = & 6 + 3 (-\frac{1}{3} + \gamma_b - 2\gamma_x) + 6 (-\frac{1}{3} + \gamma_x) \nonumber \\ & + & 2 (-\frac{1}{3} -4 \gamma_b + 13\gamma_x) + 4 (-\frac{2}{3} - 4\gamma_x) . \label{Rtraces}
\eea 

We then use

\be
a = \frac{3}{32} \left(3 Tr(U(1)^3_R) - Tr(U(1)_R)\right) \quad , \quad c = \frac{1}{32} \left(9 Tr(U(1)^3_R) - 5 Tr(U(1)_R)\right) \label{aandc}
\ee

to compute the central charges, notably $a$. The statement of a maximization is that the superconformal R-symmetry is the one that maximizes $a$. We can use this to fix the coefficients $\gamma_b$ and $\gamma_x$ that determines the mixing between $U(1)_R$, $U(1)_b$ and $U(1)_x$. We find that $\gamma_b \approx -0.224$, $\gamma_x \approx -0.071$. 

This result should give the superconformal R-symmetry in the IR. However, this may fail if there are accidental $U(1)$ symmetries arising in the IR, that can then mix with the R-symmetry \cite{KPS}. It is generally very difficult to completely rule this out, but one consistency check one can do is to verify that all BPS operator dimensions, expected from their superconformal R-charge, are above the unitarity bound. Operators below the unitarity bound are inconsistent in SCFTs suggesting that the IR theory cannot be an SCFT with that superconformal R-symmetry. It is generally thought that in these cases the violating operators decouple, and become free fields along the flow. This leads to additional symmetries that then mix with the R-symmetry. Returning to the case at hand, with the R-symmetry we found all gauge invariant operators are above the unitarity bound and so it seems plausible that this theory flows to an interacting SCFT. Specifically, the gauge invariant $B^2$ has dimension bigger than $1$. 

The resulting superconformal R-charges differ from those of the $\mathcal{N}=3$ theory we seek, as $U(1)_R$ is not the superconformal R-symmetry. We can try to force $U(1)_R$ to be the superconformal R-symmetry by introducing superpotentials that break the additional $U(1)$ groups that mixed with it. As we do not want $U(1)_R$ to be broken, these superpotentials must have R-charge $2$. It is also important that the superpotentials be relevant with respect to the fixed point we found previously, as if these are irrelevant we expect the theory to flow back to the same fixed point. In that case the superpotential interaction should flow to zero in the IR, leading to the symmetry broken by the superpotential reemerging at low-energies. Therefore we next consider the possible superpotentials terms with charge $2$ under $U(1)_R$.

Looking at the quiver, we find three possible terms: $A C^2$, $A F^2$, and $B^2 C F$. In the first term, two $C$ fields are contracted symmetrically so that the triplet indices under $SU(2)_2$ are contracted to form a singlet, while the doublet indices under $SU(2)_1$ are contracted so as to form an $SU(2)_1$ triplet. This can them be contracted with the $SU(2)_1$ triplet indices in $A$ to form a singlet. Similarly, in the second term,  the doublet indices of $F$ are contracted so as to form a triplet, that is then cotracted with the triplet index in $A$ to form a singlet. Finally, in the third term, we contract $B$ symmetrically so as to form a bi-triplet under the two $SU(2)$ groups, and similarly contract the $SU(2)_1$ doublet indices of $F$ with the $SU(2)_1$ doublet indices of $C$ to form a bi-triplet. These are then contracted together to form a gauge invariant.

We next need to consider whether these are relevant or irrelevant at the IR fixed point, by looking at their R-charges. This follows as in $\mathcal{N}=1$ SCFTs, the dimension of chiral operators, like superpotential terms, is determined by their R-charge. Specifically, chiral operators with R-charge less than $2$ correspond to relevant operators, while ones with R-charge greather than $2$ correspond to irrelevant ones. The results for the charge are summarized in table \ref{MOdim}. We thus see that $A C^2$ and $A F^2$ are relevant, while $B^2 C F$ is irrelevant. We can then consider deforming the fixed point by turning on the relevant superpotential deformations. To mitigate potential problems due to changes in the relevancy of operators during the flow, it is convenient to introduce them in steps. This is done by first turning on only the most relevant one, performing a maximization and checking whether the other remains relevant also with respect to the new superconformal R-symmetry, and if so, then we can continue and introduce also the second term.    

\begin{table}[htbp]
\begin{center}
	\begin{tabular}{|c||c|c|c||c|}
		\hline
		Operator & $U(1)_b$ & $U(1)_x$ & $U(1)_{R}$ & $U(1)^{spcr}_R$ \\
		\hline
		$A C^2$ & $1$ & $0$ & $2$ & $\approx 1.776$ \\
		
		$A F^2$ & $-7$ & $24$ & $2$ & $\approx 1.864$ \\
		
		$B^2 C F$ & $-4$ & $6$ & $2$ & $\approx 2.47$ \\
		
		\hline
	\end{tabular}
\end{center}
\caption{Possible superpotential terms with $U(1)_R$ charge $2$. The last entry, $U(1)^{spcr}_R$, stands for the approximate R-charge under the $U(1)_R$ symmetry expected to be the superconformal one in the IR when there is no superpotential.} 
\label{MOdim}
\end{table}

Therefore, we consider deforming the fixed point corresponding to the theory with matter content as in figure \ref{quiverN3}, by the superpotential $A C^2$. This breaks $U(1)_b$, and as it is relevant, we expect a flow to a new IR theory. We can determine its superconformal R-symmetry using a maximization. We again define a trial R-symmetry $U(1)^{trial}_R = U(1)_R + \gamma_{x} U(1)_{x}$, and using a maximization we find: $\gamma_{x} = \frac{121-\sqrt{27001}}{3090} \approx -0.014$.

We can now repeat the analysis we did previously on this new putative fixed point. First, we confirm that there are no operators going below the unitarity bound. As a result, there is no contradiction with this theory being an SCFT with the found R-symmetry as the superconformal one, again up to the usual caveats regarding accidental symmetries. Next, we examine the other two superpotential terms with R-charge $2$, and find that they are now both relevant with respect to this fixed point. we can then turn any one of these on, and expect a flow to a new IR theory. This breaks $U(1)_x$, and as there are no other $U(1)$ groups remaining, $U(1)_R$ must be the superconformal R-symmetry at the IR, again baring the possibility of accidental symmetries. However, here we do have one field, $B^2$, that hits the unitarity bound, and so we expect it to decouple becoming a free chiral field. Note that this also implies an accidental symmetry acting only on it. However, as the R-charge of this field is locked to $\frac{2}{3}$, this additional symmetry will not mix with the R-symmetry. 

\begin{figure}
\center
\includegraphics[width=0.75\textwidth]{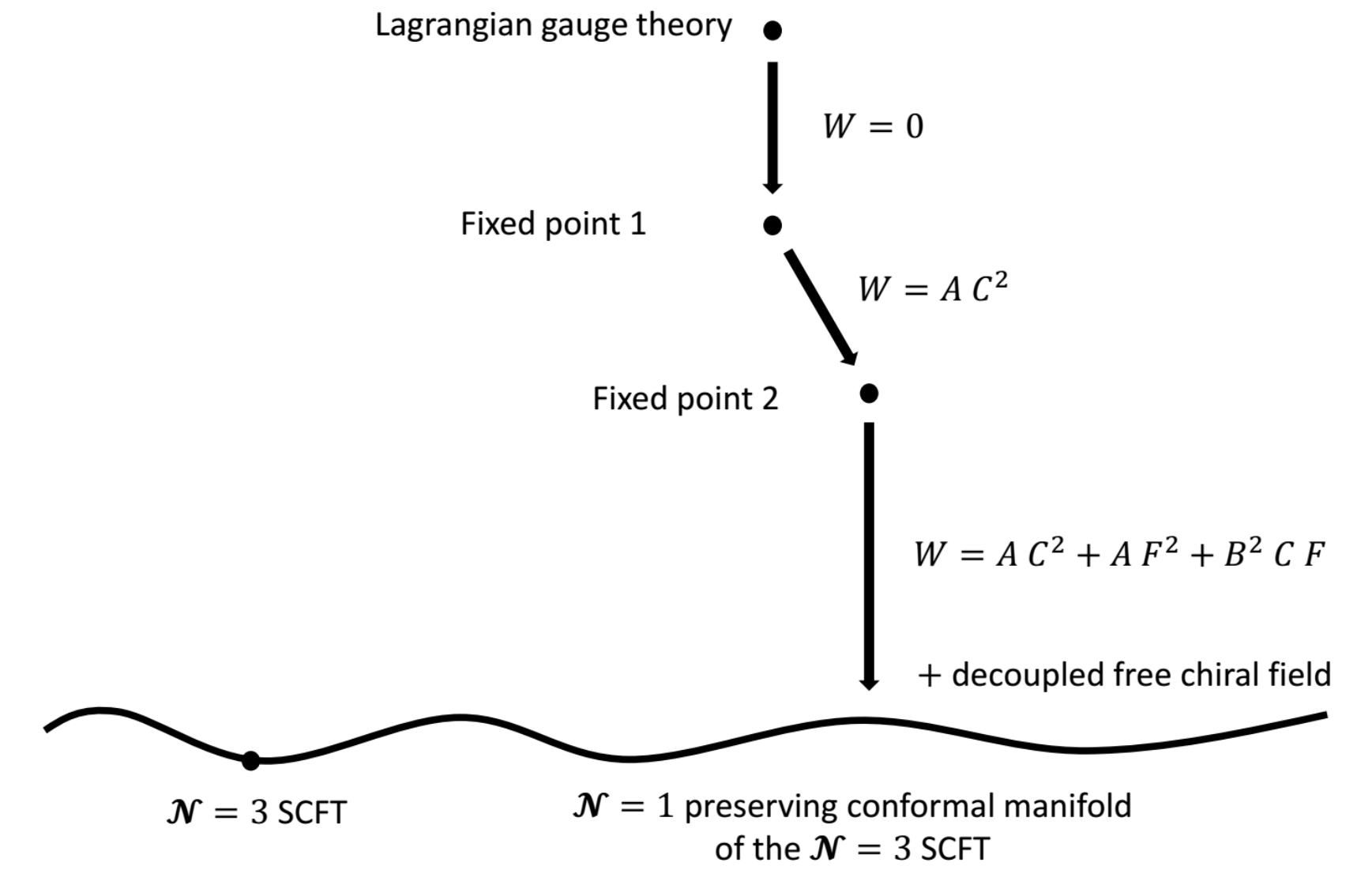} 
\caption{A pictorial summary of the proposed relation between the Lagrangian gauge theory studied in this section and the $\mathcal{N}=3$ SCFT. Here the arrows represent RG flows, with the initial one being triggered by the asymptotically free gauge couplings, and the others by superpotential terms. Here at the end point of the flow there is also a decoupled free chiral field, as indicated by the text there.}
\label{ExFlow}
\end{figure}

We are therefore lead to conclude that the resulting IR theory contains a decoupled free chiral field plus a potential interacting part. We can evaluate the $a$ and $c$ central charges of this additional part finding $a=c=\frac{5}{4}$, as previously shown. We claim that this remaining fixed point is the rank $1$ $\mathcal{N}=3$ SCFT with dimension $3$ Coulomb branch operator. To be precise, our claim is that the interacting part of the end point of the flow we mentioned is said $\mathcal{N}=3$ SCFT on a generic point on its one dimensional $\mathcal{N}=1$ only preserving conformal manifold. Regarding the latter, as the SCFT has a dimension three Coulomb branch operator, it has marginal operators preserving only $\mathcal{N}=1$ supersymmetry, associated with this operator, as well as ones related to it by $\mathcal{N}=3$ SUSY. As shown in \ref{N3app}, by decomposing the $\mathcal{N}=3$ multiplets containing these operators to $\mathcal{N}=2$ multiplets, one finds that these give a one dimensional $\mathcal{N}=1$ only preserving conformal manifold on a generic point of which the $SU(2)\times U(1)$, which is the global symmetry from the $\mathcal{N}=1$ viewpoint, is completely broken. The expected flow pattern is summerized in figure \ref{ExFlow}. We next present evidence for this claim.

\subsection{Index}

As a first test we consider the superconformal index of the theory\cite{Index}. As the index is invariant under RG flows, the index of the IR and UV theories should be equal, up to the proper identification of symmetries, notably the superconformal R-symmetry. As a result we should be able to evaluate the superconformal index of the end point of the flow from the UV Lagrangian evaluated using the R-symmetry that we expect should be the IR superconformal R-symmetry. This can be compared against the expected index of the $\mathcal{N}=3$ SCFT, and so constitutes an important test for our conjecture.

We expect the end point of the flow to be the $\mathcal{N}=3$ SCFT plus a free chiral field, so in order to evaluate the index of the former, we need to remove the contribution of the latter. A convenient way to do this is through the procedure of flipping \cite{BG}. In this method we introduce an additional singlet field $M$ and couple it to the gauge invariant $B^2$, which we expect to decouple in the IR, through the superpotential $W = M B^2$. Here $M$ has charge $\frac{4}{3}$ under $U(1)_R$, which is the only global symmetry that remains unbroken. The idea here is that the F term relation of the field $M$ eliminates the the gauge invariant $B^2$ from the chiral ring, and thus it should no longer be present in the IR. This can also be seen physically as the superpotential becomes a mass term for the fields $M$ and $B^2$, once the latter decouples. This makes the index computation easier, and is also conceptually convenient as it circumvents having free fields in the IR and the additional global symmetries associated to them.   

We can then evaluate the index with the flip fields finding:

\bea \label{N3Lindex}
I & = & 1 + 2(p q)^{\frac{2}{3}} - (p q)^{\frac{1}{3}}(p+q) + p q + (p q)^{\frac{2}{3}}(p+q) - (p q)^{\frac{1}{3}}(p^2+q^2) \\ & + & (p q)^{\frac{2}{3}}(p^2+q^2) - (p q)^{\frac{1}{3}}(p^3+q^3) + (p q)^{\frac{4}{3}} (p+q) - p q (4p^2+5p q+4q^2) + ... , \nonumber
\eea
where we use the standard notations \cite{DO,RR} for the index.
 
We next wish to compare this against the index expected from the $\mathcal{N}=3$ SCFT. While the full superconformal index of this theory is unknown, the fact that it has $\mathcal{N}=3$ supersymmetry put stringent limitations on the form the superconformal index can have, especially for the first few terms in the expansion in terms of $p$ and $q$. Additionally, as analyzed in the previous section, the knowledge of the moduli space of the theory also allows us to infer various multiplets expected in the theory, which can be used to form a minimal guess for the superconformal index.

We are then lead to the following strategy to check whether the terms in the index are consistent with $\mathcal{N}=3$ supersymmetry. From  our knowledge of this $\mathcal{N}=3$ SCFT we can identify three operators that must be present. The first is the $\mathcal{N}=3$ energy-momentum tensor, $B_1 \bar{B}_1 [0;0]^{(1,1;0)}_2$. The other two contain the dimension three Coulomb and Higgs branch operators and so their vevs generate the moduli space. These are the $B_1 \bar{B}_1 [0;0]^{(3,0;6)}_3$ type multiplet and its complex conjugate, $B_1 \bar{B}_1 [0;0]^{(0,3;-6)}_3$, that must also be present. We can use the decompositions of these multiplets into $\mathcal{N}=1$ superconformal multiplets given in section \eqref{subsubsec:N3opdec}, together with the contributions of the various $\mathcal{N}=1$ superconformal multiplets to the index given in section \eqref{subsec:sindex}, to determine the expected contributions of these multiplets to the index. These are given by:

\bea
& & I_{B_1 \bar{B}_1 [0;0]^{(1,1;0)}_2} (p,q) = \frac{1}{(1-p)(1-q)}\left(\frac{1}{v}\chi[\bold{2}]_{SU(2)} (p q)^{\frac{2}{3}} - p q (1+\chi[\bold{3}]_{SU(2)}) - \frac{1}{v^2}(p q)^{\frac{1}{3}}(p+q) \right. \nonumber \\ & + & \left. \frac{1}{v}\chi[\bold{2}]_{SU(2)}(p q)^{\frac{2}{3}}(p+q) + v\chi[\bold{2}]_{SU(2)}(p q)^{\frac{4}{3}} - p q (p+q)\right) ,
\eea

\bea
& & I_{B_1 \bar{B}_1 [0;0]^{(3,0;6)}_3} (p,q) = \frac{1}{(1-p)(1-q)}\left(v^3 \chi[\bold{4}]_{SU(2)} p q - v^2\chi[\bold{3}]_{SU(2)}(p q)^{\frac{2}{3}}(p+q) \right. \nonumber \\ & - & \left. (p q)^{\frac{4}{3}}(v^4\chi[\bold{3}]_{SU(2)} - v\chi[\bold{2}]_{SU(2)}) + v^3\chi[\bold{2}]_{SU(2)} p q (p+q) - v^2 (p q)^{\frac{5}{3}}\right) ,
\eea

\be
I_{B_1 \bar{B}_1 [0;0]^{(0,3;-6)}_3} (p,q) = \frac{1}{(1-p)(1-q)}\left(\frac{1}{v^6} p q - \frac{1}{v^5}\chi[\bold{2}]_{SU(2)}(p q)^{\frac{4}{3}} + \frac{1}{v^4} (p q)^{\frac{5}{3}}\right) .
\ee

Here for completeness we have specified their $U(1)\times SU(2)$ charges, which is the flavor symmetry from the $\mathcal{N}=1$ viewpoint, where we use the fugacity $v$ for the $U(1)$ and the notation $\chi[\bold{d}]_{SU(2)}$ for the $d$-dimensional representation under the $SU(2)$. Nevertheless, when evaluating the contributions here we shall unrefine with respect to these symmetries as these are not seen in the Lagrangian theories, presumably since these are broken on the conformal manifold.

Additionally, we also expect to have operators of the form $B_1 \bar{B}_1 [0;0]^{(3a+b,3c+b;6(a-c))}_{3(a+c)+2b}$, that are generated from products of the three operators. These can be taken into account by taking the plethystic exponent\footnote{The plethystic exponent is defined as:
\be
PE[f(x)] = e^{\sum^{\infty}_{k=1} \frac{1}{k}f(x^k)} \nonumber
\ee.} of the contribution of the three operators: 

\be
PE[I_{B_1 \bar{B}_1 [0;0]^{(1,1;0)}_2} (p,q) + I_{B_1 \bar{B}_1 [0;0]^{(3,0;6)}_3} (p,q) + I_{B_1 \bar{B}_1 [0;0]^{(0,3;-6)}_3} (p,q)].
\ee

We note that this takes into account all possible symmetric products of the three operators. While we expect the highest component, corresponding to the $B_1 \bar{B}_1 [0;0]^{(3a+b,3c+b;6(a-c))}_{3(a+c)+2b}$ type operators, of each product to be present, there are other type of multiplets that can appear in these products. These may or may not be present, which is something that we may need to account for later. 

We can next compare the two expressions finding:

\be
\Delta_I = - 3 (p q)^{\frac{5}{3}} + .... .
\ee 

We then see that the index is consistent with that expected from the three $\mathcal{N}=3$ basic multiplets up to order of $(p q)^{\frac{5}{3}}$. The deviation at this order has a straightforward interpretation as due to the plethystic exponent taking all possible symmetric products. Notably, order $(p q)^{\frac{5}{3}}$ receives contributions from the product of the energy-momentum tensor multiplet $B_1 \bar{B}_1 [0;0]^{(1,1;0)}_2$ with either $B_1 \bar{B}_1 [0;0]^{(3,0;6)}_3$ or $B_1 \bar{B}_1 [0;0]^{(0,3;-6)}_3$ multiplets. For simplicity we consider only the product with $B_1 \bar{B}_1 [0;0]^{(3,0;6)}_3$, as the other one can then be recovered by complex conjugation. Consider taking the product of the ground state of each. This gives a dimension $5$ scalar operator with charge $6$ under the $\mathcal{N}=3$ $U(1)_R$ symmetry, and in the representation $\bold{8}_{SU(3)_R}\otimes \bold{10}_{SU(3)_R}$ of $SU(3)_R$. As the latter is a reducible representation, it in fact gives four different ground stats with $SU(3)_R$ charges: $\bold{8}_{SU(3)_R}\otimes \bold{10}_{SU(3)_R} = \bold{35}_{SU(3)_R} \oplus \bold{27}_{SU(3)_R} \oplus \bold{8}_{SU(3)_R} \oplus \bold{10}_{SU(3)_R}$. Going over the shortening conditions in \cite{CID}, we see we must have that\footnote{The full representation is generated by acting on the ground state with the supercharges $Q, \bar{Q}$ and the translation generators $P$. As the latter are bosonic, the dimension of each representation is infinite, and when taking products we need to also consider the higher order states. As a result there are additional terms in the product coming from these. However, as these contribute to the index only at higher orders, we need not consider these here.}:

\bea
& & B_1 \bar{B}_1 [0;0]^{(1,1;0)}_2 \otimes B_1 \bar{B}_1 [0;0]^{(3,0;6)}_3 \\ \nonumber & = & B_1 \bar{B}_1 [0;0]^{(4,1;6)}_5 \oplus A_2 \bar{B}_1 [0;0]^{(2,2;6)}_5 \oplus L \bar{A}_{\bar{2}} [0;0]^{(1,1;6)}_5 \oplus L \bar{A}_{\bar{2}} [0;0]^{(3,0;6)}_5 \oplus ... .
\eea

We can next consider the contribution of each one of these to the index. Considering that these contain operators of dimension $\Delta\geq 5$, and using the contribution to the index of the various $\mathcal{N}=1$ multiplets, summarized in appendix \ref{subsec:sindex}, we see that the only $\mathcal{N}=1$ multiplets that can both appear in the above $\mathcal{N}=3$ multiplets and contribute to the index at order $(p q)^{\frac{5}{3}}$ are the $L \bar{B}_1 [0;0]^{(\frac{10}{3})}_5$ and $\bar{L} B_1 [0;0]^{(-\frac{10}{3})}_5$ multiplets. The latter can contribute to the product with $B_1 \bar{B}_1 [0;0]^{(0,3;-6)}_3$, which is the complex conjugate of the above product.

Performing the decomposition into $\mathcal{N}=1$ multiplets, as explained in the appendix, we find that the $B_1 \bar{B}_1 [0;0]^{(4,1;6)}_5$ type multiplet contains the operators $L \bar{B}_1 [0;0]^{(\frac{10}{3}),(4;2)}_5 \oplus \bar{L} B_1 [0;0]^{(-\frac{10}{3}),(1;7)}_5$ plus shorter representations that contribute at higher orders. The $A_2 \bar{B}_1 [0;0]^{(2,2;6)}_5$ type multiplet contains the operator $L \bar{B}_1 [0;0]^{(\frac{10}{3}),(2;2)}_5$ plus shorter representations that contribute at higher orders. The rest only contain multiplets that contribute at higher orders. As previously stated, we expect the multiplet $B_1 \bar{B}_1 [0;0]^{(4,1;6)}_5$ and its complex conjugate to appear in the index as some of the products of the three basic $\mathcal{N}=3$ multiplets expected in the $\mathcal{N}=3$ SCFT. However, the $A_2 \bar{B}_1 [0;0]^{(2,2;6)}_5$ may or may not appear, and the deviation in the index can be interpreted as pointing out that it does not appear. This comes about as it precisely contributes $3 (p q)^{\frac{5}{3}}$ plus higher order terms, and so its absence will exactly account for the difference observed in the index to that order.  

To summarize, we have shown that the superconformal index of the $SU(2)\times SU(2)$ gauge theory we introduced, after the removal of the free field, is consistent with being equal to the index of the rank $1$ $\mathcal{N}=3$ SCFT with dimension three Coulomb branch operator, at least to order $(p q)^{\frac{5}{3}}$. Specifically, the consistency refers to both the constraints placed by $\mathcal{N}=3$ supersymmetry on the index, as well as containing operators expected of the specific SCFT. As previously stated, we expect the $SU(2)\times SU(2)$ gauge theory, with the appropriate superpotential, to flow in the IR to the $\mathcal{N}=3$ SCFT deformed by an $\mathcal{N}=1$ only preserving marginal deformation. This is also consistent with the fact that we do not observe the $SU(2)\times U(1)$ symmetry that is the commutant of the $\mathcal{N}=1$ $U(1)_R$ in the $\mathcal{N}=3$ R-symmetry. Since the index is invariant under continuous deformations, the index must be equal to that of the $\mathcal{N}=3$ SCFT, unrefined with respect to the $SU(2)\times U(1)$ symmetry. However, this does allow various recombinations that would otherwise won't occur. For instance, the $SU(2)\times U(1)$ conserved currents recombine with some of the marginal operators.  

We can in principle continue to compare the indices to higher orders However, this becomes increasingly complicated due to the need to evaluate the contribution of various multiplets that can appear in the product of the basic multiplets, but may not be present in the $\mathcal{N}=3$ SCFT. Here we shall content ourselves with going up to order $(p q)^{\frac{5}{3}}$, reserving a study to higher orders for future work.

\subsubsection{Schur index}

We can consider an interesting limit of the superconformal index called the Schur index. This limit was introduced in \cite{GRRYpol} as one of several limits of the index for $\mathcal{N}=2$ SCFTs. As the superconfomal index we consider here should be equal to that of an $\mathcal{N}=3$ SCFT, these limits could also be considered for it. However, most of the limits considered in \cite{GRRYpol}, requires specialization of fugacities associated with the $\mathcal{N}=2$ R-symmetry. Since we see only the $U(1)$ subgroup, which is the $\mathcal{N}=1$ R-symmetry, most of these limits cannot be taken for the unrefined index we have. Nevertheless, as first pointed out in \cite{RZCM} (see also \cite{RZAS}), it is actually possible to take the Schur limit even for theories only manifesting the $U(1)$ subgroup associated with the $\mathcal{N}=1$ R-symmetry, and that limit should equal the Schur index of the $\mathcal{N}=2$ SCFT.

This limit is taken by setting $q=p^2$. We can next evaluate this limit for the gauge theory considered here, and if our conjecture is correct, it should be equal to the Schur limit of the index of the $\mathcal{N}=3$ SCFT. Evaluating the Schur index for the case at hand we find\footnote{Here $p$ is equal to the fugacity $\rho$ used in \cite{GRRYpol} for the Schur index.}:

\be
I_{Schur} = 1 + p^2 + p^4 + 2 p^6 - 2 p^7 + 3 p^8 - 2 p^9 + 4 p^{10} - 4 p^{11} + 6 p^{12} + ... .
\ee

There are several reasons why to consider this limit. One is that this limit leads to a simplification of the resulting expression, and as such is easier to compute and present. Another reason is the relation between the Schur limit of the index and the chiral algebra associated with an $\mathcal{N}=2$ SCFT that was introduced in \cite{BLLPRR}. The chiral algebras associated with $\mathcal{N}=3$ SCFTs were examined in \cite{NT,LLMM,BMRl}. It would be interesting if such studies of the chiral algebra can be used to perform futher checks on our proposal.

\subsection{Mass deformations}

There is one additional piece of information known about the rank $1$ $\mathcal{N}=3$ SCFT with dimension three Coulomb branch operator which we can use to test our proposal. Specifically, this piece of information is the behavior of the theory under SUSY preserving mass deformations. It was determined in \cite{ALLM} that this $\mathcal{N}=3$ SCFT has a mass deformation sending it to the $\mathcal{N}=2$ IR free gauge theory given by an $\mathcal{N}=2$ $SU(2)$ vector multiplet with a half-hyper in the $\bold{4}$ dimensional representation of the $SU(2)$ (see also \cite{ZafTW}). 

We next want to use this to test our proposal. We do this by studying SUSY preserving mass deformations of the Lagrangian theory and matching them against those expected from the $\mathcal{N}=3$ SCFT. Before looking at the mass deformations of the Lagrangian theory, we should consider our expectations from the $\mathcal{N}=3$ SCFT. Notably, we need to consider the fact that the Lagrangian theory is claimed to flow to the $\mathcal{N}=3$ SCFT deformed by an $\mathcal{N}=1$ only preserving marginal deformation, and so we need to consider its effect on the flow caused by the mass deformation. 

For this it is convenient to study this flow in the limit where the marginal deformation is taken to be very small, but non-zero. In that limit we expect its effect to be minute and the flow to proceed as if it was not turned on at least until energies much smaller than the scale set by the mass deformation. At that point the effect of the marginal coupling may become important due to potential changes in its marginality property that can occur during the flow. Therefore, we expect in this limit to get the $\mathcal{N}=2$ IR free gauge theory given by an $SU(2)$ gauge theory with a half-hyper in the $\bold{4}$ of $SU(2)$, deformed by an $\mathcal{N}=1$ only preserving superpotential. This superpotential should implement the effect caused by the presence of a non-trivial $\mathcal{N}=1$ only preserving marginal deformation at the UV. 

Here we note that the marginal deformation broke all the flavor symmetries at the UV, besides the $\mathcal{N}=1$ $U(1)$ R-symmetry that is in turn broken by the mass deformation. As such there is no symmetry that can restrict what superpotential terms can appear in the low-energy $\mathcal{N}=2$ $SU(2)$ gauge theory, and we would expect all to be generated. However, almost all of these are irrelevant and so won't change the IR behavior. The only one that is relevant, and so changes the IR behavior, is the dimension two Coulomb branch operator of the $SU(2)$ gauge theory. Also we note that the original marginal deformation we turned on contained the dimension three Coulomb branch operator of the rank $1$ $\mathcal{N}=3$ SCFT, and so it is reasonable to expect it to descend to a deformation by the Coulomb branch operator of the low-energy theory. As a result, we expect the flow to lead to the $\mathcal{N}=2$ $SU(2)$ gauge theory with a half-hyper in the $\bold{4}$ of $SU(2)$, deformed by the dimension two Coulomb branch operator. The latter is just a mass term for the adjoint chiral in the $\mathcal{N}=2$ $SU(2)$ vector multiplet. Therefore, we expect the low-energy theory to be just an $\mathcal{N}=1$ $SU(2)$ vector multiplet with a chiral field in the $\bold{4}$ of the $SU(2)$. The latter is expected to flow to an interacting $\mathcal{N}=1$ SCFT\footnote{This theory has a rather interesting history, with the result stated here being arrived at by the results of \cite{ISS,BCI,IntMA,Vart}, see also \cite[Sec.\ 7.5]{TachiRev} for a summary of the historical development.}. 

We can use the unique anomaly free R-symmetry, again under the assumption of no accidental symmetries, to show that there are no relevant operators in this $\mathcal{N}=1$ SCFT, and so it should be the end of the flow. To summarize the discussion so far, we determined that it is reasonable that the $\mathcal{N}=3$ SCFT in question, deformed by a combination of a relevant and a marginal deformation, flows to the $\mathcal{N}=1$ SCFT describing the low-energy dynamics of an $SU(2)$ gauge theory with  a single chiral field in its $\bold{4}$ dimensional representation.  

So far we have not specified the relevant deformation we used, but next we turn to consider this. Looking at the index of the $\mathcal{N}=3$ SCFT, we see that there are two possible relevant deformations, which forms a doublet of the $SU(2)$ subgroup of the $SU(3)$ R-symmetry group, which is a global symmetry from the $\mathcal{N}=1$ viewpoint. This implies that these should give equivalent deformations, and so we can consider either of them. Both of them preserve $\mathcal{N}=2$ supersymmetry, but embedded differently in the $\mathcal{N}=3$ supersymmetry so together they preserve only $\mathcal{N}=1$ supersymmetry.

The two mass deformations can be represented by the operators $Tr(A^2)$ and $M$. Next we consider the effect of each of these in turn. 

\subsubsection*{$M$}

We first consider the effect of the mass deformation represented by the operator $M$, that is we consider adding to the superpotential a term linear in $M$. The resulting theory still has the matter content shown in figure \ref{quiverN3}, with the flip field $M$, but with the superpotential:

\be \label{WMD1}
W = A F^2 + A C^2 + C F B^2 + M B^2 + M .
\ee  

\begin{figure}
\center
\includegraphics[width=0.75\textwidth]{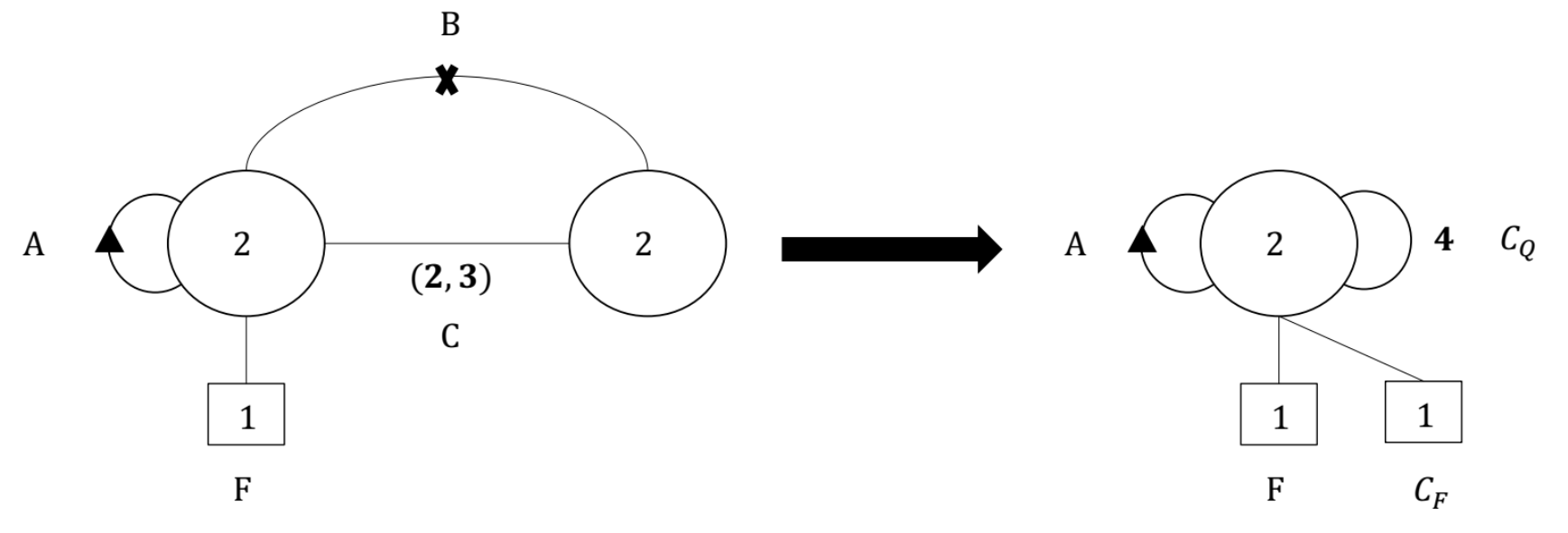} 
\caption{The effect of the mass deformation given by the flip field $M$. On the left we have the initial quiver with the flip field $M$ flipping $B^2$, here represented by the $X$ on the field $B$. The mass deformation causes the field $B^2$ to acquire an expectation value, leading to the Higgsing of the two $SU(2)$ groups down to the diagonal $SU(2)$. This leads to the quiver on the right. The fields $A$ and $F$ there come from the same ones on the right side, while the fields $C_Q$ and $C_F$ come from the field $C$. Here the field $C_Q$ is in the $\bold{4}$ of the $SU(2)$.}
\label{MD1}
\end{figure}

Here the first three terms are necessary for the gauge theory to flow to the $\mathcal{N}=3$ SCFT on its conformal manifold, the fourth one is the term flipping $B^2$ and the fifth is the added mass deformation. The F-term relation of $M$ now forces the gauge invariant $B^2$ to acquire an expectation value. This causes the two $SU(2)$ groups to be identified leading to the theory shown on the right of figure \ref{MD1}. The theory also inherits the following superpotential from the original superpotential \eqref{WMD1}:

\be
W = A F^2 + A C^2_F + A C^2_Q + C_F F .
\ee

The gauge theory itself is IR free so the only source of interesting IR behavior can come from the superpotential. The first three terms are marginal, so by themselves cannot generate interesting IR behavior, but the last term is relevant. We note that it is a mass term for the two fundamental fields, leading to them being integrated out in the IR. After that we get the matter content of an $\mathcal{N}=2$ $SU(2)$ gauge theory with a half-hyper in the $\bold{4}$ of $SU(2)$, and furthermore the third term in the superpotential is precisely the $\mathcal{N}=2$ preserving superpotential. However, we must also consider the effect of the first two superpotentials coupling the adjoint and fundamental chirals. After the fundamental chirals are integrated out, it should lead to effective interactions of $A$ with itself, mediated through the massive fundamental fields. As there is no symmetry forbidding them, we expect all possible such terms to be generated, specifically the mass term $Tr(A^2)$. This should lead to it being integrated out at low energies. We then end up with an $\mathcal{N}=1$ $SU(2)$ gauge theory with a chiral field in the $\bold{4}$ of the $SU(2)$, which is expected to flow to an $\mathcal{N}=1$ SCFT at low energies. Overall, we see that this flow indeed gives a result consistent with our proposal.    

\subsubsection*{$Tr(A^2)$}

We next consider the effect of the mass deformation represented by the operator $Tr(A^2)$, that is we consider changing the superpotential by adding the term $Tr(A^2)$. The new superpotential can now be written schematically as:    

\be \label{WMD2}
W = A F^2 + A C^2 + C F B^2 + M B^2 + A^2 .
\ee

The added term makes the fields associated with the adjoint chiral massive and we expect to flow to the same theory, but without it, and with a modified superpotntial:

\be \label{WMD3}
W = F^2 C^2 + C^4 + C F B^2 + M B^2 ,
\ee
which can be generated by inserting the F-term relation imposed by the field $A$, $2 A = - F^2 - C^2$ , into the superpotential \eqref{WMD2}. Here the $F^4$ term was dropped as it is impossible to make it gauge invariant, due to symmetrization requirements. 

After integrating out the adjoint, the left $SU(2)$ gauge group in figure \ref{quiverN3} now see only six fundamental chirals. This combination is known to be Seiberg dual to free fields, given by the $15$ gauge invariants one can build from the six fundamental chirals \cite{SeiDul}. As a result we expect the theory to flow to the one after the duality. The additional $15$ gauge invariants are the singlet $B^2$, the two adjoints $F C$ and $C^2$, the fundamental $F B$, and the fundamental and $\bold{4}$ coming from $B C$. We note that the suerpotential \eqref{WMD3}, gives all of them a mass, save for the chiral in the $\bold{4}$ \footnote{There are also superpotential terms coming from the Seiberg duality, but these will not be important here.}. Overall, we expect to end up with an $SU(2)$ gauge theory with a single chiral field in the $\bold{4}$, which is expected to flow to an $\mathcal{N}=1$ SCFT at low energies. Like in the previous case, there may be additional superpotential terms generated along the flow, but these will all be irrelevant from the viewpoint of the low energy $\mathcal{N}=1$ SCFT. Therefore, we see that the result of this flow is also consistent with our proposal.         

\section{Generalizations}
\label{sec:gen}

There are several interesting questions that arise from the model we presented. Specifically, whether it can be generalized to other $\mathcal{N}=3$ SCFTs, and whether it can be explained, for instance, by a string theory model. We next consider these two questions, and while we will not provide a satisfactory answer to these questions, we will see that these considerations will lead us to an $\mathcal{N}=1$ model that has the correct characteristics to be dual to a certain $\mathcal{N}=3$ SCFT. 

\begin{figure}
\center
\includegraphics[width=0.4\textwidth]{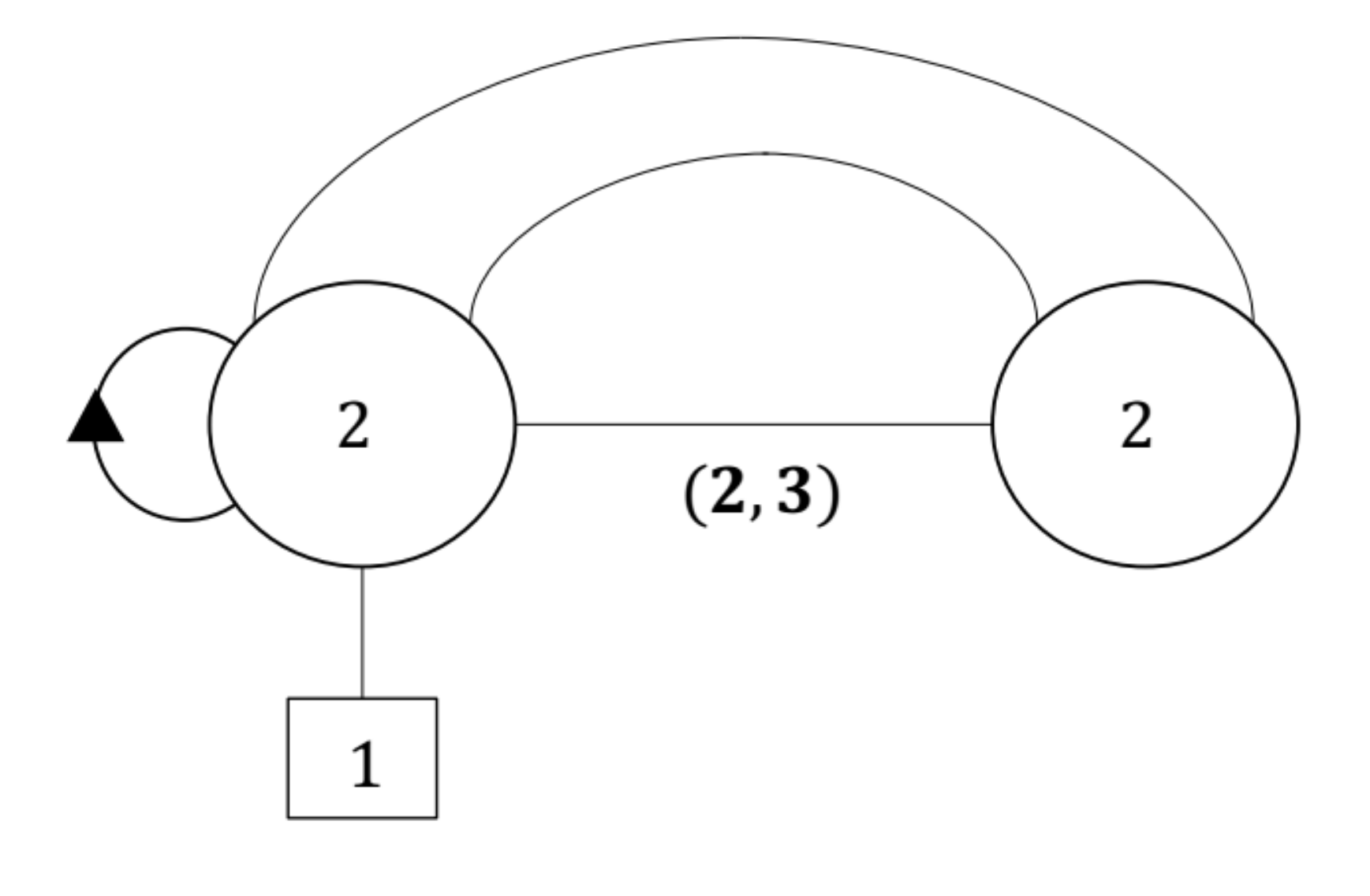} 
\caption{A modification of the in figure \ref{quiverN3} by the addition of an extra bifundamentl.}
\label{quiverN3imp}
\end{figure}

We begin by trying to cast the previous model in a manner that looks like something that may come from a string theory constructions. For that it is convenient to introduce an additional bifundamental between the two $SU(2)$ gauge groups, leading to the model depicted in figure \ref{quiverN3imp}. In this model the beta function for both $SU(2)$ groups vanishes at zero coupling. We can then get back to the previous model by integrating out one of the bifundamentals and introduce the superpotentials necessary for the flow.

The curious thing about the model in figure \ref{quiverN3} is that it can be cast as being made of class S components. Specifically, the model can be recast as two $SU(2)$ groups, one with an adjoint chiral and one without, gauging two $T_2$ theories. The $T_2$ theory is just a chiral field in the trifundamental of its $SU(2)^3$ global symmetry. One of them gives the two bifundamentals, corresponding to gauging two of its $SU(2)$ global symmetry groups by the two $SU(2)$ gauge groups. The second $T_2$ gives rise to the $C$ and $F$ fields, where here we gauge all three of te $SU(2)$ global symmetry groups by the two $SU(2)$ gauge groups lading to: $(\bold{2},\bold{2},\bold{2}) \rightarrow (\bold{2}_1,\bold{2}_2\otimes \bold{2}_2)\rightarrow (\bold{2}_1,\bold{1}_2) \oplus (\bold{2}_1,\bold{3}_2)$.     

This model can now be immediately generalized to arbitrary $N$, given by changing $SU(2)$ to $SU(N)$, one $T_2$ to $T_N$, and the other one to the $SU(N)\times SU(N)$ bifundmental hyper\footnote{Here we have a choice regarding the generalization of the ungauged puncture. The simplest generalization is to a minimal puncture, which is the one we consider here. It will be interesting to also consider the other possibilities, though we will not do so here.}. Like the $N=2$ case, the model has the property that the beta functions of the two $SU(N)$ groups vanish.

We next want to consider the mass deformation we used in the $N=2$ case, that is a mass to one of the bifundamental chirals in the hyper. However, such a mass term is not possible for $N>2$. This ends the possibility of a straightforward generalization. Nevertheless, a curious thing happens for $N=3$. In that case the central charges turn up to be:

\be \label{ACT3m}
a = c = \frac{21}{4} .
\ee  

This raises the possibility that this theory is dual to an $\mathcal{N}=3$ SCFT. Specifically, there is the $\mathcal{N}=3$ SCFT with moduli space $(\mathbb{C}^3)^3/G(3,3,3)$ discovered in \cite{ATsf}. This SCFT is known to be of rank $3$ with Coulomb branch operators of dimensions: $3$, $3$ and $6$. Using \eqref{env}, this indeed gives \eqref{ACT3m}. The moduli space of this $\mathcal{N}=3$ SCFT was discussed in section \ref{N3mds}. As it possesses dimension three Coulomb branch operators, from our previous observations it also has an $\mathcal{N}=1$ only preserving conformal manifold. From the results in \ref{N3app}, we conclude that the conformal manifold is six dimensional. 

It is then possible that this $\mathcal{N}=3$ SCFT and the model presented here for $N=3$ sit on the same conformal manifold. As we shall see next, this proposal passes some non-trivial tests, although not as strongly as the previous case, partially due to our lack of knowledge regarding that specific $\mathcal{N}=3$ SCFT.

Before turning out to analyze the model, we want to make a few comments: 

\begin{enumerate}

\item The remaining cases do not appear to be dual to $\mathcal{N}=3$ SCFTs. Furthermore, it is still not clear why the theories presented here should be related to $\mathcal{N}=3$ SCFTs. We do note the role played in these construction by the simultaneous gauging of three maximal punctures of class S theories. It is interesting if this has an interpretation in class S. We postpone further dealing with these issues to future study. 

\item The dual we consider here contain the $T_3$ theory as a building block. As such, it is not actually a Lagrangian dual. However, while the $T_3$ theory has no Lagrangian description manifesting the $\mathcal{N}=2$ supersymmetry, it does have at least two Lagrangians manifesting $\mathcal{N}=1$ supersymmetry. The first, given in \cite{GRW}, based on the results in \cite{E6Index}, involves a gauging of a symmetry that only emerges at strong coupling. The second one, given in \cite{ZafE6} involves a Lagrangian gauge theory flowing in the IR to the $T_3$ theory. However, both of these Lagrangians do not manifest the full $E_6$ symmetry, and also not its $SU(3)^3$ subgroup, at the UV, where the Lagrangian is valid. As a result we can not use these to turn the construction to a fully Lagrangian one, but rather we end up with a Lagrangian in the sense of \cite{GRW} (see also \cite{RVZ,AMS}). These can still be used to compute RG invariant quantities allowing us to perform all of these computations as if the theory was completely Lagrangian. 

\end{enumerate}

\subsection{The model}

We begin with some general considerations regarding the model. The first thing to note is that the construction we gave previously does not uniquely fixes the model, and in fact there are two distinct models sharing the same matter content. The difference is in how one performs the gaugings inside the $T_3$ theory. Specifically, when we gauge an $SU(3)$ global symmetry we have the choice of whether the gauging is done such that $\bold{3}\rightarrow \bold{3}$ or $\bold{3}\rightarrow \overline{\bold{3}}$. This is not important in an individual gauging as one can change between the two by redefining the $SU(3)$ generators, but is when there are multiple gaugings using the same groups. 

For the case at hand there are two distinct possibilities, which can be seen as follows. We can represent this theory as an $SU(3)$ gauge group gauging the $SU(3)^3$ subgroup of the $4d$ SCFT represented by the compactification of the $6d$ $A_2$ $(2,0)$ theory on a sphere with three maximal punctures and one minimal puncture. The possibilities are then manifested by the choice of embedding for each of the three gauging modulo two symmetries. One is global charge conjugation while the other is the symmetry related to the exchange of the three punctures implied by the class S construction. Taking these into account, we find two distinct possibilities: one given by taking $\bold{3}\rightarrow \bold{3}$ for all three embeddings (or the complex conjugate), while for the other we take one $\bold{3}\rightarrow \overline{\bold{3}}$ embedding and the rest are taken to be $\bold{3}\rightarrow \bold{3}$ (or ones reated to them by permutations or complex conjugation).

While the matter content for both cases is similar, as we shall soon see, the two models have different indices. In fact they also have different global symmetries. This comes about by considering the gauge group of the model as a group rather than just as an algebra. Specifically, consider the representations we have under the two $SU(3)$ gauge groups. First we have an adjoint, which is invariant under the center of $SU(3)$. Second we have the bifundamental fields which are invariant under the diagonal center, so has far as the perturbative states are concerned, the gauge group is consistent with being $\frac{SU(3)\times SU(3)}{\mathbb{Z}_3}$. Finally we need to consider the $T_3$ theory. The basic operators in it charged under the global symmetry are the moment maps of its $E_6$ global symmetry. These are in the adjoint of $E_6$, and when the $E_6$ is decomposed to its $SU(3)^3$ subgroup, they give an operator in the adjoint of each $SU(3)$ group, and an operator in the trifundamental plus its complex conjugate. The latter are sensitive to the center if we use the non-symmetric embedding, but are invariant if the symmetric embedding is used. Therefore we see that for the case with the symmetric embedding, the matter spectrum is invariant under the diagonal $\mathbb{Z}_3$ center, and we expect there to be an additional $\mathbb{Z}_3$ $1$-form symmetry acting on the line operators charged in the fundamental of each group \footnote{This corresponds to choosing the gauge group to be $SU(3)\times SU(3)$. In this case it is also possible to choose it to be $\frac{SU(3)\times SU(3)}{\mathbb{Z}_3}$ instead. In that case, we still have a $\mathbb{Z}_3$ $1$-form symmetry, but now acting on the magnetic 't Hooft line operators that exist in the theory. The two choices are related by gauging the $1$-form symmetry\cite{GKSW}.}. In the case using the non-symmetric embedding, however, there are operator charged under the entire $SU(3)\times SU(3)$ gauge group and we do not expect such $1$-form symmetries. Additionally, both models have a $U(1)^2$ zero-form global symmetry\footnote{The models may also have various discrete zero-form symmetries, but we shall not consider this in detail here.}.

Our claim is that the model with the symmetric embedding might be dual to a specific $\mathcal{N}=3$ SCFT. The duality implies the following:         

\begin{enumerate}

\item Matching of symmetries generically preserved on the conformal manifold, and their anomalies. As almost all symmetries can be broken on the conformal manifold, we are mostly left with the $\mathcal{N}=1$ superconformal symmetry. This still leaves two non-trivial tests. One is that anomalies of these symmetries, which reduce to the $a$ and $c$ central charges, matches the values expected for the $\mathcal{N}=3$ SCFT. The second is that the model indeed possesses $\mathcal{N}=1$ superconformal symmetry, which is not guaranteed for $\mathcal{N}=1$ theories with vanishing beta function at the free point \cite{LS,GKSTW}. We already noted the matching of the central charges, and would consider the second point in the next subsection. Additionally, as it is hard to break $1$-form symmetries, our proposal also suggests that the $\mathcal{N}=3$ SCFT should have a $\mathbb{Z}_3$ $1$-form symmetry. It will be interesting to check this prediction, though we shall not do so here. 

\item The superconformal indices, refined only with respect to symmetries generically preserved on the conformal manifold, must agree. In the next subsection, we shall show that the index is consistent with our proposal, where for the $\mathcal{N}=3$ SCFT we use a minimal guess of the index based on the operator spectrum expected from the moduli space of the theory. We shall also show that the dimension of the conformal manifold agrees between the two theories. This is the major evidence motivating our proposal.

\end{enumerate}

\subsection{The index}

\begin{figure}
\center
\includegraphics[width=0.5\textwidth]{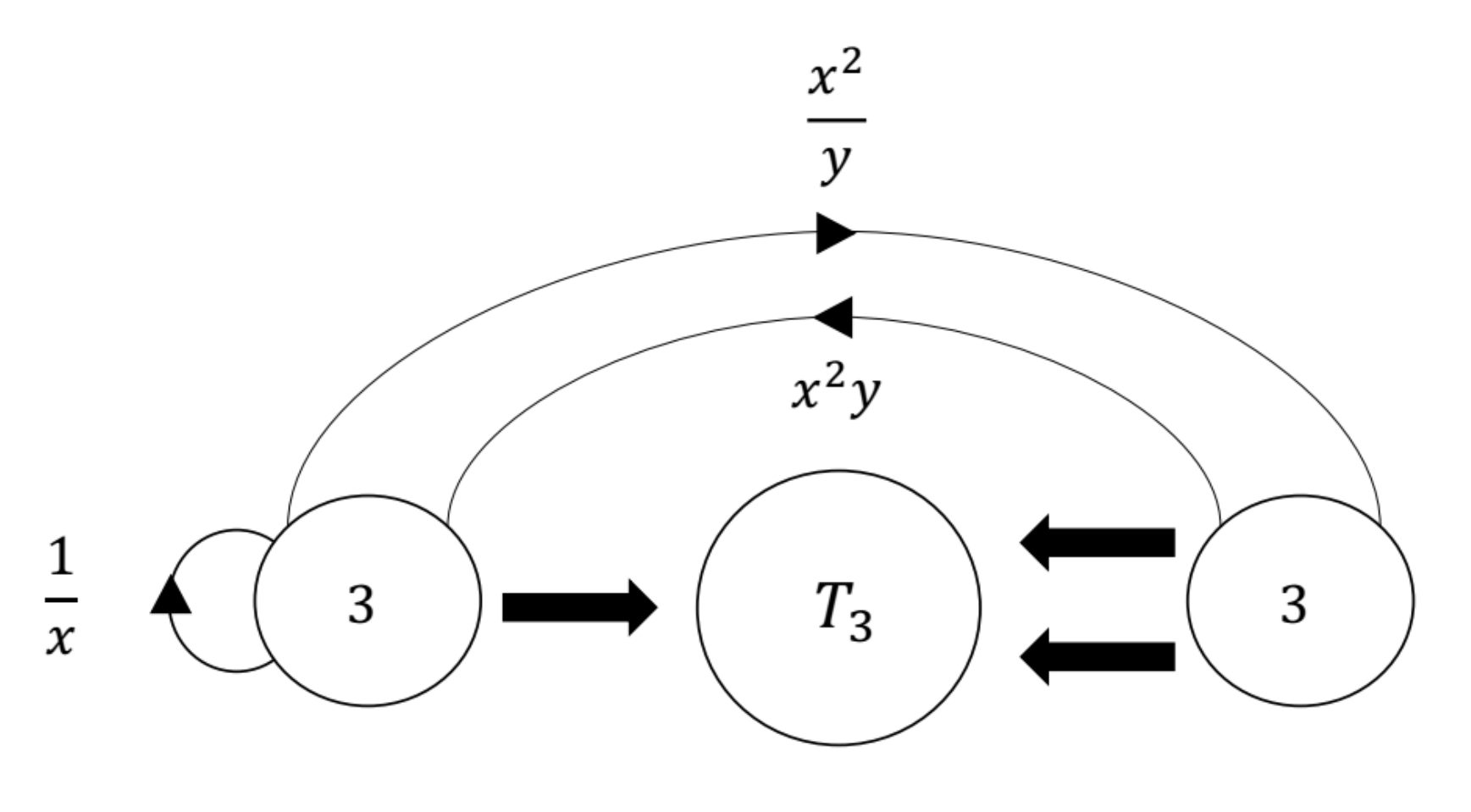} 
\caption{The model proposed to be dual to the $\mathcal{N}=3$ SCFT with moduli space $(\mathbb{C}^3)^3/G(3,3,3)$ with the charges under the gobal symmetries written in terms of fugacities. We also note that $U(1)_x$ acts on the $T_3$ theory, where it acts as $U(1)_g$.}
\label{quiverG333}
\end{figure}

Here we shall study the index of the model presented previously. First we consider the index refined by the $U(1)^2$ continuous global symmetry of the model at the 'free' point. We have written the charges under the two $U(1)$ groups using fugacities in figure \ref{quiverG333}. One of the symmetries, $U(1)_y$, is the one acting on the two bifundamentals, and can be identified with the $U(1)$ associated with the minimal puncture in the class S construction. Additionally, we have $U(1)_x$ which is the remaining anomaly free combination of the symmetries acting on the perturbative fields and $U(1)_g$, the commutant of the $\mathcal{N}=1$ R-symmetry in the $\mathcal{N}=2$ R-symmetry, of the $T_3$ theory. The combination is given such that the charge of the dimension $3$ Coulomb branch operator under $U(1)_x$ is $6$.

Evaluating the superconformal index, we find:

\be \label{IndRef}
I = 1 + (p q)^{\frac{2}{3}} (x^4 + \frac{1}{x^2}) - \frac{1}{x} (p q)^{\frac{1}{3}} (p+q) + p q (x^6 + x^3 + \frac{2}{x^3} + x^6 y^3 +\frac{x^6}{y^3} + y + \frac{1}{y} - 2) + ... .
\ee

We will be mostly interested here in the terms at order $p q$. The reason for this is that the theory is an $\mathcal{N}=1$ gauge theory with vanishing $1$-loop beta function at weak coupling, where we regard the $T_3$ as exotic matter for the gauge theory, and for these types of theories to actually be SCFTs there are must be a conformal manifold. The condition for the existence of the latter is that there will be marginal operators with a non-trivial Kahler quotient under the global symmetry of the theory \cite{GKSTW}. To determine this we need to know the marginal operators in the theory and their charges under the global symmetries.

This brings us back to the $p q$ order terms in the index. The positive terms in the expression are the marginal operators, while the negative ones are the conserved currents of the global symmetry \cite{GB}. Indeed, the negative terms in the $p q$ order of \eqref{IndRef} are just the conserved currents of the $U(1)^2$ global symmetry. Looking at the positive terms, we see that there are eight marginal operators, and their charges are such that there is Kahler quotient. As there are operators charged under both $U(1)$ groups, on a generic point of the expected conformal manifold all continuous global symmetries should be broken. This gives a six dimensional conformal manifold, where two of the eight marginal operators combine with the broken currents and become marginally irrelevant.

On the $\mathcal{N}=3$ SCFT side, we expect to have marginal operators coming from the multiplets containing the dimension three moduli space spanning operators. For the SCFT at hand, we actually have two of these. As noted in the appendix, each of them contains $5$ $\mathcal{N}=1$ SUSY preserving marginal operators that have a non-trivial Kahler quotient under the $SU(2)\times U(1)$ symmetry, which is the commutant of the $\mathcal{N}=1$ R-symmetry in the $\mathcal{N}=3$ R-symmetry. This gives a six dimensional conformal manifold on a generic point of which no symmetry is preserved. Therefore, we see that the conformal manifolds of the two theories have similar properties and it is possible that the two theories are different special points on the same conformal manifold.

To further check this we need to compare the indices of the two theories. If the theories indeed share the same conformal manifold, then they must have the same indices, as it should be possible to move from one theory to the other by turning on marginal operators. However, as the marginal operators break symmetries, only the unrefined index must match. As a result, and to save computational power we shall unrefine with respect to the $U(1)^2$ global symmetry. We next consider this index for the $\mathcal{N}=1$ model, finding:

\bea \label{Ind2}
& & I = 1 + 2(p q)^{\frac{2}{3}} - (p q)^{\frac{1}{3}} (p+q) + 6 p q - 2 (p q)^{\frac{2}{3}} (p+q) + 6 (p q)^{\frac{4}{3}} - (p q)^{\frac{1}{3}} (p^2+q^2) \nonumber \\ &+& p q (p+q) + 8 (p q)^{\frac{5}{3}} - 2(p q)^{\frac{2}{3}} (p^2+q^2) - 9(p q)^{\frac{4}{3}}(p+q) - (p q)^{\frac{1}{3}} (p^3+q^3) + ... .
\eea

We next want to compare the index against that expected from the $\mathcal{N}=3$ SCFT. As we do not know the index of this theory we cannot make a precise comparison. However, from the structure of the moduli space we do know some of the operators that are expected to be present, which we can use to formulate a conjecture for the index. The simplest such conjecture is to take the contributions of the energy-momentum tensor multiplet, the two dimension three Coulomb branch operators and their complex conjugate. This leads to:

\be \label{N3G333mg}
PE[I_{B_1 \bar{B}_1 [0;0]^{(1,1;0)}_2} (p,q) + 2I_{B_1 \bar{B}_1 [0;0]^{(3,0;6)}_3} (p,q) + 2I_{B_1 \bar{B}_1 [0;0]^{(0,3;-6)}_3} (p,q)].
\ee

Comparing the resulting expression with \eqref{Ind2}, we see that: 

\be
\Delta_I = 9 (p q)^{\frac{4}{3}} + .... .
\ee 

This is expected as we noted in section \ref{N3mds}, when we examined the moduli space of the expected $\mathcal{N}=3$ SCFT, that there should also be three additional $B_1 \bar{B}_1 [0,0]^{(2,2;0)}_4$ multiplets that we did not include in \eqref{N3G333mg}. Using the decomposition into $\mathcal{N}=1$ multiplets, as explained in the appendix, we find that the $B_1 \bar{B}_1 [0,0]^{(2,2;0)}_4$ type multiplet contains the operators $L \bar{B}_1 [0;0]^{(\frac{8}{3}),(2;-2)}_4$ plus multiplet that contribute to the index only at higher orders. These give the precise index contribution necessary to match the indices also to order $(p q)^{\frac{4}{3}}$.

It is possible to continue and check the indices to higher orders, but this becomes increasingly complicated as new operators keep appearing. We receive this to future research.  

\subsubsection{The Schur index}

Finally, we can consider the Schur limit of the index, taken by setting $q=p^2$. This limit should be equal to the Schur limit of the $\mathcal{N}=3$ SCFT. In this limit the contribution of the $T_3$ theory reduces to its Schur index, which can be computed from its class S description using the results of \cite{GRRY}. In this limit we find:

\be
I_{Schur} = 1 + p^2 + 2p^3 + 2 p^4 + 2 p^5 + 3 p^6 + 4 p^7 + 5 p^8 + ... .
\ee

It would be interesting if it is possible to check this against the expression expected from the $2d$ chiral algebra of this theory conjectured in \cite{BMRl}. 

\subsection{The other model}

Finally, we return to the second model given by the asymmetric embedding. Particularly, we can consider its superconformal index. For the case refined by the fugacities for the two $U(1)$ global symmetries we find:

\be \label{IndRef3}
I = 1 + (p q)^{\frac{2}{3}} (x^4 + \frac{1}{x^2}) - \frac{1}{x} (p q)^{\frac{1}{3}} (p+q) + p q (x^6 + x^3 + \frac{2}{x^3} + x^6 y^3 +\frac{x^6}{y^3} - 2) + ... ,
\ee
Which is similar to \eqref{IndRef} save for two marginal operators. The difference in the index again heightens the fact that it defines a different theory. Despite the lack of the two marginal operators, there is still a Kahler quotient under the remaining marginal operators, suggesting that this model as well gives an SCFT. Here, however, the conformal manifold is only four dimensional, where again on a generic point of which all the continuous global symmetry is broken. We can evaluate the index on the conformal manifold finding:

\bea \label{Ind3}
& & I = 1 + 2(p q)^{\frac{2}{3}} - (p q)^{\frac{1}{3}} (p+q) + 4 p q - 2 (p q)^{\frac{2}{3}} (p+q) + 6 (p q)^{\frac{4}{3}} - (p q)^{\frac{1}{3}} (p^2+q^2) \nonumber \\ &+& 3 p q (p+q) - 2(p q)^{\frac{2}{3}} (p^2 + p q + q^2) - 7(p q)^{\frac{4}{3}}(p+q) - (p q)^{\frac{1}{3}} (p^3+q^3) + ... .
\eea

We can also consider the schur limit, finding: 

\be
I_{Schur} = 1 + p^2 + 4 p^4 - 4 p^5 + 13 p^6 - 20 p^7 + 45 p^8 + ... .
\ee

This model also possess some of the properties expected from a model sharing the same conformal manifold as an $\mathcal{N}=3$ SCFT. Notably, its $a$ and $c$ central charges are equal, and the first few terms in the index match the contribution expected from the $\mathcal{N}=3$ energy-momentum tensor. The conformal manifold, though, cannot be reproduces by an $\mathcal{N}=3$ SCFT using only the dimension three moduli space spanning operator. Furthermore, we are not aware of any known $\mathcal{N}=3$ SCFT that can match these properties. It remains an interesting question then whether this model is related to an $\mathcal{N}=3$ SCFT or not.  

\section*{Acknowledgments}
We thank Shlomo Razamat and Yuji Tachikawa for relevant discussions. GZ is supported in part by the ERC-STG grant 637844-HBQFTNCER and by the INFN.

\appendix

\section{Relations between $4d$ SUSY}
\label{sec:prop}

In this appendix we consider some of the relations between $4d$ SUSY theories. Specifically, theories with extended supersymmetry can be regarded as theories with less SUSY, but then the presence of the extra supersymmetry has some implications for various properties of the theory. Here, we wish to consider some of these implications. Specifically, the implications of $\mathcal{N}=3$ SUSY are of great use to us in this article. Many of the results here have appeared before in the literature, notably \cite{NT,LLMM,AoMe,BPP,Evt1,Evt2,ABMm1,ABMm2}.

\subsection{Relations between symmetries of different SUSY theories}

We begin by considering the global symmetries of theories with a given amount of supersymmetry, when these are viewed as theories with less supersymmetry. 

\subsubsection{Relations between $\mathcal{N}=2$ and $\mathcal{N}=1$}

In $\mathcal{N}=1$ we have a single supercharge $Q$. The R-symmetry is $U(1)$ and it is convenient to normalize it such that $Q$ has charge $-1$. In $\mathcal{N}=2$ we have two supercharge $Q_1$ and $Q_2$. The R-symmetry is $U(1)\times SU(2)$ and it is convenient to normalize it such that $Q_1$ and $Q_2$ form a doublet under the $SU(2)$ and have charge $-1$ under the $U(1)$. This confirms with the notations in \cite{CID}. We can consider an $\mathcal{N}=2$ SCFT has an $\mathcal{N}=1$ SCFT, and we want to consider how the $\mathcal{N}=1$ $U(1)$ R-symmetry is embedded in the $\mathcal{N}=2$ $U(1)\times SU(2)$. This was done for instance in \cite{BTW}, but for latter convenience we shall repeat the analysis here.

First we need to choose which of the two supercharges is the $\mathcal{N}=1$ supercharge. We shall take that to be $Q_2$, which we take to also be the lowest component of the $SU(2)_R$ doublet. We do not explicitly see the $SU(2)_R$ symmetry in the $\mathcal{N}=1$ viewpoint, but we do explicitly see its Cartan element. We shall refer to that element, defined so that the states in the doublet have charges $\pm 1$, as $U(1)_{SU(2)_R}$. The group $U(1)^{\mathcal{N}=1}_R$, which is the superconformal R-symmetry from the $\mathcal{N}=1$ viewpoint, is then given by some linear combination of $U(1)_{SU(2)_R}$ and $U(1)^{\mathcal{N}=2}_R$, the abelian component in the $\mathcal{N}=2$ superconformal R-symmetry. 

The exact combination can be determined as follows. First, we know that under $U(1)^{\mathcal{N}=1}_R$, $Q_2$ must have charge $-1$. To fully fix the relation, we need one additional constraint. For this, it is useful to consider the $\mathcal{N}=2$ vector multiplet. This multiplet contains a scalar as its ground state, with the gaugino and the vector given by acting on it once or twice with the $Q$'s, respectively \cite{CID}. The vector must be uncharged under the R-symmetry, as it should be related to the connection of a gauge symmetry. These considerations force the scalar in the vector multiplet to have $U(1)^{\mathcal{N}=2}_R$ charge $2$, and be neutral under $SU(2)_R$. Alternatively, from the $\mathcal{N}=1$ viewpoint, it is just part of a free chiral field, and so should have $U(1)^{\mathcal{N}=1}_R$ charge of $\frac{2}{3}$. These two constraints fix the relation to be:

\be \label{N2tN11}
U(1)^{\mathcal{N}=1}_R = \frac{1}{3} U(1)^{\mathcal{N}=2}_R + \frac{2}{3} U(1)_{SU(2)_R} ,
\ee  
which is also the one found in \cite{BTW}. The other combination of $U(1)^{\mathcal{N}=2}_R$ and $U(1)_{SU(2)_R}$ becomes a standard global symmetry from the $U(1)^{\mathcal{N}=1}_R$ viewpoint. In Lagrangian theories, it can be defined as the one that acts on chiral fields in hypermultiplets with charge $-1$ and on chiral fields in vector multiplets with charge $2$. From a Lagrangian independent viewpoint, the important thing about the symmetry is that the supercharge be uncharged under it. Therefore, calling that symmetry $U(1)_g$, it must be given by the combination of $U(1)^{\mathcal{N}=2}_R$ and $U(1)_{SU(2)_R}$ such that $Q_2$ has charge $0$ under it. This fixes:

\be \label{N2tN12}
U(1)_g = \alpha (U(1)^{\mathcal{N}=2}_R - U(1)_{SU(2)_R}),
\ee  
where $\alpha$ is a normalization dependent constant. The $U(1)_g$ that we defined previously from a Lagrangian viewpoint use $\alpha=1$, which is also the choice used in \cite{BTW}. However, here we will use the choice $\alpha=\frac{1}{2}$.

\subsubsection{Relations between $\mathcal{N}=3$, $\mathcal{N}=2$ and $\mathcal{N}=1$}

Next we wish to consider an $\mathcal{N}=3$ SCFT. We can think of such a theory also as an $\mathcal{N}=2$ or an $\mathcal{N}=1$ SCFT. We next determine the relations between the R-symmetries of the different descriptions.

An $\mathcal{N}=3$ SCFT has three supercharges: $Q_1$, $Q_2$ and $Q_3$. It also has a $U(1)\times SU(3)$ R-symmetry, under which the three supercharges transform in the fundamental of $SU(3)$ and with charge $-1$ under the $U(1)$. Let us treat this theory as an $\mathcal{N}=2$ SCFT with supercharges $Q_1$ and $Q_2$. From this viewpoint a $U(1)\times SU(2)$ subgroup of $U(1)\times SU(3)$ is the superconformal R-symmetry, while the commutant, which is $U(1)$, is seen as a global symmetry. To find the relation, it is convenient to decompose the $SU(3)$ to $U(1)_{SU(3)_R}\times SU(2)$ such that: $\bold{3}_{SU(3)_R}\rightarrow \bold{2}^{-1}_{SU(2)} + \bold{1}^{2}_{SU(2)}$. The $SU(2)$ then is the non-abelian part in the $\mathcal{N}=2$ superconformal R-symmetry. The abelian part, $U(1)^{\mathcal{N}=2}_R$, should then be a linear combination of $U(1)_{SU(3)_R}$ and $U(1)^{\mathcal{N}=3}_R$. To determine it we again consider two constraints. One is that $Q_1$ and $Q_2$ must have charge $-1$ under it. 

For the second, we again consider the vector multiplet. Like in the previous case, its ground state is given by scalars. The rest of the multiplet is build by acting on it with the $Q$'s and $\bar{Q}$'s, where the vector is given by acting with two $Q$ operators on the ground state\footnote{To be precise, this gives the self-dual part of the field strength associated with the vector. The multiplet itself is not CPT invariant and so we must include also the conjugate which contains the anti self-dual part.} \cite{CID}. As the vector needs to be an R-symmetry singlet, we see that the scalars need to be in the fundamental of the $SU(3)$ and with charge $2$ under $U(1)^{\mathcal{N}=3}_R$. When decomposed in terms of $U(1)_{SU(3)_R}\times SU(2)$ representations, these give a doublet and a singlet of the $SU(2)$. From the $\mathcal{N}=2$ viewpoint, the former corresponds to the scalars in the hypermultiplets while the latter corresponds to the scalar in the vector multiplet. The former must be neutral under $U(1)^{\mathcal{N}=2}_R$, which provides the additional constraint, while the latter must have charge $2$, which in turn already follows from the previous constraint. These two constraints fix the relation to be:

\be \label{N3tN21}
U(1)^{\mathcal{N}=2}_R = \frac{1}{3} U(1)^{\mathcal{N}=3}_R + \frac{2}{3} U(1)_{SU(3)_R} .
\ee

Additionally, there is a $U(1)$ global symmetry from the $\mathcal{N}=2$ viewpoint that we shall call $U(1)_G$. Again, it is given by a different combination of $U(1)^{\mathcal{N}=3}_R$ and $U(1)_{SU(3)_R}$, defined by the requirement that the supercharges, $Q_1$ and $Q_2$, are uncharged under it. This gives:

\be \label{N3tN22}
U(1)_G = \alpha (U(1)^{\mathcal{N}=3}_R - U(1)_{SU(3)_R}),
\ee  
where $\alpha$ is again a normalization dependent constant. Here we shall usually take $\alpha=\frac{1}{3}$.

Finally, we can consider the theory as an $\mathcal{N}=1$ SCFT with supercharge $Q_1$. From this viewpoint, we have a $U(1)$ R-symmetry, and its commutant in $U(1)\times SU(3)$, which in this case is $U(1)\times SU(2)$, is seen as a global symmetry. We again seek to find the relation between the $\mathcal{N}=1$ $U(1)$ R-symmetry, the $U(1)\times SU(2)$ global symmetry from the $\mathcal{N}=1$ viewpoint and the $\mathcal{N}=3$ $U(1)\times SU(3)$ R-symmetry. For this we again decompose the $SU(3)$ to $U(1)\times SU(2)$ as done previously, but now the $SU(2)$ singlet in the decomposition of the $SU(3)$ fundamental acts on $Q_1$, while the doublet acts on $Q_2$ and $Q_3$. This implies that the $SU(2)$ in the decomposition is a global symmetry from the $\mathcal{N}=1$ viewpoint. The abelian part of the global symmetry, which we denote as $U(1)_v$, is again expressible as a combination of $U(1)^{\mathcal{N}=3}_R$ and $U(1)_{SU(3)_R}$, defined by the requirement that the supercharge $Q_1$ is uncharged under it. This gives:

\be \label{N3tN12}
U(1)_v = \alpha (2U(1)^{\mathcal{N}=3}_R + U(1)_{SU(3)_R}),
\ee  
where $\alpha$ is again a normalization dependent constant, which we shall take to be $\alpha=\frac{1}{3}$.

To find the expression for $U(1)^{\mathcal{N}=1}_R$, we again use the constraint that under it $Q_1$ must have charge $1$, supplemented by the constraint that the scalars in the hyper and vector multiplets, that we previously determined, have R-charge $\pm\frac{2}{3}$. The demand that all three will be obeyed for some sign fixes: 

\be \label{N3tN11}
U(1)^{\mathcal{N}=1}_R = \frac{1}{9} U(1)^{\mathcal{N}=3}_R - \frac{4}{9} U(1)_{SU(3)_R} .
\ee

\subsubsection{Relations between $\mathcal{N}=4$, $\mathcal{N}=3$, $\mathcal{N}=2$ and $\mathcal{N}=1$}

Finally, we consider the case of $\mathcal{N}=4$ SCFTs. Like the previous cases we can consider them as $\mathcal{N}=3$, $\mathcal{N}=2$ or $\mathcal{N}=1$ SCFTs, and we inquire as to the mapping of the symmetries. For $\mathcal{N}=4$ SCFTs the R-symmetry is $SU(4)$, and there is no abelian part\footnote{This follows as the superconformal algebra in this case is $psu(2,2|4)$ rather then $su(2,2|4)$. It appears that the later cannot actually be the superconformal algebra of an SCFT \cite{CID}.}. This makes determining the mapping of the symmetries easier. The four supercharges, $Q_i$, transform in the fundamental representation of the $SU(4)$.

First, we consider the theory as an $\mathcal{N}=3$ SCFT, with the supercharges being $Q_{1-3}$. In that case, it is convenient to decompose the $SU(4)$ R-symmetry group to $U(1)\times SU(3)$ such that $\bold{4}_{SU(4)}\rightarrow \bold{3}^{-1}_{SU(3)} + \bold{1}^{3}_{SU(3)}$. These then become the $\mathcal{N}=3$ R-symmetry. Next, we can consider the theory as an $\mathcal{N}=2$ SCFT, with the supercharges being $Q_{1, 2}$. In that case, it is convenient to decompose the $SU(4)$ R-symmetry group to $U(1)\times SU(2)^2$ such that $\bold{4}_{SU(4)}\rightarrow \bold{2}^{-1}_{SU(2)_1} + \bold{2}^{1}_{SU(2)_2}$. The $U(1)$ and one of the $SU(2)$ groups become the $\mathcal{N}=2$ R-symmetry. The other $SU(2)$ then becomes a global symmetry from the $\mathcal{N}=2$ viewpoint. In Lagrangian $\mathcal{N}=4$ SCFTs, this is the symmetry rotating the adjoint hyper.

Finally, we consider the theory as an $\mathcal{N}=1$ SCFT, with the supercharge being $Q_{1}$. In that case, it is again convenient to decompose the $SU(4)$ R-symmetry group to $U(1)\times SU(3)$, but now we define the $U(1)$ such that $\bold{4}_{SU(4)}\rightarrow \bold{3}^{\frac{1}{3}}_{SU(3)} + \bold{1}^{-1}_{SU(3)}$. The $U(1)$ then becomes the $\mathcal{N}=1$ R-symmetry, while the $SU(3)$ is seen as a global symmetry from the $\mathcal{N}=1$ viewpoint. In Lagrangian $\mathcal{N}=4$ SCFTs, this is the symmetry rotating the three adjoint chirals. 

\subsection{Anomaly polynomials of $4d$ SCFTs}

Next, we consider the form of the anomaly polynomial for $4d$ SCFTs.

\subsubsection{$\mathcal{N}=4$}

We begin with the case of $\mathcal{N}=4$ SCFTs. These are known to have only an $SU(4)$ R-symmetry as their global symmetry. As a result, there is only one non-trivial anomaly, $Tr(SU(4)^3)$, and correspondingly only one possible term in the anomaly polynomial, $C_3 (SU(4))$. These anomalies in turn are proportional to the conformal central charges, which for $\mathcal{N}=4$ SCFTs obey $a=c$. The anomaly polynomial of $\mathcal{N}=4$ SCFTs then takes the form:

\be
I_{\mathcal{N}=4} = 2a C_3 (SU(4))_{\bold{4}} .
\ee 

This in turn implies that: $Tr(SU(4)^3)=4a$, where we use the normalization such that the fundamental contributes $1$.

\subsubsection{$\mathcal{N}=3$}

We next consider the case of $\mathcal{N}=3$ SCFTs. As previously stated, these have a $U(1)\times SU(3)$ R-symmetry, and no other global symmetries. As a result, there are naively four non-trivial anomalies and correspondingly four possible terms in the anomaly polynomial. However, it turns out that these are not all independent, with one vanishing and with the remaining three all being related to one another and to the conformal central charges, which again obey $a=c$ \cite{AoMe}. We next wish to determine how the anomaly polynomial looks like for generic $\mathcal{N}=3$ SCFTs. The simplest way to do that is to use the fact that all $\mathcal{N}=4$ SCFTs are also $\mathcal{N}=3$ SCFTs, and that in both cases there is only one independent anomaly. As a result, we can determine the general form of the anomaly polynomial by starting with the anomaly polynomial of $\mathcal{N}=4$ SCFTs, and decompose the $SU(4)$ to $U(1)\times SU(3)$ as done in the previous sections\footnote{One might worry that this method will miss the potential linear anomaly of $U(1)_R$. However, such anomaly does not exist as it is inconsistent with the anomaly polynomial of generic $\mathcal{N}=2$ SCFTs.}. We then find that the anomaly polynomial of $\mathcal{N}=3$ SCFTs takes the form:

\be
I_{\mathcal{N}=3} = 2a C_3 (SU(3))_{\bold{3}} - 4a C_1 (U(1)_R) C_2 (SU(3))_{\bold{3}} - 16a C^3_1 (U(1)_R).
\ee 

This in turn implies that: $Tr(SU(3)^3)=4a$, $Tr(U(1)_R\times SU(3)^2)=4a$ and $Tr(U(1)^3_R)=-96a$.

\subsubsection{$\mathcal{N}=2$}

We next consider the case of $\mathcal{N}=2$ SCFTs. As previously stated, these have a $U(1)\times SU(2)$ R-symmetry, but unlike the previous cases, may have additional global symmetries. The major constraint on the anomaly polynomial is that all non-trivial anomalies must be quadratic in all symmetries save for $U(1)_R$. The general form of the anomaly polynomial for an $\mathcal{N}=2$ SCFT with global symmetry $G$ is\footnote{The constraint on the anomalies for $\mathcal{N}=2$ SCFTs was first derived in \cite{KuTh}. This was then used to constrain the anomaly polynomial in many cases, for instance \cite{BTW,TSZ}.}:  

\bea
& & I_{\mathcal{N}=2} = - 8(c-a) C^3_1 (U(1)_R) + 2(c-a)p_1 (T) C_1 (U(1)_R) \nonumber \\  & - & 4(2a-c) C_1 (U(1)_R) C_2 (R) + \frac{k_G}{2} C_1 (U(1)_R) C_2 (G). 
\eea 

Here the number $k_G$ is usually refereed to as the central charge associated with the flavor symmetry $G$.

\subsection{Operator structure of $4d$ SCFTs}

In this section we consider the spectrum of protected operators of $4d$ SCFTs. Specifically, say we have a $4d$ SCFT, then the operators in it must be in unitary representations of the corresponding superconformal group. This in turn limit the possible operators that can appear in the theory. Of specific importance are short multiplets as these obey various restrictions that can prove useful in the study of SCFTs. Notably, these are counted by the superconformal index which is an RG invariant.  

\subsubsection{$\mathcal{N}=1$}

We begin with the case of $\mathcal{N}=1$ SCFTs. Here we use the conventions of \cite{CID}, so we take $Q$ to have charges $[1;0]^{(-1)}_{\frac{1}{2}}$, where following \cite{CID}, we use the notation $[j_1;j_2]^{(r)}_{\Delta}$ to signal an operator with in the representation $[j_1;j_2]$ under the Lorentz group, charge $r$ under $U(1)_R$ and dimension $\Delta$. The multiplets are denoted based on the shortening conditions obeyed by them with respect to $Q$ and $\bar{Q}$, where $B$ is the shortest and $L$ the longest. In addition the charges of the ground state are given, so for instance, $L \bar{L} [j_1;j_2]^{(r)}_{\Delta}$ denote the long multiplet whose ground state is in the $[j_1;j_2]$ representation under the Lorentz group, has $U(1)_R$ charge $r$ and dimension $\Delta$. Some $\mathcal{N}=1$ superconformal multiplets of special interest are:

\subsubsection*{$A_l \bar{B}_1 [j;0]^{(r)}_{\Delta}$}

For these multiplets we have that $l = 1$ if $j\geq 1$ and $l = 2$ if $j=0$, and $\Delta=\frac{3 r}{2}$, $r=\frac{1}{3}(j+2)$. This type of multiplets are related to free fields. Specifically, $j=0$ are free chiral fields while $j=1$ are free vector fields. The cases $j>1$ describe higher spin free fields.  

\subsubsection*{$L \bar{B}_1 [j;0]^{(r)}_{\Delta}$}

These obey $\Delta=\frac{3 r}{2}$, $r>\frac{1}{3}(j+2)$. This type of multiplets describe chiral fields with R-charge $r$ and spin $j$.   

\subsubsection*{$A_l \bar{A}_{\bar{l}} [j;\bar{j}]^{(r)}_{\Delta}$}

For these multiplets we have that $l = 1$ if $j\geq 1$ and $l = 2$ if $j=0$, and similarly for $\bar{l}$ and $\bar{j}$. Also $\Delta=2+\frac{1}{2}(j+\bar{j})$, $r=\frac{1}{3}(j-\bar{j})$. This type of multiplets are related to various conserved currents. Specifically, the case of $j=\bar{j}=0$ gives flavor conserved currents, the cases $\bar{j}=0, j=1$ and $\bar{j}=1, j=0$ describe extra SUSY currents, and the case $j=\bar{j}=1$ describes a stress tensor multiplet. The other cases describe higher spin currents.  

\subsubsection{$\mathcal{N}=2$}

We next consider the case of $\mathcal{N}=2$ SCFTs. We again use the conventions of \cite{CID}, so we take $Q$ to have charges $[1;0]^{(1;-1)}_{\frac{1}{2}}$, where now the charges in the superscript, $(R;r)$, refer to the representation under $SU(2)_R$ and the charge under the $\mathcal{N}=2$ $U(1)$ R-symmetry. We shall decompose $Q$ to two components, where we shall take the lower $SU(2)$ component to be the $\mathcal{N}=1$ supercharge so as to confirm with equations \eqref{N2tN11}, \eqref{N2tN12}. We then have the $\mathcal{N}=1$ supercharge, $Q_2$, with charges $[1;0]^{(-1),0}_{\frac{1}{2}}$ and the other one, $Q_1$, with charges $[1;0]^{(\frac{1}{3}),-1}_{\frac{1}{2}}$, where the second superscript denotes the $U(1)_g$ charge (here we use $\alpha=\frac{1}{2}$). Like in the previous case, there are two types of shortening conditions, where both can apply to either $Q$ or $\bar{Q}$. The first, denoted as $B$, imply that the ground state is killed by acing with $Q$ if one forms the fully symmetric product in $SU(2)_R$. The second, denoted as $A$, is a bit more involve and we refer to \cite{CID} for the details regarding it. 

To understand how the $\mathcal{N}=2$ multiplets decompose in terms of $\mathcal{N}=1$ multiplets, we consider how the various states decompose. First, consider the case with ground state $[j;\bar{j}]^{(R;r)}_{\Delta}$, then due to the $SU(2)_R$ charge, this single ground state becomes $R+1$ different $\mathcal{N}=1$ ground states. Second, $Q$ decomposes to two supercharges, one of which builds the $\mathcal{N}=1$ multiplet. The other generates new ground states that can then build new $\mathcal{N}=1$ multiplets by applying the $\mathcal{N}=1$ supercharge. As a result, the $\mathcal{N}=2$ $L \bar{L}$ multiplet, which is just the standard long multiplet, with ground state $[j;\bar{j}]^{(R;r)}_{\Delta}$ ($j,\bar{j}\geq 1$) gives $16(R+1)$ $\mathcal{N}=1$ $L \bar{L}$ multiplets, coming from acing with $Q_1$ on the $R+1$ different $SU(2)_R$ components.

The heart of determining the different decompositions is understanding how the shortening conditions transform between the different amounts of supersymmetry. Let us first consider the $B$ type condition. First, this implies that the fully symmetrized product of $Q$ and the ground state vanishes, which in turn implies that we can relate the acting with $Q_1$ on the ground state to acting with $Q_2$ instead. As a result, we can ignore the action of $Q_1$, and only need to consider the different $SU(2)_R$ components of the ground state. In addition the bottom $SU(2)_R$ component is killed by $Q_2$, as that would lower the $SU(2)_R$ Cartan charge, and likewise for the top component with $\bar{Q}_2$ in $\bar{B}$ type conditions. Similarly, the next-to-bottom component will be killed by $Q^2_2$ and likewise for the next-to-top component with $\bar{Q}^2_2$, in $\bar{B}$ type conditions. The remaining components are unrestricted. As a result, in $B$ type conditions the bottom and next-to-bottom components acquire the $B$ and $A$ shortening conditions, the top and next-to-top components acquire the $\bar{B}$ and $\bar{A}$ shortening conditions in in $\bar{B}$ type conditions, and all the rest get $L$ and $\bar{L}$.

In the same vein, for $A$ type conditions, the bottom component acquires the $A$ shortening condition, and the top component acquires the $\bar{A}$ shortening condition for $\bar{A}$ type conditions. All other components get the $L$ and $\bar{L}$ conditions. Finally, like in $B$ type conditions, for $A$ type conditions we can relate the application of some combination of $Q_1$'s to applications involving $Q_2$ instead, implying that we need not consider all possible applications of $Q_2$. The structure generated by the independent application of $Q_1$ then looks like the $\mathcal{N}=1$ multiplet obeying the same shortening conditions.      

Next, we present the decomposition rules for selected $\mathcal{N}=2$ multiplets:

\subsubsection*{$B_1 \bar{B}_1$}

The shortest multiplet type is the $B_1 \bar{B}_1$, whose ground state carry the charges $[0;0]^{(R;0)}_{R}$. This type of BPS operators are also known as the Higgs branch chiral ring operators of dimension $R$. The $R=0$ case describes the vacuum state, while the $R=1$ case describes the free hyper. The case of $R=2$ describes the flavor conserved current multiplet, and decomposes as:

\be \nonumber
B_1 \bar{B}_1 [0;0]^{(2;0)}_{2} \rightarrow B_1 \bar{L} [0;0]^{(-\frac{4}{3}),1}_{2} + A_2 \bar{A}_{\bar{2}} [0;0]^{(0),0}_{2} + L \bar{B}_1 [0;0]^{(\frac{4}{3}),-1}_{2}.
\ee 

Here, the two chiral fields are the moment map operators, while the $A_2 \bar{A}_{\bar{2}}$ is the $\mathcal{N}=1$ conserved current multiplet. The term in the superscript after the $\mathcal{N}=1$ R-charge denotes the $U(1)_g$ charge. The remaining $B_1 \bar{B}_1$ multiplets decompose as:

\bea \nonumber
&& B_1 \bar{B}_1 [0;0]^{(R;0)}_{R} \rightarrow B_1 \bar{L} [0;0]^{(-\frac{2}{3}R),\frac{R}{2}}_{R} + L \bar{B}_1 [0;0]^{(\frac{2}{3}R),-\frac{R}{2}}_{R} + A_2 \bar{L} [0;0]^{(-\frac{(2R-4)}{3}),\frac{R-2}{2}}_{R} \\ \nonumber &+& L \bar{A}_{\bar{2}} [0;0]^{(\frac{(2R-4)}{3}),-\frac{(R-2)}{2}}_{R} + \sum^{R-2}_{i=2} L \bar{L} [0;0]^{(\frac{2}{3}(R-2i)),-\frac{(R-2i)}{2}}_{R}.
\eea

In Lagrangian theories, these type of multiplets describe operators made from $R$ hypermultiplets. 

\subsubsection*{$A_l \bar{B}_1$}

The next type of short multiplets is $A_l \bar{B}_1$, whose ground state carry the charges $[j;0]^{(R;j+2)}_{1+R+\frac{j}{2}}$. There are also the $B_1 \bar{A}_{\bar{l}}$ multiplet, which can be generated by taking the complex conjugate. When $R=0$, these describe free fields. Notably, the case of $j=0$ is the free vector while $j>0$ is associated with free higher-spin fields. The case of $R=1$ and $j=0$ describes extra SUSY currents and decomposes as:

\be \nonumber
A_2 \bar{B}_1 [0;0]^{(1;2)}_{2} \rightarrow A_2 \bar{A}_{\bar{2}} [0;0]^{(0),\frac{3}{2}}_{2} + L \bar{B}_1 [0;0]^{(\frac{4}{3}),\frac{1}{2}}_{2} + A_1 \bar{A}_{\bar{2}} [1;0]^{(\frac{1}{3}),\frac{1}{2}}_{\frac{5}{2}} + L \bar{B}_1 [1;0]^{(\frac{5}{3}),-\frac{1}{2}}_{\frac{5}{2}}.
\ee 

Here, the the first term is the extra conserved current associated with the larger R-symmetry, while the second one is a relevant operator, which in Lagrangian theories is given by the product of a chiral in an hypermultiplet with the adjoint chiral in the vector multiplet. The third term is the $\mathcal{N}=1$ extra SUSY current multiplet. Cases with $j>0$ describe higher-spin conserved currents. The remaining $A_l \bar{B}_1$ multiplets decompose as:

\bea \nonumber
&& A_l \bar{B}_1 [j;0]^{(R;j+2)}_{1+R+\frac{j}{2}} \rightarrow L \bar{B}_1 [j;0]^{(\frac{j+2+2R}{3}),\frac{j-R+2}{2}}_{1+R+\frac{j}{2}} + A_1 \bar{L} [j;0]^{(\frac{j+2-2R}{3}),\frac{j+R+2}{2}}_{1+R+\frac{j}{2}} + L \bar{A}_{\bar{2}} [j;0]^{(\frac{j+2R-2}{3}),\frac{j-R+4}{2}}_{1+R+\frac{j}{2}} \\ \nonumber &+& \sum^{R-1}_{i=2} L \bar{L} [j;0]^{(\frac{j+2R+2-4i}{3}),\frac{j-R+2+2i}{2}}_{1+R+\frac{j}{2}} + L \bar{B}_1 [j+1;0]^{(\frac{j+3+2R}{3}),\frac{j-R}{2}}_{1+R+\frac{j+1}{2}} + A_1 \bar{L} [j+1;0]^{(\frac{j+3-2R}{3}),\frac{j+R}{2}}_{1+R+\frac{j+1}{2}} \\ \nonumber &+& L \bar{A}_{\bar{2}} [j+1;0]^{(\frac{j+2R-1}{3}),\frac{j-R+2}{2}}_{1+R+\frac{j+1}{2}} + \sum^{R-1}_{i=2} L \bar{L} [j+1;0]^{(\frac{j+2R+3-4i}{3}),\frac{j-R+2i}{2}}_{1+R+\frac{j+1}{2}}.
\eea

In Lagrangian theories, the cases with $j=0$ correspond to operators made from $R$ chirals in the hypermultiplets and one chiral in the vector multiplet. 

\subsubsection*{$L \bar{B}_1$}

The next type of short multiplets is $L \bar{B}_1$, whose ground state carry the charges $[j;0]^{(R;r)}_{R+\frac{r}{2}}$. Here, $r$ is not fixed but must obey $r>j+2$. For $R=j=0$, these are the so called Coulomb branch chiral ring operators of dimension $\frac{r}{2}$, and decompose as:

\be \nonumber
L \bar{B}_1 [0;0]^{(0;r)}_{\frac{r}{2}} \rightarrow L \bar{B}_1 [0;0]^{(\frac{r}{3}),\frac{r}{2}}_{\frac{r}{2}} + L \bar{B}_1 [1;0]^{(\frac{r+1}{3}),\frac{r-2}{2}}_{\frac{r+1}{2}} + L \bar{B}_1 [0;0]^{(\frac{r+2}{3}),\frac{r-4}{2}}_{\frac{r+2}{2}}.
\ee 

In Lagrangian theories, $r$ must be even, and the first term is the chiral operator associated with the $\frac{r}{2}$ product of the chiral fields in the vector multiplet.

For generic values, we have the decomposition: 

\bea \nonumber
&& L \bar{B}_1 [j;0]^{(R;r)}_{R+\frac{r}{2}} \rightarrow L \bar{B}_1 [j;0]^{(\frac{r+2R}{3}),\frac{r-R}{2}}_{R+\frac{r}{2}} + L \bar{A}_{\bar{2}} [j;0]^{(\frac{r+2R-4}{3}),\frac{r-R+2}{2}}_{R+\frac{r}{2}} + \sum^{R}_{i=2} L \bar{L} [j;0]^{(\frac{r+2R-4i}{3}),\frac{r-R+2i}{2}}_{R+\frac{r}{2}} \\ \nonumber &+& L \bar{B}_1 [j\pm 1;0]^{(\frac{r+1+2R}{3}),\frac{r-R-2}{2}}_{R+\frac{r+1}{2}} + L \bar{A}_{\bar{2}} [j\pm 1;0]^{(\frac{r+2R-3}{3}),\frac{r-R}{2}}_{R+\frac{r+1}{2}} + \sum^{R}_{i=2} L \bar{L} [j\pm 1;0]^{(\frac{r+2R+1-4i}{3}),\frac{r-R+2i-2}{2}}_{R+\frac{r+1}{2}} \\ \nonumber &+& L \bar{B}_1 [j;0]^{(\frac{r+2R+2}{3}),\frac{r-R-4}{2}}_{R+\frac{r+2}{2}} + L \bar{A}_{\bar{2}} [j;0]^{(\frac{r+2R-2}{3}),\frac{r-R-2}{2}}_{R+\frac{r+2}{2}} + \sum^{R}_{i=2} L \bar{L} [j;0]^{(\frac{r+2R+2-4i}{3}),\frac{r-R-4+2i}{2}}_{R+\frac{r+2}{2}},
\eea
where $r> j+2$, but otherwise unconstrained.

\subsubsection*{$A_l \bar{A}_{\bar{l}}$}

The next type of short multiplets is $A_l \bar{A}_{\bar{l}}$, whose ground state carry the charges $[j;\bar{j}]^{(R;j-\bar{j})}_{2+R+\frac{1}{2}(j+\bar{j})}$. The case of $R=j=\bar{j}=0$ contains a conserved symmetric tensor and so corresponds to the energy-momentum tensor multiplet. It decomposes as:

\be \nonumber
A_2 \bar{A}_{\bar{2}} [0;0]^{(0;0)}_{2} \rightarrow A_2 \bar{A}_{\bar{2}} [0;0]^{(0),0}_{2} + A_1 \bar{A}_{\bar{2}} [1;0]^{(\frac{1}{3}),-1}_{\frac{5}{2}} + A_2 \bar{A}_{\bar{1}} [0;1]^{(-\frac{1}{3}),1}_{\frac{5}{2}} + A_1 \bar{A}_{\bar{1}} [1;1]^{(0),0}_{3}.
\ee 

Other cases with $R=0$ contain higher-spin conserved currents.

For generic values, we have the decomposition: 

\bea \nonumber
&& A_l \bar{A}_{\bar{l}} [j;\bar{j}]^{(R;j-\bar{j})}_{2+R+\frac{1}{2}(j+\bar{j})} \rightarrow L \bar{A}_{\bar{l}} [j;\bar{j}]^{(\frac{j-\bar{j}+2R}{3}),\frac{j-\bar{j}-R}{2}}_{2+R+\frac{1}{2}(j+\bar{j})} + A_l \bar{L} [j;\bar{j}]^{(\frac{j-\bar{j}-2R}{3}),\frac{j-\bar{j}+R}{2}}_{2+R+\frac{1}{2}(j+\bar{j})} \\ \nonumber &+& \sum^{R-1}_{i=1} L \bar{L} [j;\bar{j}]^{(\frac{j-\bar{j}+2R-4i}{3}),\frac{j-\bar{j}-R+2i}{2}}_{2+R+\frac{1}{2}(j+\bar{j})} + L \bar{A}_{\bar{l}} [j+1;\bar{j}]^{(\frac{j-\bar{j}+2R+1}{3}),\frac{j-\bar{j}-R-2}{2}}_{2+R+\frac{1}{2}(1+j+\bar{j})} + A_l \bar{L} [j+1;\bar{j}]^{(\frac{j-\bar{j}-2R+1}{3}),\frac{j-\bar{j}+R-2}{2}}_{2+R+\frac{1}{2}(1+j+\bar{j})} \\ \nonumber &+& \sum^{R-1}_{i=1} L \bar{L} [j+1;\bar{j}]^{(\frac{j-\bar{j}+2R+1-4i}{3}),\frac{j-\bar{j}-R-2+2i}{2}}_{2+R+\frac{1}{2}(1+j+\bar{j})} + L \bar{A}_{\bar{l}} [j;\bar{j}+1]^{(\frac{j-\bar{j}+2R-1}{3}),\frac{j-\bar{j}-R+2}{2}}_{2+R+\frac{1}{2}(1+j+\bar{j})} \\ \nonumber &+& A_l \bar{L} [j;\bar{j}+1]^{(\frac{j-\bar{j}-2R-1}{3}),\frac{j-\bar{j}+R+2}{2}}_{2+R+\frac{1}{2}(1+j+\bar{j})} + \sum^{R-1}_{i=1} L \bar{L} [j;\bar{j}+1]^{(\frac{j-\bar{j}+2R-1-4i}{3}),\frac{j-\bar{j}-R+2+2i}{2}}_{2+R+\frac{1}{2}(1+j+\bar{j})} \\ \nonumber &+& L \bar{A}_{\bar{l}} [j+1;\bar{j}+1]^{(\frac{j-\bar{j}+2R}{3}),\frac{j-\bar{j}-R}{2}}_{3+R+\frac{1}{2}(j+\bar{j})} + A_l \bar{L} [j+1;\bar{j}+1]^{(\frac{j-\bar{j}-2R}{3}),\frac{j-\bar{j}+R}{2}}_{3+R+\frac{1}{2}(j+\bar{j})} \\ \nonumber &+& \sum^{R-1}_{i=1} L \bar{L} [j+1;\bar{j}+1]^{(\frac{j-\bar{j}+2R-4i}{3}),\frac{j-\bar{j}-R+2i}{2}}_{3+R+\frac{1}{2}(j+\bar{j})}.
\eea

Additionally, there are also the $L \bar{A}_{\bar{l}}$ type of short multiplets. It is not difficult to work out their decomposition, though we won't write it explicitly here. 

\paragraph{Applications}

So far we have worked out the decomposition of $\mathcal{N}=2$ superconformal multiplets into $\mathcal{N}=1$ ones. We next want to use this to say something about the spectrum of SUSY preserving relevant and marginal operators. If one wish to maintain the full $\mathcal{N}=2$ SUSY, then this was already considered in \cite{CIDrmi}. Here, however, we will be more interested in deformations preserving also $\mathcal{N}=1$ supersymmetry. The only $\mathcal{N}=1$ relevant or marginal deformations preserving $\mathcal{N}=1$ SUSY reside in $L\bar{B}_1 [0;0]^{(r)}_{\frac{3}{2}r}$ type multiplets, for $r\leq 2$. As a result, by using the decompositions above, we can determine which multiplets contain $\mathcal{N}=1$ SUSY preserving relevant or marginal deformations. Also, from the results of \cite{GB}, the resulting relevant deformations are absolutely protected, while the marginal ones can only fail to be protected through recombination with a conserved current multiplet \cite{GKSTW}.

\subparagraph*{Relevant operators with dimension $1<\Delta<2$}

The only $\mathcal{N}=2$ multiplets containing $L\bar{B}_1 [0;0]^{(r)}_{\frac{3}{2}r}$ type multiplets with $\frac{2}{3}<r<\frac{4}{3}$ are the $L\bar{B}_1 [0;0]^{(0;r')}_{\frac{r'}{2}}$ type ones, for $2<r'<4$. These describe Coulomb branch operators of dimension $\frac{r'}{2}$, and carry $U(1)_g$ charge $\frac{r'}{2}$. Due to the results of \cite{BNP}, these operators cannot be charged under any flavor symmetries.  

\subparagraph*{Relevant operators with dimension $\Delta=2$}

There are several types of $\mathcal{N}=2$ multiplets containing $L\bar{B}_1 [0;0]^{(\frac{4}{3})}_{2}$ type multiplets. First there are the $L\bar{B}_1 [0;0]^{(0;4)}_{2}$ type multiplets, corresponding to dimension $2$ Coulomb branch operators. These carry $U(1)_g$ charge $2$, and cannot be charged under any $\mathcal{N}=2$ flavor symmetry. Second, there are the $A_2 \bar{B}_1 [0;0]^{(1;2)}_{2}$ type multiplets, corresponding to extra SUSY currents. These carry $U(1)_g$ charge $\frac{1}{2}$. Finally, there are $B_1\bar{B}_1 [0;0]^{(2;0)}_{2}$ type multiplets, corresponding to dimension $2$ Higgs branch operators. These carry $U(1)_g$ charge $-1$, and must be in the adjoint representation of the $\mathcal{N}=2$ flavor symmetry group.  

We note that the $L\bar{B}_1 [0;0]^{(0;4)}_{2}$ type multiplets also contain $L\bar{B}_1 [0;0]^{(2)}_{3}$ type multiplets, which are in fact the only possible $\mathcal{N}=2$ preserving marginal deformations. We also note that $A_2 \bar{B}_1 [0;0]^{(1;2)}_{2}$ type multiplets contain extra SUSY currents, while the $B_1\bar{B}_1 [0;0]^{(2;0)}_{2}$ type multiplets contain flavor conserved currents. As a result, we have that

\be \nonumber
n_{\Delta=2} = d_{G_F} + d_{CM_{\mathcal{N}=2}} + N_{SUSY} - 2 ,
\ee 
where $n_{\Delta=2}$ is the number of SUSY preserving dimension two operators, $d_{G_F}$ is the dimension of the $\mathcal{N}=2$ global symmetry, $d_{CM_{\mathcal{N}=2}}$ is the dimension of the $\mathcal{N}=2$ preserving conformal manifold and $N_{SUSY}$ is the number of supercharges, assumed to be at least two.

\subparagraph*{Relevant operators with dimension $2<\Delta<3$}

There are several $\mathcal{N}=2$ multiplets containing $L\bar{B}_1 [0;0]^{(r)}_{\frac{3}{2}r}$ type multiplets with $\frac{4}{3}<r<2$. First, there are the $L\bar{B}_1 [0;0]^{(0;r')}_{\frac{r'}{2}}$ type ones, for $4<r'<6$ or $2<r'<4$. These describe Coulomb branch operators of dimension $\frac{r'}{2}$. For $4<r'<6$, the required $L\bar{B}_1 [0;0]^{(r)}_{\frac{3}{2}r}$ type multiplet is in the ground state and directly corresponds to the Coulomb branch operator. The operator carry $U(1)_g$ charge $\frac{r'}{2}$. When $2<r'<4$, however, the required $L\bar{B}_1 [0;0]^{(r)}_{\frac{3}{2}r}$ type multiplet is not the Coulomb branch spanning operator itself, but rather one related to it by the $\mathcal{N}=2$ SUSY. These are $U(1)_g$ singlets. Both operators must also be singlets under the $\mathcal{N}=2$ global symmetry.

Additionally, the $L\bar{B}_1 [1;0]^{(0;r')}_{\frac{r'}{2}}$ and $L\bar{B}_1 [0;0]^{(1;r')}_{\frac{r'}{2}}$ type multiplets contain such operators, with $3<r'<5$ for the first case and $2<r'<4$ for the second one. The former are the so called spinning Coulomb branch operators, which can not appear in physical theories\cite{Man}, and thus need not be considered here. For the $L\bar{B}_1 [0;0]^{(1;r')}_{\frac{r'}{2}}$ type operator, the relevant operator is the ground state. It carries $U(1)_g$ charge $\frac{r'-1}{2}$. It can also be charged under flavor symmetries. The simplest example of such operators is the product of a Coulomb branch operator of dimension less than two and a free hyper.    

\subparagraph*{Marginal operators}

Finally we consider marginal operators. By going over the list, we find that there are precisely four types of possible $\mathcal{N}=1$ preserving marginal operators, differing by their $U(1)_g$ charge. These are:

\begin{enumerate}

\item - $L\bar{B}_1 [0;0]^{(2),-\frac{3}{2}}_{3}$ : This type of multiplets can only come from the $B_1\bar{B}_1 [0;0]^{(3;0)}_{3}$ type multiplets, and as such are always associated with dimension $3$ Higgs branch operators. These can also be charged under flavor symmetries, and as they also contain $B_1\bar{L} [0;0]^{(-2),\frac{3}{2}}_{3}$ type multiplets, must be charged in a real (potentially reducable) representation of the $\mathcal{N}=2$ flavor symmetry. In Lagrangian theories, these are given by gauge invariant combinations of three scalars in hypermultiplets.

\item - $L\bar{B}_1 [0,0]^{(2),3}_{3}$ : This type of multiplets can only come from the $L\bar{B}_1 [0,0]^{(0;6)}_{3}$ type multiplets, and as such are always associated with dimension $3$ Coulomb branch operators. These can not be charged under $\mathcal{N}=2$ flavor symmetries. In Lagrangian theories, these are given by gauge invariant combinations of three scalars in vector multiplets.

\item - $L\bar{B}_1 [0;0]^{(2),\frac{3}{2}}_{3}$ : This type of multiplets can in principle come from two types of $\mathcal{N}=2$ multiplets. One are the $L\bar{B}_1 [0;0]^{(1;4)}_{3}$ type multiplets, and the other are the $L\bar{B}_1 [1;0]^{(0;5)}_{\frac{5}{2}}$ type multiplets. The latter are spinning Coulomb branch operators, and so cannot appear in unitary $\mathcal{N}=2$ SCFTs\cite{Man}. This means that in physical theories the only source of these operators is $L\bar{B}_1 [0;0]^{(1;4)}_{3}$ type multiplets, which correspond to mixed branch operators. These can also be charged under flavor symmetries. In Lagrangian theories, these are given by gauge invariant combinations of a scalar in a hypermultiplet and two scalars in vector multiplets.

\item - $L\bar{B}_1 [0;0]^{(2),0}_{3}$ : This type of multiplets can come from two types of $\mathcal{N}=2$ multiplets. One are the $L\bar{B}_1 [0;0]^{(0;4)}_{2}$ type multiplets, and the other are the $A_2\bar{B}_1 [0;0]^{(2;2)}_{3}$ type multiplets. Both can appear in $\mathcal{N}=2$ SCFTs. While, both give the same $\mathcal{N}=1$ marginal operator, they differ by their effect on the $\mathcal{N}=2$ supersymmetry. The marginal operator in the $L\bar{B}_1 [0;0]^{(0;4)}_{2}$ type multiplets comes from the top component of the multiplets and as such preserves the full $\mathcal{N}=2$ supersymmetry. Also these cannot be charged under flavor symmetries. In contrary, the marginal operator in the $A_2\bar{B}_1 [0;0]^{(2;2)}_{3}$ type multiplets comes from the ground state and so preserves only $\mathcal{N}=1$ supersymmetry. These can also be charged under flavor symmetries.  The $L\bar{B}_1 [0;0]^{(0;4)}_{2}$ type multiplets correspond to dimension two Coulomb branch operators, and the associated marginal operator is given by the gauge coupling constant. The $A_2\bar{B}_1 [0;0]^{(2;2)}_{3}$ type multiplets correspond to mixed branch operators. In Lagrangian theories, these are given by gauge invariant combinations of two scalars in a hypermultiplet and a scalar in a vector multiplet.

\end{enumerate}

These lead to the following observations regarding conformal manifolds of $\mathcal{N}=2$ SCFTs. These statements mirror similar ones made in \cite{RZCM}, which are based on analysis of Lagrangian theories. The discussion here extends many of these statements to general $\mathcal{N}=2$ SCFTs. We first note that it is possible to have an $\mathcal{N}=2$ preserving conformal manifold whose complex dimension must be equal to the number of dimension two Coulomb branch operators. In addition to the $\mathcal{N}=2$ SUSY, also the $\mathcal{N}=2$ flavor symmetry must be preserved on the conformal manifold\footnote{The $\mathcal{N}=2$ flavor symmetry comes from $B_1\bar{B}_1 [0;0]^{(2;0)}_{2}$ type multiplets, whose ground state is an $SU(2)_R$ triplet of scalar fields. When we refer to the global symmetry of an $\mathcal{N}=2$ SCFT, we mean global symmetries whose currents come from such multiplets. Nevertheless, there are other sources of conserved flavor currents in $\mathcal{N}=2$ SCFTs. Notably, the energy-momentum tensor multiplet, $A_2 \bar{A}_{\bar{2}} [0;0]^{(0;0)}_2$, also contains a conserved flavor current. In interacting SCFTs, this is the $U(1)_g$ current, which is part of the $\mathcal{N}=2$ R-symmetry. However, there can be cases with additional such multiplets when free fields are involved. This follows from the following recombination rule of the $\mathcal{N}=2$ superconformal algebra:

\be
L\bar{L} [0;0]^{(0;0)}_{\Delta\rightarrow 2} \rightarrow A_2 \bar{A}_{\bar{2}} [0;0]^{(0;0)}_2 \oplus A_2\bar{B}_1 [0;0]^{(2;2)}_{3} \oplus B_1 \bar{A}_{\bar{2}} [0;0]^{(2;-2)}_{3} \oplus B_1\bar{B}_1 [0;0]^{(4;0)}_{4}.
\ee

As a result, it is possible for long multiplets to decompose in such a way at special points on the conformal manifold. This is common in $\mathcal{N}=2$ Lagrangian SCFTs in the weak coupling limit, which develop additional $\mathcal{N}=1$ global symmetries due to the vanishing of the $\mathcal{N}=2$ preserving superpotential, which are precisely the $A_2\bar{B}_1 [0;0]^{(2;2)}_{3}$ type multiplets that appear in the decomposition. As a result, while the $\mathcal{N}=2$ flavor symmetry does not change on the $\mathcal{N}=2$ preserving conformal manifold, the $\mathcal{N}=1$ flavor symmetry can enhance at special points on it, see for instance \cite{GRW,AMS}.}.

Additionally, it is possible to have an $\mathcal{N}=1$ only preserving conformal manifold as there are several multiplets containing $\mathcal{N}=1$ only preserving marginal operators. However, in order to have such conformal manifolds it is necessary that there is a Kahler quotient under the $\mathcal{N}=1$ flavor symmetry. This includes both the $\mathcal{N}=2$ flavor symmetry as well as $U(1)_g$. Going over the list of operators we see that to get a $U(1)_g$ quotient requires either the $A_2\bar{B}_1 [0;0]^{(2;2)}_{3}$ mixed branch operators or a combination of the $B_1\bar{B}_1 [0;0]^{(3;0)}_{3}$ Higgs branch operators and either the $L\bar{B}_1 [0;0]^{(0;6)}_{3}$ Coulomb branch operators or the $L\bar{B}_1 [0;0]^{(1;4)}_{3}$ mixed branch operators. The former allows $\mathcal{N}=1$ only preserving conformal manifolds that preserve $U(1)_g$, while the latter option gives ones that break this symmetry, assuming they can form a quotient under the $\mathcal{N}=2$ flavor symmetry. 

\subsubsection{$\mathcal{N}=3$}
\label{subsubsec:N3opdec}

We next consider the case of $\mathcal{N}=3$ SCFTs. We again use the conventions of \cite{CID}, so we take $Q$ to have charges $[1;0]^{(1,0;-1)}_{\frac{1}{2}}$, where the superscript now denoting the representation under the $SU(3)_R$, followed by the charge under the $U(1)$ R-symmetry. We shall consider here the decomposition into both $\mathcal{N}=2$ and $\mathcal{N}=1$ superconformal multiplets. For the former, we shall decompose the $SU(3)$ such that $\bold{3}\rightarrow \bold{2}^{-1} + \bold{1}^{2}$, where the doublet becomes the $\mathcal{N}=2$ supercharges. In this way equation \eqref{N3tN21} holds. For the decomposition into $\mathcal{N}=1$ superconformal multiplets, we shall instead use the singlet in the decomposition to be the $\mathcal{N}=1$ supercharge. In this way equation \eqref{N3tN11} holds. The global $U(1)$ we take to be defined by \eqref{N3tN22} for $\mathcal{N}=2$ and \eqref{N3tN12} for $\mathcal{N}=1$ with $\alpha=\frac{1}{3}$ for both.

Like in the previous cases, there are two types of shortening conditions, where both can apply to either $Q$ or $\bar{Q}$. The first, denoted as $B$, implies that the ground state is killed by acing with $Q$ if one forms the fully symmetric product in $SU(3)_R$. The second, denoted as $A$, is a bit more involved and we again refer to \cite{CID} for the details. The decomposition to $\mathcal{N}=2$ multiplets can be analyzed in a similar manner to as done in the previous case. Like there, the ground state of the $\mathcal{N}=3$ multiplet splits into ground states of multiple $\mathcal{N}=2$ multiplets governed by the decomposition of the $SU(3)$ R-symmetry. Additionally, we need to take into account the action of the remaining $\mathcal{N}=3$ supercharge that is not part of the $\mathcal{N}=2$ SUSY we are considering. Application of it on the ground states leads to additional $\mathcal{N}=2$ multiplets. Similarly to the previous cases, these form the structure of an $\mathcal{N}=1$ multiplet obeying the same shortening conditions. 

Once the decomposition into $\mathcal{N}=2$ multiplets is determined, it is straightforward to workout the decomposition to $\mathcal{N}=1$ multiplets by using the $\mathcal{N}=2\rightarrow \mathcal{N}=1$ decomposition we determined previously.     

Next, we present the decomposition rules for selected $\mathcal{N}=3$ multiplets:

\subsubsection*{$B_1 \bar{B}_1$}

The shortest multiplet type is the $B_1 \bar{B}_1$, whose ground state carry the charges $[0;0]^{(R_1,R_2;2(R_1-R_2))}_{R_1 + R_2}$. Like the previous case, the $R_1=R_2=0$ case describes the vacuum state, while the $R_1=1, R_2 = 0$ case, and its complex conjugate, is the $\mathcal{N}=3$ free vector. The case of $R_1=2, R_2=0$, and its complex conjugate, describes the extra SUSY current multiplet, and the case of $R_1= R_2=1$ describes the energy-momentum tensor multiplet. Additional cases of special physical interest are the $R_1=3, R_2=0$ case, $R_1=2, R_2=1$ case and their complex conjugates, which contain SUSY preserving marginal operators.

We next consider the decomposition of these multiplets under the $\mathcal{N}=2$ subgroup:

\be \nonumber
B_1 \bar{B}_1 [0;0]^{(2,0;4)}_{2} \rightarrow B_1 \bar{B}_1 [0;0]^{(2;0),2}_{2} + A_2 \bar{B}_{1} [0;0]^{(1;2),1}_{2} + L \bar{B}_1 [0;0]^{(0;4),0}_{2}.
\ee 

Here, the number following the $\mathcal{N}=2$ R-symmetry charges is the $U(1)_G$ charge. The first term is the dimension two Higgs branch chiral ring operator, the second is the $\mathcal{N}=2$ extra SUSY current operator, while the third is the dimension two Coulomb branch chiral ring operator. In terms of $\mathcal{N}=1$ multiplets, it decomposes as:

\bea \nonumber
&& B_1 \bar{B}_1 [0;0]^{(2,0;4)}_{2} \rightarrow L \bar{B}_1 [0;0]^{(\frac{4}{3}),(2;2)}_{2} + B_1 \bar{L} [0;0]^{(-\frac{4}{3}),(0;4)}_{2} + A_2 \bar{A}_2 [0;0]^{(0),(1;3)}_{2} \\ \nonumber & + & A_1 \bar{A}_{\bar{2}} [1;0]^{(\frac{1}{3}),(0;2)}_{\frac{5}{2}} + L \bar{B}_1 [1;0]^{(\frac{5}{3}),(1;1)}_{\frac{5}{2}} + L \bar{B}_1 [0;0]^{(2),(0;0)}_{3},
\eea
where the numbers following the $\mathcal{N}=1$ R-symmetry charge denote the representation under the $SU(2)$ followed by the $U(1)_v$ charge, both being the commutant of the $\mathcal{N}=1$ R-symmetry in the $\mathcal{N}=3$ R-symmetry.

Next we consider the EM tensor multiplet:

\be \nonumber
B_1 \bar{B}_1 [0;0]^{(1,1;0)}_{2} \rightarrow B_1 \bar{B}_1 [0;0]^{(2;0),0}_{2} + A_2 \bar{B}_{1} [0;0]^{(1;2),-1}_{2} + B_1 \bar{A}_{\bar{2}} [0;0]^{(1;-2),1}_{2} + A_2 \bar{A}_{\bar{2}} [0;0]^{(0;0),0}_{2}.
\ee 

These are, in the order listed, the conserved current of the global $U(1)_G$, the two $\mathcal{N}=2$ extra SUSY currents and the $\mathcal{N}=2$ EM tensor multiplet. In terms of $\mathcal{N}=1$ multiplets, it decomposes as:

\bea \nonumber
&& B_1 \bar{B}_1 [0;0]^{(1,1;0)}_{2} \rightarrow L \bar{B}_1 [0;0]^{(\frac{4}{3}),(1;-1)}_{2} + B_1 \bar{L} [0;0]^{(-\frac{4}{3}),(1;1)}_{2} + A_2 \bar{A}_2 [0;0]^{(0),(2;0)}_{2} + A_2 \bar{A}_2 [0;0]^{(0),(0;0)}_{2} \\ \nonumber & + & L \bar{B}_1 [1;0]^{(\frac{5}{3}),(0;-2)}_{\frac{5}{2}} + B_1 \bar{L} [0;1]^{(-\frac{5}{3}),(0;2)}_{\frac{5}{2}} + A_1 \bar{A}_{\bar{2}} [1;0]^{(\frac{1}{3}),(1;-1)}_{\frac{5}{2}} + A_2 \bar{A}_{\bar{1}} [0;1]^{(-\frac{1}{3}),(1;1)}_{\frac{5}{2}} \\ \nonumber & + & A_1 \bar{A}_1 [1;1]^{(0),(0;0)}_{3}.
\eea

We next consider the $B_1 \bar{B}_1 [0;0]^{(3,0;6)}_{3}$ multiplet:

\be \nonumber
B_1 \bar{B}_1 [0;0]^{(3,0;6)}_{3} \rightarrow B_1 \bar{B}_1 [0;0]^{(3;0),3}_{3} + A_2 \bar{B}_{1} [0;0]^{(2;2),2}_{3} + L \bar{B}_{1} [0;0]^{(1;4),1}_{3} + L \bar{B}_{1} [0;0]^{(0;6),0}_{3}.
\ee 

In terms of $\mathcal{N}=1$ multiplets, it decomposes as:

\bea \nonumber
&& B_1 \bar{B}_1 [0;0]^{(3,0;6)}_{3} \rightarrow L \bar{B}_1 [0;0]^{(2),(3;3)}_{3} + L \bar{A}_{\bar{2}} [0;0]^{(\frac{2}{3}),(2;4)}_{3} + A_2 \bar{L} [0;0]^{(-\frac{2}{3}),(1;5)}_{3} + B_1 \bar{L} [0;0]^{(-2),(0;6)}_{3} \\ \nonumber & + & L \bar{B}_1 [1;0]^{(\frac{7}{3}),(2;2)}_{\frac{7}{2}} + L \bar{A}_{\bar{2}} [1;0]^{(1),(1;3)}_{\frac{7}{2}} + A_1 \bar{L} [1;0]^{(-\frac{1}{3}),(0;4)}_{\frac{7}{2}} + L \bar{B}_{1} [0;0]^{(\frac{8}{3}),(1;1)}_{4} \\ \nonumber & + & L \bar{A}_{\bar{2}} [0;0]^{(\frac{4}{3}),(0;2)}_{4}.
\eea

We next consider the $B_1 \bar{B}_1 [0;0]^{(2,1;2)}_{3}$ multiplet:

\bea \nonumber
&& B_1 \bar{B}_1 [0;0]^{(2,1;2)}_{3} \rightarrow B_1 \bar{B}_1 [0;0]^{(3;0),1}_{3} + A_2 \bar{B}_{1} [0;0]^{(2;2),0}_{3} + B_1 \bar{A}_{\bar{2}} [0;0]^{(2;-2),2}_{3} \\ \nonumber & + & A_2 \bar{A}_{\bar{2}} [0;0]^{(1;0),1}_{3} + L \bar{B}_{1} [0;0]^{(1;4),-1}_{3} + L \bar{A}_{\bar{2}} [0;0]^{(0;2),0}_{3}.
\eea 

In terms of $\mathcal{N}=1$ multiplets, it decomposes as:

\bea \nonumber
&& B_1 \bar{B}_1 [0;0]^{(2,1;2)}_{3} \rightarrow L \bar{B}_1 [0;0]^{(2),(2;0)}_{3} + L \bar{A}_{\bar{2}} [0;0]^{(\frac{2}{3}),(3;1)}_{3} + L \bar{A}_{\bar{2}} [0;0]^{(\frac{2}{3}),(1;1)}_{3} + A_2 \bar{L} [0;0]^{(-\frac{2}{3}),(2;2)}_{3} \\ \nonumber & + & A_2 \bar{L} [0;0]^{(-\frac{2}{3}),(0;2)}_{3} + B_1 \bar{L} [0;0]^{(-2),(1;3)}_{3} + L \bar{B}_1 [1;0]^{(\frac{7}{3}),(1;-1)}_{\frac{7}{2}} + L \bar{A}_{\bar{2}} [1;0]^{(1),(2;0)}_{\frac{7}{2}} + L \bar{A}_{\bar{2}} [1;0]^{(1),(0;0)}_{\frac{7}{2}} \\ \nonumber & + & A_1 \bar{L} [1;0]^{(-\frac{1}{3}),(1;1)}_{\frac{7}{2}} + B_1 \bar{L} [0;1]^{(-\frac{7}{3}),(0;4)}_{\frac{7}{2}} + L \bar{A}_{\bar{1}} [0;1]^{(\frac{1}{3}),(2;2)}_{\frac{7}{2}} \\ \nonumber & + & A_2 \bar{L} [0;1]^{(-1),(1;3)}_{\frac{7}{2}} + L \bar{B}_1 [0;0]^{(\frac{8}{3}),(0;-2)}_{4} + L \bar{A}_{\bar{2}} [0;0]^{(\frac{4}{3}),(1;-1)}_{4} + L \bar{A}_{\bar{1}} [1;1]^{(\frac{2}{3}),(1;1)}_{4} + A_1 \bar{L} [1;1]^{(-\frac{2}{3}),(0;2)}_{4}.
\eea

\subsubsection*{$A_l \bar{B}_1$}

The next type of short multiplets is $A_l \bar{B}_1$, whose ground state carry the charges $[j;0]^{(R_1,R_2;6+3j+2(R_1-R_2))}_{1+R_1+R_2+\frac{j}{2}}$. When $R_1=R_2=0$, these multiplets contain free higher-spin fields. The cases of $R_1=1, R_2=0$ and $R_1=0, R_2=1$ contains higher-spin conserved currents. The decompsitions into $\mathcal{N}=2$ multiplets for all three cases are: 

\be \nonumber
A_l \bar{B}_1 [j;0]^{(0,0;3j+6)}_{1+\frac{j}{2}} \rightarrow A_l \bar{B}_1 [j;0]^{(0;j+2),j+2}_{1+\frac{j}{2}} + A_1 \bar{B}_1 [j+1;0]^{(0;j+3),j+1}_{\frac{3+j}{2}} ,
\ee 

\bea \nonumber
&& A_l \bar{B}_1 [j;0]^{(1,0;3j+8)}_{2+\frac{j}{2}} \rightarrow A_l \bar{B}_1 [j;0]^{(1;j+2),j+3}_{2+\frac{j}{2}} + L \bar{B}_1 [j;0]^{(0;j+4),j+2}_{2+\frac{j}{2}} + A_1 \bar{B}_1 [j+1;0]^{(1;j+3),j+2}_{\frac{5+j}{2}} \\ \nonumber & + & L \bar{B}_1 [j+1;0]^{(0;j+5),j+1}_{\frac{5+j}{2}} ,
\eea 

\bea \nonumber
&& A_l \bar{B}_1 [j;0]^{(0,1;3j+4)}_{2+\frac{j}{2}} \rightarrow A_l \bar{B}_1 [j;0]^{(1;j+2),j+1}_{2+\frac{j}{2}} + A_l \bar{A}_{\bar{2}} [j;0]^{(0;j),j+2}_{2+\frac{j}{2}} + A_1 \bar{B}_1 [j+1;0]^{(1;j+3),j}_{\frac{5+j}{2}} \\ \nonumber & + & A_1 \bar{A}_{\bar{2}} [j+1;0]^{(0;j+1),j+1}_{\frac{5+j}{2}} .
\eea 

In terms of $\mathcal{N}=1$ multiplets, the decomposition is:

\be \nonumber
A_l \bar{B}_1 [j;0]^{(0,0;3j+6)}_{1+\frac{j}{2}} \rightarrow A_l \bar{B}_1 [j;0]^{(\frac{j+2}{3}),(0;2j+4)}_{1+\frac{j}{2}} + A_1 \bar{B}_1 [j+1;0]^{(\frac{j+3}{3}),(1;2j+3)}_{\frac{3+j}{2}} + A_1 \bar{B}_1 [j+2;0]^{(\frac{j+4}{3}),(0;2j+2)}_{2+\frac{j}{2}} ,
\ee 

\bea \nonumber
&& A_l \bar{B}_1 [j;0]^{(1,0;3j+8)}_{2+\frac{j}{2}} \rightarrow L \bar{B}_1 [j;0]^{(\frac{4+j}{3}),(1;2j+5)}_{2+\frac{j}{2}} + A_l \bar{A}_{\bar{2}} [j;0]^{(\frac{j}{3}),(0;2j+6)}_{2+\frac{j}{2}} + L \bar{B}_1 [j+1;0]^{(\frac{5+j}{3}),(2;2j+4)}_{\frac{j+5}{2}} \\ \nonumber & + & L \bar{B}_1 [j\pm 1;0]^{(\frac{5+j}{3}),(0;2j+4)}_{\frac{j+5}{2}} + A_1 \bar{A}_{\bar{2}} [j+1;0]^{(\frac{1+j}{3}),(1;2j+5)}_{\frac{5+j}{2}} + A_1 \bar{A}_{\bar{2}} [j+2;0]^{(\frac{2+j}{3}),(0;2j+4)}_{\frac{6+j}{2}} \\ \nonumber & + & L \bar{B}_1 [j+2;0]^{(\frac{6+j}{3}),(1;2j+3)}_{3+\frac{j}{2}} + L \bar{B}_1 [j;0]^{(\frac{6+j}{3}),(1;2j+3)}_{3+\frac{j}{2}} + L \bar{B}_1 [j+1;0]^{(\frac{7+j}{3}),(0;2j+2)}_{\frac{7+j}{2}} ,
\eea 

\bea \nonumber
&& A_l \bar{B}_1 [j;0]^{(0,1;3j+4)}_{2+\frac{j}{2}} \rightarrow L \bar{B}_1 [j;0]^{(\frac{4+j}{3}),(0;2j+2)}_{2+\frac{j}{2}} + A_l \bar{A}_{\bar{2}} [j;0]^{(\frac{j}{3}),(1;2j+3)}_{2+\frac{j}{2}} + L \bar{B}_1 [j+1;0]^{(\frac{5+j}{3}),(1;2j+1)}_{\frac{j+5}{2}} \\ \nonumber & + &  A_1 \bar{A}_{\bar{2}} [j+1;0]^{(\frac{1+j}{3}),(2;2j+2)}_{\frac{5+j}{2}} + A_1 \bar{A}_{\bar{2}} [j+1;0]^{(\frac{1+j}{3}),(0;2j+2)}_{\frac{5+j}{2}} + L \bar{B}_1 [j+2;0]^{(\frac{6+j}{3}),(0;2j)}_{3+\frac{j}{2}} \\ \nonumber & + & A_l \bar{A}_{\bar{1}} [j;1]^{(\frac{j-1}{3}),(0;2j+4)}_{\frac{5+j}{2}} + A_1 \bar{A}_{\bar{2}} [j+2;0]^{(\frac{2+j}{3}),(1;2j+1)}_{\frac{6+j}{2}} + A_1 \bar{A}_{\bar{1}} [j+1;1]^{(\frac{j}{3}),(1;2j+3)}_{\frac{6+j}{2}} \\ \nonumber & + & A_1 \bar{A}_{\bar{1}} [j+2;1]^{(\frac{1+j}{3}),(0;2j+2)}_{\frac{7+j}{2}}.
\eea

Next we present the decomposition for several other multiplets, first in terms of $\mathcal{N}=2$ multiplets:

\bea \nonumber
&& A_l \bar{B}_1 [j;0]^{(2,0;3j+10)}_{3+\frac{j}{2}} \rightarrow A_l \bar{B}_1 [j;0]^{(2;j+2),j+4}_{3+\frac{j}{2}} + L \bar{B}_1 [j;0]^{(1;j+4),j+3}_{3+\frac{j}{2}} + L \bar{B}_1 [j;0]^{(0;j+6),j+2}_{3+\frac{j}{2}} \\ \nonumber & + & A_1 \bar{B}_1 [j+1;0]^{(2;j+3),j+3}_{\frac{7+j}{2}} + L \bar{B}_1 [j+1;0]^{(1;j+5),j+2}_{\frac{7+j}{2}} + L \bar{B}_1 [j+1;0]^{(0;j+7),j+1}_{\frac{7+j}{2}} ,
\eea 

\bea \nonumber
&& A_l \bar{B}_1 [j;0]^{(1,1;3j+6)}_{3+\frac{j}{2}} \rightarrow A_l \bar{B}_1 [j;0]^{(2;j+2),j+2}_{3+\frac{j}{2}} + L \bar{A}_{\bar{2}} [j;0]^{(0;j+2),j+2}_{3+\frac{j}{2}} + L \bar{B}_1 [j;0]^{(1;j+4),j+1}_{3+\frac{j}{2}} \\ \nonumber & + & A_l \bar{A}_{\bar{2}} [j;0]^{(1;j),j+3}_{3+\frac{j}{2}} + A_1 \bar{B}_1 [j+1;0]^{(2;j+3),j+1}_{\frac{7+j}{2}} + L \bar{A}_{\bar{2}} [j+1;0]^{(0;j+3),j+1}_{\frac{7+j}{2}} + L \bar{B}_1 [j+1;0]^{(1;j+5),j}_{\frac{7+j}{2}} \\ \nonumber & + & A_1 \bar{A}_{\bar{2}} [j+1;0]^{(1;j+1),j+2}_{\frac{7+j}{2}} ,
\eea 

\bea \nonumber
&& A_l \bar{B}_1 [j;0]^{(0,2;3j+2)}_{3+\frac{j}{2}} \rightarrow A_l \bar{B}_1 [j;0]^{(2;j+2),j}_{3+\frac{j}{2}} + A_l \bar{A}_{\bar{2}} [j;0]^{(1;j),j+1}_{3+\frac{j}{2}} + A_l \bar{L} [j;0]^{(0;j-2),j+2}_{3+\frac{j}{2}} \\ \nonumber & + & A_1 \bar{B}_1 [j+1;0]^{(2;j+3),j-1}_{\frac{7+j}{2}} + A_1 \bar{A}_{\bar{2}} [j+1;0]^{(1;j+1),j}_{\frac{7+j}{2}} + A_1 \bar{L} [j+1;0]^{(0;j-1),j+1}_{\frac{7+j}{2}} ,
\eea 

and next in terms of $\mathcal{N}=1$ multiplets:

\bea \nonumber
&& A_l \bar{B}_1 [j;0]^{(2,0;3j+10)}_{3+\frac{j}{2}} \rightarrow L \bar{B}_1 [j;0]^{(\frac{6+j}{3}),(2;2j+6)}_{3+\frac{j}{2}} + L \bar{A}_{\bar{2}} [j;0]^{(\frac{2+j}{3}),(1;2j+7)}_{3+\frac{j}{2}} + A_1 \bar{L} [j;0]^{(\frac{j-2}{3}),(0;2j+8)}_{3+\frac{j}{2}} \\ \nonumber & + & L \bar{B}_1 [j+1;0]^{(\frac{7+j}{3}),(3;2j+5)}_{\frac{7+j}{2}} + L \bar{B}_1 [j+1;0]^{(\frac{7+j}{3}),(1;2j+5)}_{\frac{7+j}{2}} + L \bar{B}_1 [j-1;0]^{(\frac{7+j}{3}),(1;2j+5)}_{\frac{7+j}{2}} \\ \nonumber & + & L \bar{A}_{\bar{2}} [j+1;0]^{(\frac{3+j}{3}),(2;2j+6)}_{\frac{7+j}{2}} + L \bar{A}_{\bar{2}} [j+1;0]^{(\frac{3+j}{3}),(0;2j+6)}_{\frac{7+j}{2}} + L \bar{A}_{\bar{2}} [j-1;0]^{(\frac{3+j}{3}),(0;2j+6)}_{\frac{7+j}{2}} \\ \nonumber & + & A_1 \bar{L} [j+1;0]^{(\frac{j-1}{3}),(1;2j+7)}_{\frac{7+j}{2}} + L \bar{B}_1 [j+2;0]^{(\frac{8+j}{3}),(2;2j+4)}_{\frac{8+j}{2}} + L \bar{A}_{\bar{2}} [j+2;0]^{(\frac{4+j}{3}),(1;2j+5)}_{\frac{8+j}{2}} \\ \nonumber & + & A_1 \bar{L} [j+2;0]^{(\frac{j}{3}),(0;2j+6)}_{\frac{8+j}{2}} + L \bar{B}_1 [j;0]^{(\frac{8+j}{3}),(2;2j+4)}_{\frac{8+j}{2}} + L \bar{B}_1 [j;0]^{(\frac{8+j}{3}),(0;2j+4)}_{\frac{8+j}{2}} + L \bar{A}_{\bar{2}} [j;0]^{(\frac{4+j}{3}),(1;2j+5)}_{\frac{8+j}{2}} \\ \nonumber & + & L \bar{B}_1 [j+1;0]^{(\frac{9+j}{3}),(1;2j+3)}_{\frac{9+j}{2}} + L \bar{A}_{\bar{2}} [j+1;0]^{(\frac{5+j}{3}),(0;2j+4)}_{\frac{9+j}{2}},
\eea 

\bea \nonumber
&& A_l \bar{B}_1 [j;0]^{(1,1;3j+6)}_{3+\frac{j}{2}} \rightarrow L \bar{B}_1 [j;0]^{(\frac{6+j}{3}),(1;2j+3)}_{3+\frac{j}{2}} + L \bar{A}_{\bar{2}} [j;0]^{(\frac{2+j}{3}),(2\oplus 0;2j+4)}_{3+\frac{j}{2}} + A_1 \bar{L} [j;0]^{(\frac{j-2}{3}),(1;2j+5)}_{3+\frac{j}{2}} \\ \nonumber & + & L \bar{B}_1 [j+1;0]^{(\frac{7+j}{3}),(2\oplus 0;2j+2)}_{\frac{7+j}{2}} + L \bar{B}_1 [j-1;0]^{(\frac{7+j}{3}),(0;2j+2)}_{\frac{7+j}{2}} + L \bar{A}_{\bar{2}} [j+1;0]^{(\frac{3+j}{3}),(3;2j+3)}_{\frac{7+j}{2}} \\ \nonumber & + & 2 L \bar{A}_{\bar{2}} [j+1;0]^{(\frac{3+j}{3}),(1;2j+3)}_{\frac{7+j}{2}} + L \bar{A}_{\bar{2}} [j-1;0]^{(\frac{3+j}{3}),(1;2j+3)}_{\frac{7+j}{2}} + A_1 \bar{L} [j+1;0]^{(\frac{j-1}{3}),(2\oplus 0;2j+4)}_{\frac{7+j}{2}} \\ \nonumber & + & L \bar{A}_{\bar{2}} [j;1]^{(\frac{1+j}{3}),(1;2j+5)}_{\frac{7+j}{2}} + A_1 \bar{L} [j;1]^{(\frac{j-3}{3}),(0;2j+6)}_{\frac{7+j}{2}} + L \bar{B}_1 [j+2;0]^{(\frac{8+j}{3}),(1;2j+1)}_{\frac{8+j}{2}} + L \bar{A}_{\bar{2}} [j+2;0]^{(\frac{4+j}{3}),(2;2j+2)}_{\frac{8+j}{2}} \\ \nonumber & + & L \bar{A}_{\bar{2}} [j+2;0]^{(\frac{4+j}{3}),(0;2j+2)}_{\frac{8+j}{2}} + A_1 \bar{L} [j+2;0]^{(\frac{j}{3}),(1;2j+3)}_{\frac{8+j}{2}} + L \bar{B}_1 [j;0]^{(\frac{8+j}{3}),(1;2j+1)}_{\frac{8+j}{2}} + L \bar{A}_{\bar{2}} [j;0]^{(\frac{4+j}{3}),(2\oplus 0;2j+2)}_{\frac{8+j}{2}} \\ \nonumber & + & L \bar{A}_{\bar{1}} [j+1;1]^{(\frac{2+j}{3}),(2\oplus 0;2j+4)}_{\frac{8+j}{2}} + A_1 \bar{L} [j+1;1]^{(\frac{j-2}{3}),(1;2j+5)}_{\frac{8+j}{2}} + L \bar{A}_{\bar{1}} [j-1;1]^{(\frac{2+j}{3}),(0;2j+4)}_{\frac{8+j}{2}} \\ \nonumber & + & L \bar{B}_1 [j+1;0]^{(\frac{9+j}{3}),(0;2j)}_{\frac{9+j}{2}} + L \bar{A}_{\bar{2}} [j+1;0]^{(\frac{5+j}{3}),(1;2j+1)}_{\frac{9+j}{2}} + L \bar{A}_{\bar{1}} [j+2;1]^{(\frac{3+j}{3}),(1;2j+3)}_{\frac{9+j}{2}} \\ \nonumber & + & L \bar{A}_{\bar{1}} [j;1]^{(\frac{3+j}{3}),(1;2j+3)}_{\frac{9+j}{2}} + A_1 \bar{L} [j+2;1]^{(\frac{j-1}{3}),(0;2j+4)}_{\frac{9+j}{2}} + L \bar{A}_{\bar{1}} [j+1;1]^{(\frac{4+j}{3}),(0;2j+2)}_{\frac{10+j}{2}},
\eea

\bea \nonumber
&& A_l \bar{B}_1 [j;0]^{(0,2;3j+2)}_{3+\frac{j}{2}} \rightarrow L \bar{B}_1 [j;0]^{(\frac{6+j}{3}),(0;2j)}_{3+\frac{j}{2}} + L \bar{A}_{\bar{2}} [j;0]^{(\frac{2+j}{3}),(1;2j+1)}_{3+\frac{j}{2}} + A_1 \bar{L} [j;0]^{(\frac{j-2}{3}),(2;2j+2)}_{3+\frac{j}{2}} \\ \nonumber & + & L \bar{B}_1 [j+1;0]^{(\frac{7+j}{3}),(1;2j-1)}_{\frac{7+j}{2}} + L \bar{A}_{\bar{2}} [j+1;0]^{(\frac{3+j}{3}),(2\oplus 0;2j)}_{\frac{7+j}{2}} + A_1 \bar{L} [j+1;0]^{(\frac{j-1}{3}),(3\oplus 1;2j+1)}_{\frac{7+j}{2}} \\ \nonumber & + & L \bar{A}_{\bar{1}} [j;1]^{(\frac{1+j}{3}),(0;2j+2)}_{\frac{7+j}{2}} + A_l \bar{L} [j;1]^{(\frac{j-3}{3}),(1;2j+3)}_{\frac{7+j}{2}} + L \bar{B}_1 [j+2;0]^{(\frac{8+j}{3}),(0;2j-2)}_{\frac{8+j}{2}} + L \bar{A}_{\bar{2}} [j+2;0]^{(\frac{4+j}{3}),(1;2j-1)}_{\frac{8+j}{2}} \\ \nonumber & + & A_1 \bar{L} [j+2;0]^{(\frac{j}{3}),(2;2j)}_{\frac{8+j}{2}} + L \bar{A}_{\bar{1}} [j+1;1]^{(\frac{2+j}{3}),(1;2j+1)}_{\frac{8+j}{2}} + A_1 \bar{L} [j+1;1]^{(\frac{j-2}{3}),(2\oplus 0;2j+2)}_{\frac{8+j}{2}} \\ \nonumber & + & A_l \bar{L} [j;0]^{(\frac{j-4}{3}),(0;2j+4)}_{\frac{8+j}{2}} + L \bar{A}_{\bar{1}} [j+2;1]^{(\frac{3+j}{3}),(0;2j)}_{\frac{9+j}{2}} + A_1 \bar{L} [j+2;1]^{(\frac{j-1}{3}),(1;2j+1)}_{\frac{9+j}{2}} \\ \nonumber & + & A_1 \bar{L} [j+1;0]^{(\frac{j-3}{3}),(1;2j+3)}_{\frac{9+j}{2}} + A_1 \bar{L} [j+2;0]^{(\frac{j-2}{3}),(0;2j+2)}_{\frac{10+j}{2}}.
\eea

\subsubsection*{$L \bar{B}_1$}

The next type of short multiplets is $L \bar{B}_1$, whose ground state carry the charges $[j;0]^{(R_1,R_2;r)}_{\frac{2}{3}(R_1+2R_2)+\frac{r}{6}}$. Here, $r$ is not fixed but must obey $r>6+3j+2(R_1-R_2)$. The decompositions for some selected cases are:

\be \nonumber
L \bar{B}_1 [j;0]^{(0,0;r>3j+6)}_{\frac{r}{6}} \rightarrow L \bar{B}_1 [j;0]^{(0;\frac{r}{3}),\frac{r}{3}}_{\frac{r}{6}} + L \bar{B}_1 [j\pm 1;0]^{(0;\frac{3+r}{3}),\frac{r-3}{3}}_{\frac{3+r}{6}} + L \bar{B}_1 [j;0]^{(0;\frac{r+6}{3}),\frac{r-6}{3}}_{1+\frac{r}{6}} ,
\ee 

\bea \nonumber
& & L \bar{B}_1 [j;0]^{(1,0;r>3j+8)}_{\frac{r+4}{6}} \rightarrow L \bar{B}_1 [j;0]^{(1;\frac{r-2}{3}),\frac{r+1}{3}}_{\frac{r+4}{6}} + L \bar{B}_1 [j;0]^{(0;\frac{r+4}{3}),\frac{r-2}{3}}_{\frac{r+4}{6}} + L \bar{B}_1 [j\pm 1;0]^{(1;\frac{1+r}{3}),\frac{r-2}{3}}_{\frac{7+r}{6}} \\ \nonumber & + & L \bar{B}_1 [j\pm 1;0]^{(0;\frac{7+r}{3}),\frac{r-5}{3}}_{\frac{7+r}{6}} + L \bar{B}_1 [j;0]^{(1;\frac{r+4}{3}),\frac{r-5}{3}}_{\frac{r+10}{6}} + L \bar{B}_1 [j;0]^{(0;\frac{r+10}{3}),\frac{r-8}{3}}_{\frac{r+10}{6}},
\eea

\bea \nonumber
& & L \bar{B}_1 [j;0]^{(0,1;r>3j+4)}_{\frac{r+8}{6}} \rightarrow L \bar{B}_1 [j;0]^{(1;\frac{r+2}{3}),\frac{r-1}{3}}_{\frac{r+8}{6}} + L \bar{A}_{\bar{2}} [j;0]^{(0;\frac{r-4}{3}),\frac{r+2}{3}}_{\frac{r+8}{6}} + L \bar{B}_1 [j\pm 1;0]^{(1;\frac{r+5}{3}),\frac{r-4}{3}}_{\frac{r+11}{6}} \\ \nonumber & + & L \bar{A}_{\bar{2}} [j\pm 1;0]^{(0;\frac{r-1}{3}),\frac{r-1}{3}}_{\frac{r+11}{6}} + L \bar{B}_1 [j;0]^{(1;\frac{r+8}{3}),\frac{r-7}{3}}_{\frac{r+14}{6}} + L \bar{A}_{\bar{2}} [j;0]^{(0;\frac{r+2}{3}),\frac{r-4}{3}}_{\frac{r+14}{6}},
\eea

in terms of $\mathcal{N}=2$ multiplets, and:

\bea \nonumber
& & L \bar{B}_1 [j;0]^{(0,0;r>3j+6)}_{\frac{r}{6}} \rightarrow L \bar{B}_1 [j;0]^{(\frac{r}{9}),(0;\frac{2r}{3})}_{\frac{r}{6}} + L \bar{B}_1 [j\pm 1;0]^{(\frac{r+3}{9}),(1;\frac{2r-3}{3})}_{\frac{3+r}{6}} + L \bar{B}_1 [j\pm 2;0]^{(\frac{r+6}{9}),(0;\frac{2r-6}{3})}_{1+\frac{r}{6}} \\ \nonumber & + & L \bar{B}_1 [j;0]^{(\frac{r+6}{9}),(2;\frac{2r-6}{3})}_{1+\frac{r}{6}} + \epsilon^{0 j} L \bar{B}_1 [j;0]^{(\frac{r+6}{9}),(0;\frac{2r-6}{3})}_{1+\frac{r}{6}} + L \bar{B}_1 [j\pm 1;0]^{(\frac{r+9}{9}),(1;\frac{2r-9}{3})}_{\frac{9+r}{6}} \\ \nonumber & + & L \bar{B}_1 [j;0]^{(\frac{r+12}{9}),(0;\frac{2r-12}{3})}_{\frac{r+12}{6}} ,
\eea

\bea \nonumber
& & L \bar{B}_1 [j;0]^{(1,0;r>3j+8)}_{\frac{r+4}{6}} \rightarrow L \bar{B}_1 [j;0]^{(\frac{r+4}{9}),(1;\frac{2r-1}{3})}_{\frac{r+4}{6}} + L \bar{A}_{\bar{2}} [j;0]^{(\frac{r-8}{9}),(0;\frac{2r+2}{3})}_{\frac{r+4}{6}} + L \bar{B}_1 [j\pm 1;0]^{(\frac{r+7}{9}),(2\oplus 0;\frac{2r-4}{3})}_{\frac{7+r}{6}} \\ \nonumber & + & L \bar{A}_{\bar{2}} [j\pm 1;0]^{(\frac{r-5}{9}),(1;\frac{2r-1}{3})}_{\frac{7+r}{6}} + L \bar{B}_1 [j\pm 2;0]^{(\frac{r+10}{9}),(1;\frac{2r-7}{3})}_{\frac{r+10}{6}} + L \bar{B}_1 [j;0]^{(\frac{r+10}{9}),(3\oplus 1;\frac{2r-7}{3})}_{\frac{r+10}{6}} \\ \nonumber & + & \epsilon^{0 j} L \bar{B}_1 [j;0]^{(\frac{r+10}{9}),(1;\frac{2r-7}{3})}_{\frac{r+10}{6}} + L \bar{A}_{\bar{2}} [j\pm 2;0]^{(\frac{r-2}{9}),(0;\frac{2r-4}{3})}_{\frac{r+10}{6}} + L \bar{A}_{\bar{2}} [j;0]^{(\frac{r-2}{9}),(2;\frac{2r-4}{3})}_{\frac{r+10}{6}} \\ \nonumber & + & \epsilon^{0 j} L \bar{A}_{\bar{2}} [j;0]^{(\frac{r-2}{9}),(0;\frac{2r-4}{3})}_{\frac{r+10}{6}} + L \bar{B}_1 [j\pm 1;0]^{(\frac{r+13}{9}),(2\oplus 0;\frac{2r-10}{3})}_{\frac{13+r}{6}} + L \bar{A}_{\bar{2}} [j\pm 1;0]^{(\frac{r+1}{9}),(1;\frac{2r-7}{3})}_{\frac{r+13}{6}} \\ \nonumber & + & L \bar{B}_1 [j;0]^{(\frac{r+16}{9}),(1;\frac{2r-13}{3})}_{\frac{r+16}{6}} + L \bar{A}_{\bar{2}} [j;0]^{(\frac{r+4}{9}),(0;\frac{2r-10}{3})}_{\frac{r+16}{6}} ,
\eea

\bea \nonumber
& & L \bar{B}_1 [j;0]^{(0,1;r>3j+4)}_{\frac{r+8}{6}} \rightarrow L \bar{B}_1 [j;0]^{(\frac{r+8}{9}),(0;\frac{2r-2}{3})}_{\frac{r+8}{6}} + L \bar{A}_{\bar{2}} [j;0]^{(\frac{r-4}{9}),(1;\frac{2r+1}{3})}_{\frac{r+8}{6}} + L \bar{B}_1 [j\pm 1;0]^{(\frac{r+11}{9}),(1;\frac{2r-5}{3})}_{\frac{r+11}{6}} \\ \nonumber & + & L \bar{A}_{\bar{2}} [j\pm 1;0]^{(\frac{r-1}{9}),(2\oplus 0;\frac{2r-2}{3})}_{\frac{r+11}{6}} + L \bar{A}_{\bar{1}} [j;1]^{(\frac{r-7}{9}),(0;\frac{2r+4}{3})}_{\frac{r+11}{6}} + L \bar{B}_1 [j\pm 2;0]^{(\frac{r+14}{9}),(0;\frac{2r-8}{3})}_{\frac{r+14}{6}} \\ \nonumber & + & L \bar{B}_1 [j;0]^{(\frac{r+14}{9}),(2;\frac{2r-8}{3})}_{\frac{r+14}{6}} + \epsilon^{0 j} L \bar{B}_1 [j;0]^{(\frac{r+14}{9}),(0;\frac{2r-8}{3})}_{\frac{r+14}{6}} + L \bar{A}_{\bar{2}} [j\pm 2;0]^{(\frac{r+2}{9}),(1;\frac{2r-5}{3})}_{\frac{r+14}{6}} \\ \nonumber & + & L \bar{A}_{\bar{2}} [j;0]^{(\frac{r+2}{9}),(3\oplus 1;\frac{2r-5}{3})}_{\frac{r+14}{6}} + \epsilon^{0 j} L \bar{A}_{\bar{2}} [j;0]^{(\frac{r+2}{9}),(1;\frac{2r-5}{3})}_{\frac{r+14}{6}} + L \bar{A}_{\bar{1}} [j\pm 1;1]^{(\frac{r-4}{9}),(1;\frac{2r+1}{3})}_{\frac{r+14}{6}} \\ \nonumber & + & L \bar{B}_1 [j\pm 1;0]^{(\frac{r+17}{9}),(1;\frac{2r-11}{3})}_{\frac{r+17}{6}} + L \bar{A}_{\bar{2}} [j\pm 1;0]^{(\frac{r+5}{9}),(2\oplus 0;\frac{2r-8}{3})}_{\frac{r+17}{6}} +  L \bar{A}_{\bar{1}} [j;1]^{(\frac{r-1}{9}),(2;\frac{2r-2}{3})}_{\frac{r+17}{6}} \\ \nonumber & + & \epsilon^{0 j} L \bar{A}_{\bar{1}} [j;1]^{(\frac{r-1}{9}),(0;\frac{2r-2}{3})}_{\frac{r+17}{6}} +  L \bar{A}_{\bar{1}} [j\pm 2;1]^{(\frac{r-1}{9}),(0;\frac{2r-2}{3})}_{\frac{r+17}{6}} + L \bar{B}_1 [j;0]^{(\frac{r+20}{9}),(0;\frac{2r-14}{3})}_{\frac{r+20}{6}} \\ \nonumber & + & L \bar{A}_{\bar{2}} [j;0]^{(\frac{r+1}{9}),(1;\frac{2r-11}{3})}_{\frac{r+20}{6}} +  L \bar{A}_{\bar{1}} [j\pm 1;1]^{(\frac{r+2}{9}),(1;\frac{2r-5}{3})}_{\frac{r+20}{6}} +  L \bar{A}_{\bar{1}} [j;1]^{(\frac{r+5}{9}),(0;\frac{2r-8}{3})}_{\frac{r+23}{6}} ,
\eea

in terms of $\mathcal{N}=1$ multiplets.

We will not have need for the decomposition of other multiplets.

\paragraph{Applications}
\label{N3app}

Like in the $\mathcal{N}=2$ case, we can consider the implications of the decompositions we found on the possible SUSY preserving relevant and marginal deformations. As we are dealing with $\mathcal{N}=3$ SCFTs, there are no flavor symmetries, and all the operators are only charged under the $\mathcal{N}=3$ superconformal symmetry.

\subparagraph*{Relevant operators with dimension $1<\Delta<2$}

The only $\mathcal{N}=3$ multiplets containing $L\bar{B}_1 [0;0]^{(r)}_{\frac{3}{2}r}$ type multiplets with $\frac{2}{3}<r<\frac{4}{3}$ are the $L\bar{B}_1 [0;0]^{(0,0;r')}_{\frac{r'}{6}}$ type ones, for $6<r'<12$. However, looking at the decomposition of these operators into $\mathcal{N}=2$ multiplets, it is apparent that these contain spinning Coulomb branch operators, and so from the results of \cite{Man}, cannot appear in physical theories. This also implies that we can ignore this type of multiplets when considering SUSY preserving operators of dimension $\Delta\geq 2$. We therefore conclude that $\mathcal{N}=3$ SCFTs have no SUSY preserving relevant operators of dimension $1<\Delta<2$. 

\subparagraph*{Relevant operators with dimension $\Delta=2$}

Ignoring multiplets containing higher spin currents or spinning Coulomb branch operators, there are two types of $\mathcal{N}=3$ multiplets containing $L\bar{B}_1 [0;0]^{(\frac{4}{3})}_{2}$ type multiplets. The first are the $B_1\bar{B}_1 [0;0]^{(1,1;0)}_{2}$ type multiplets, corresponding to the energy-momentum tensor. As such, these are always present in any $\mathcal{N}=3$ SCFT, and contain two relevant deformations. From the $\mathcal{N}=2$ viewpoint, these correspond to the moment map of the $U(1)_G$ flavor symmetry and the relevant operator in the $\mathcal{N}=2$ extra SUSY current multiplet. From the $\mathcal{N}=1$ viewpoint, these give two relevant operators that form a doublet under the $SU(2)$ and carry $U(1)_v$ charge $-1$. The $\mathcal{N}=2$ decomposition implies that one preserves the $\mathcal{N}=2$ SUSY while the other breaks it to $\mathcal{N}=1$, but the $\mathcal{N}=1$ viewpoint implies the two deformations are related by an $SU(2)$ transformation, that is non other then part of the $\mathcal{N}=3$ R-symmmetry. Taking both of these into account, we conclude that both of these deformations individually preserve $\mathcal{N}=2$, but together they preserve only $\mathcal{N}=1$. In other words, the deformations preserve different $\mathcal{N}=2$ subgroups of the $\mathcal{N}=3$ supersymmetry group.

The second type of multiplets are the $B_1\bar{B}_1 [0;0]^{(2,0;4)}_{2}$ type multiplets and their complex conjugates, corresponding to additional SUSY currents. Together these contain four relevant deformations. From the $\mathcal{N}=2$ viewpoint, these correspond to the moment maps associated with extra currents enhancing $U(1)_G$ to $SU(2)$, a dimension two Coulomb branch operator, and the relevant operator in the $\mathcal{N}=2$ extra SUSY current multiplet. From the $\mathcal{N}=1$ viewpoint, these give four relevant operators that form a triplet under the $SU(2)$ with $U(1)_v$ charge $2$, and an $SU(2)$ singlet with $U(1)_v$ charge $-4$. As a result, we see that generically

\be \nonumber
n_{\Delta=2} = 4N_{SUSY} - 10 ,
\ee 
where $n_{\Delta=2}$ is the number of SUSY preserving dimension two operators, and $N_{SUSY}$ is the number of supercharges, assumed to be at least three.

\subparagraph*{Relevant operators with dimension $2<\Delta<3$}

The only $\mathcal{N}=3$ multiplets containing $L\bar{B}_1 [0;0]^{(r)}_{\frac{3}{2}r}$ type multiplets with $\frac{4}{3}<r<3$ are the $L\bar{B}_1 [0;0]^{(0,0;r')}_{\frac{r'}{6}}$, $L\bar{B}_1 [0;0]^{(1,0;r')}_{\frac{r'+4}{6}}$ and $L\bar{B}_1 [0;0]^{(0,1;r')}_{\frac{r'+8}{6}}$ type ones, for appropriate values of $r'$. However, we have already noted that $L\bar{B}_1 [0;0]^{(0,0;r')}_{\frac{r'}{6}}$ multiplets cannot appear in physical theories. Furthermore, from the decomposition of these operators into $\mathcal{N}=2$ multiplets we see that also $L\bar{B}_1 [0;0]^{(1,0;r')}_{\frac{r'+4}{6}}$ contains spinning Coulomb branch operators and so is not allowed. This leaves us only with the $L\bar{B}_1 [0;0]^{(0,1;r')}_{\frac{r'+8}{6}}$ type multiplets for $4<r<10$. From the $\mathcal{N}=2$ viewpoint, these correspond to the relevant operator in a mixed branch operator. From the $\mathcal{N}=1$ viewpoint, this gives a single relevant operator which is an $SU(2)$ singlet with $U(1)_v$ charge $\frac{2r'-2}{3}$.    

\subparagraph*{Marginal operators}

Finally we consider marginal operators. By going over the list, ignoring cases containing higher spin currents or spinning Coulomb branch operators, we find several types of possible $\mathcal{N}=1$ preserving marginal operators, differing by their representation under the $SU(2)\times U(1)_v$ symmetry. These are:

\begin{enumerate}

\item - $L\bar{B}_1 [0;0]^{(2),(3;3)}_{3} \oplus L\bar{B}_1 [0;0]^{(2),(0;-6)}_{3}$ : This combination of multiplets can only come from the $B_1\bar{B}_1 [0;0]^{(3,0;6)}_{3}$ type multiplets together with their complex conjugate $B_1\bar{B}_1 [0;0]^{(0,3;-6)}_{3}$. These contain a dimension three Coulomb branch operator from the $\mathcal{N}=2$ viewpoint.

\item - $L\bar{B}_1 [0;0]^{(2),(2;0)}_{3} \oplus L\bar{B}_1 [0;0]^{(2),(1;-3)}_{3}$ : This combination of multiplets can only come from the $B_1\bar{B}_1 [0;0]^{(2,1;2)}_{3}$ type multiplets together with their complex conjugate $B_1\bar{B}_1 [0;0]^{(1,2;-2)}_{3}$.

\item - $L\bar{B}_1 [0;0]^{(2),(1;3)}_{3}$ : This type of multiplets can only come from the $A_2\bar{B}_1 [0;0]^{(1,1;6)}_{3}$ type multiplets.

\item - $L\bar{B}_1 [0;0]^{(2),(0;6)}_{3}$ : This type of multiplets can only come from the $L\bar{B}_1 [0;0]^{(2),(0,1;10)}_{3}$ type multiplets.

\item - $L\bar{B}_1 [0;0]^{(2),(0;0)}_{3}$ : This type of multiplets can come from one of two sources. One is the $B_1\bar{B}_1 [0;0]^{(2,0;4)}_{2}$ multiplets, where it comes from the top component. As such it preserves the full $\mathcal{N}=3$ SUSY, and in fact as these multiplets give additional SUSY currents, actually preserve $\mathcal{N}=4$. The second type is the $A_2\bar{B}_1 [0;0]^{(0,2;2)}_{3}$ type of multiplets, which only preserves $\mathcal{N}=1$ SUSY.

\end{enumerate}

These results allows us to make some statements about the possible conformal manifolds in $\mathcal{N}=3$ SCFTs. We first note that the $B_1\bar{B}_1 [0;0]^{(3,0;6)}_{3}$ type multiplets, and their complex conjugates, that contain the dimension three Coulomb branch operator, provide five marginal operators that support a Kahler quotient under the $SU(2)\times U(1)_v$ subgroup of the $\mathcal{N}=3$ R-symmetry. This comes about as the $\bold{4}$ of $SU(2)$ has a singlet in its quartic symmetric product. This gives a one dimensional conformal manifold, where four of the marginal operators combine with the broken $SU(2)\times U(1)_v$ currents to form marginally irrelevant operators. Therefore, any theory containing a dimension three Coulomb branch operator has at least a one dimensional conformal manifold along which the $SU(2)\times U(1)_v$ symmetry is broken.

Other notable multiplets are $B_1\bar{B}_1 [0;0]^{(2,1;2)}_{3}$ with its complex conjugate and $A_2\bar{B}_1 [0;0]^{(0,2;2)}_{3}$. For the former, the $L\bar{B}_1 [0;0]^{(2),(2;0)}_{3}$ type operator contained in it forms by itself a Kahler quotient under the $SU(2)\times U(1)_v$ symmetry, while $L\bar{B}_1 [0;0]^{(2),(1;-3)}_{3}$ cannot by itself form such a quotient. This leads to a one dimensional conformal manifold along which the $SU(2)$ is broken to its Cartan and $U(1)_v$ is preserved. As the operators in $L\bar{B}_1 [0;0]^{(2),(1;-3)}_{3}$ cannot form a quotient, they are marginally irrelevant, unless there are additional marginal operators. Finally, the multiplet $A_2\bar{B}_1 [0;0]^{(0,2;2)}_{3}$ contains a true singlet under $SU(2)\times U(1)_v$. As a result, in the presence of such multiplets, there is an $\mathcal{N}=1$ only preserving conformal manifold on which the $SU(2)\times U(1)_v$ remains unbroken.

The remaining multiplets do not form a Kahler quotient by themselves and so can only lead to exactly marginal deformations if they appear in conjunction with other multiplets.

\subsubsection{$\mathcal{N}=4$}

We next consider the case of $\mathcal{N}=4$ SCFTs. We again use the conventions of \cite{CID}, so we take $Q$ to have charges $[1;0]^{(1,0,0)}_{\frac{1}{2}}$. We shall first consider here the decomposition into $\mathcal{N}=3$, where for that, we shall decompose the $SU(4)$ such that $\bold{4}\rightarrow \bold{3}^{-1} + \bold{1}^{3}$, where the fundamental becomes the $\mathcal{N}=3$ supercharges. The remaining cases can in principle be recovered by applying the decompositions in the previous sections. Here we shall also write down the decomposition in terms of $\mathcal{N}=1$ multiplets as that will be useful when considering possible SUSY preserving deformations.

Next, we present the decomposition rules for selected $\mathcal{N}=4$ multiplets:

\subsubsection*{$B_1 \bar{B}_1$}

The shortest multiplet type is the $B_1 \bar{B}_1$ type, whose ground state carry the charges $[0;0]^{(R_1,R_2,R_1)}_{2R_1+R_2}$. Like the previous cases, the $R_1=R_2=0$ case describes the vacuum state, while the $R_1=0, R_2 = 1$ case is the $\mathcal{N}=4$ free vector. The case of $R1=1, R_2=0$ contains higher-spin currents, and the case of $R_1=0, R_2=2$ describes the energy-momentum tensor multiplet. Additional cases of special physical interest are the $R1=R_2=1$ and $R_1=0, R_2=3$ cases, which contain SUSY preserving marginal operators.

We next consider the decomposition of these multiplets under the $\mathcal{N}=3$ subgroup:

\be \nonumber
B_1 \bar{B}_1 [0;0]^{(0,2,0)}_{2} \rightarrow B_1 \bar{B}_1 [0;0]^{(2,0;4)}_{2} + B_1 \bar{B}_1 [0;0]^{(0,2;-4)}_{2} + B_1 \bar{B}_1 [0;0]^{(1,1;0)}_{2}.
\ee 

These are respectively, the extra SUSY currents and the $\mathcal{N}=3$ energy-momentum tensor. 

\be \nonumber
B_1 \bar{B}_1 [0;0]^{(1,0,1)}_{2} \rightarrow A_2 \bar{B}_1 [0;0]^{(0,1;4)}_{2} + B_1 \bar{A}_{\bar{2}} [0;0]^{(1,0;-4)}_{2} + B_1 \bar{B}_1 [0;0]^{(1,1;0)}_{2} + A_2 \bar{A}_{\bar{2}} [0;0]^{(0,0;0)}_{2},
\ee

\be \nonumber
B_1 \bar{B}_1 [0;0]^{(0,3,0)}_{3} \rightarrow B_1 \bar{B}_1 [0;0]^{(3,0;6)}_{3} + B_1 \bar{B}_1 [0;0]^{(0,3;-6)}_{3} + B_1 \bar{B}_1 [0;0]^{(2,1;2)}_{3} + B_1 \bar{B}_1 [0;0]^{(1,2;-2)}_{3},
\ee 

\bea \nonumber
& & B_1 \bar{B}_1 [0;0]^{(1,1,1)}_{3} \rightarrow B_1 \bar{B}_1 [0;0]^{(2,1;2)}_{3} + B_1 \bar{B}_1 [0;0]^{(1,2;-2)}_{3} + A_2 \bar{B}_1 [0;0]^{(1,1;6)}_{3} + B_1 \bar{A}_{\bar{2}} [0;0]^{(1,1;-6)}_{3} \\ \nonumber & + & A_2 \bar{B}_1 [0;0]^{(0,2;2)}_{3} + B_1 \bar{A}_{\bar{2}} [0;0]^{(2,0;-2)}_{3} + A_2 \bar{A}_{\bar{2}} [0;0]^{(1,0;2)}_{3} + A_2 \bar{A}_{\bar{2}} [0;0]^{(0,1;-2)}_{3}.
\eea

We note here that if $R_1=0$ we in general have that\cite{BPP}: 

\be \nonumber
B_1 \bar{B}_1 [0;0]^{(0,R,0)}_{R} \rightarrow \sum^{R}_{i=0} B_1 \bar{B}_1 [0;0]^{(R-i,i;2(R-2i))}_{R}.
\ee

However, cases with $R_1 \neq 0$ will in general contain $\mathcal{N}=3$ multiplets longer than $B_1 \bar{B}_1$. Given the decompositions in the previous section, these will then contain $\mathcal{N}=2$ multiplets whose ground state has spin.

Finally, we consider the decomposition in terms of $\mathcal{N}=1$ multiplets.   

\bea \nonumber
& & B_1 \bar{B}_1 [0;0]^{(0,2,0)}_{2} \rightarrow A_2 \bar{A}_{\bar{2}} [0;0]^{(0),(1,1)}_{2} + L \bar{B}_1 [0;0]^{(\frac{4}{3}),(2,0)}_{2} + B_1 \bar{L} [0;0]^{(-\frac{4}{3}),(0,2)}_{2} \\ \nonumber & + & A_1 \bar{A}_{\bar{2}} [1;0]^{(\frac{1}{3}),(0,1)}_{\frac{5}{2}} + A_2 \bar{A}_{\bar{1}} [0;1]^{(-\frac{1}{3}),(1,0)}_{\frac{5}{2}} + L \bar{B}_1 [1;0]^{(\frac{5}{3}),(1,0)}_{\frac{5}{2}} + B_1 \bar{L} [0;1]^{(-\frac{5}{3}),(0,1)}_{\frac{5}{2}} \\ \nonumber & + & L \bar{B}_1 [0;0]^{(2),(0,0)}_{3} + B_1 \bar{L} [0;0]^{(-2),(0,0)}_{3} + A_1 \bar{A}_{\bar{1}} [1;1]^{(0),(0,0)}_{3}.
\eea

\bea \nonumber
& & B_1 \bar{B}_1 [0;0]^{(0,3,0)}_{3} \rightarrow L \bar{B}_1 [0;0]^{(2),(3,0)}_{3} + B_1 \bar{L} [0;0]^{(-2),(0,3)}_{3} + L \bar{A}_{\bar{2}} [0;0]^{(\frac{2}{3}),(2,1)}_{3} \\ \nonumber & + & A_2 \bar{L} [0;0]^{(-\frac{2}{3}),(1,2)}_{3} + L \bar{B}_{\bar{1}} [1;0]^{(\frac{7}{3}),(2,0)}_{\frac{7}{2}} + B_1 \bar{L} [0;1]^{(-\frac{7}{3}),(0,2)}_{\frac{7}{2}} + L \bar{A}_{\bar{2}} [1;0]^{(1),(1,1)}_{\frac{7}{2}} + A_2 \bar{L} [0;1]^{(-1),(1,1)}_{\frac{7}{2}} \\ \nonumber & + & L \bar{A}_{\bar{1}} [0;1]^{(\frac{1}{3}),(2,0)}_{\frac{7}{2}} + A_1 \bar{L} [1;0]^{(-\frac{1}{3}),(0,2)}_{\frac{7}{2}} + L \bar{B}_1 [0;0]^{(\frac{8}{3}),(1,0)}_{4} + B_1 \bar{L} [0;0]^{(-\frac{8}{3}),(0,1)}_{4} + L \bar{A}_{\bar{2}} [0;0]^{(\frac{4}{3}),(0,1)}_{4} \\ \nonumber & + & A_2 \bar{L} [0;0]^{(-\frac{4}{3}),(1,0)}_{4} + L \bar{A}_{\bar{1}} [1;1]^{(\frac{2}{3}),(1,0)}_{4} + A_1 \bar{L} [1;1]^{(-\frac{2}{3}),(0,1)}_{4}.
\eea

\bea \nonumber
& & B_1 \bar{B}_1 [0;0]^{(1,1,1)}_{3} \rightarrow L \bar{B}_1 [0;0]^{(2),(1,1)}_{3} + B_1 \bar{L} [0;0]^{(-2),(1,1)}_{3} + L \bar{A}_{\bar{2}} [0;0]^{(\frac{2}{3}),(2,1)}_{3} \\ \nonumber & + & A_2 \bar{L} [0;0]^{(-\frac{2}{3}),(1,2)}_{3} + L \bar{A}_{\bar{2}} [0;0]^{(\frac{2}{3}),(0,2)}_{3} + A_2 \bar{L} [0;0]^{(-\frac{2}{3}),(2,0)}_{3} + L \bar{A}_{\bar{2}} [0;0]^{(\frac{2}{3}),(1,0)}_{3} \\ \nonumber & + & A_2 \bar{L} [0;0]^{(-\frac{2}{3}),(0,1)}_{3} + L \bar{B}_{\bar{1}} [1;0]^{(\frac{7}{3}),(2,0)}_{\frac{7}{2}} + B_1 \bar{L} [0;1]^{(-\frac{7}{3}),(0,2)}_{\frac{7}{2}} + L \bar{B}_{\bar{1}} [1;0]^{(\frac{7}{3}),(0,1)}_{\frac{7}{2}} \\ \nonumber & + & B_1 \bar{L} [0;1]^{(-\frac{7}{3}),(1,0)}_{\frac{7}{2}} + 2 L \bar{A}_{\bar{2}} [1;0]^{(1),(1,1)}_{\frac{7}{2}} + 2 A_2 \bar{L} [0;1]^{(-1),(1,1)}_{\frac{7}{2}} + L \bar{A}_{\bar{2}} [1;0]^{(1),(3,0)}_{\frac{7}{2}} \\ \nonumber & + & A_2 \bar{L} [0;1]^{(-1),(0,3)}_{\frac{7}{2}} + L \bar{A}_{\bar{1}} [0;1]^{(\frac{1}{3}),(2,0)}_{\frac{7}{2}} + A_1 \bar{L} [1;0]^{(-\frac{1}{3}),(0,2)}_{\frac{7}{2}} + L \bar{A}_{\bar{1}} [0;1]^{(\frac{1}{3}),(1,2)}_{\frac{7}{2}} \\ \nonumber & + & A_1 \bar{L} [1;0]^{(-\frac{1}{3}),(2,1)}_{\frac{7}{2}} + L \bar{A}_{\bar{2}} [1;0]^{(1),(0,0)}_{\frac{7}{2}} + A_2 \bar{L} [0;1]^{(-1),(0,0)}_{\frac{7}{2}} + L \bar{A}_{\bar{1}} [0;1]^{(\frac{1}{3}),(0,1)}_{\frac{7}{2}} \\ \nonumber & + & A_1 \bar{L} [1;0]^{(-\frac{1}{3}),(1,0)}_{\frac{7}{2}} + L \bar{B}_1 [0;0]^{(\frac{8}{3}),(1,0)}_{4} + B_1 \bar{L} [0;0]^{(-\frac{8}{3}),(0,1)}_{4} + ... .
\eea
where the remaining terms have dimension $\Delta\geq 4$, and we will not have need of them here. 

\subsubsection*{$A_l \bar{B}_1$}

The next type of short multiplet is $A_l \bar{B}_1$, whose ground state carry the charges $[j;0]^{(R_1,R_2,2+j+R_1)}_{1+R_1+R_2+R_3+\frac{j}{2}}$. The decompositions into $\mathcal{N}=2$ multiplets for selected cases are: 

\bea \nonumber
&& A_2 \bar{B}_1 [0;0]^{(0,0,2)}_{3} \rightarrow A_2 \bar{B}_1 [0;0]^{(0,2;2)}_{3} + A_2 \bar{A}_{\bar{2}} [0;0]^{(0,1;-2)}_{3} + A_2 \bar{L} [0;0]^{(0,0;-6)}_{3} + A_1 \bar{B}_1 [1;0]^{(0,2;5)}_{\frac{7}{2}} \\ \nonumber & + & A_1 \bar{A}_{\bar{2}} [0;0]^{(0,1;1)}_{\frac{7}{2}} + A_1 \bar{L} [1;0]^{(0,0;-3)}_{\frac{7}{2}} ,
\eea 

In terms of $\mathcal{N}=1$ multiplets, the decomposition is:

\bea \nonumber
&& A_2 \bar{B}_1 [0;0]^{(0,0,2)}_{3} \rightarrow L \bar{B}_1 [0;0]^{(2),(0,0)}_{3} + A_2 \bar{L} [0;0]^{(-\frac{2}{3}),(2,0)}_{3} + L \bar{A}_{\bar{2}} [0;0]^{(\frac{2}{3}),(1,0)}_{3} + L \bar{B}_1 [1;0]^{(\frac{7}{3}),(0,1)}_{\frac{7}{2}} \\ \nonumber & + & A_1 \bar{L} [1;0]^{(-\frac{1}{3}),(2,1)\oplus (1,0)}_{\frac{7}{2}} + A_2 \bar{L} [0;1]^{(-1),(1,1)}_{\frac{7}{2}} + L \bar{A}_{\bar{2}} [1;0]^{(1),(1,1)\oplus (0,0)}_{\frac{7}{2}} + L \bar{A}_{\bar{1}} [0;1]^{(\frac{1}{3}),(0,1)}_{\frac{7}{2}} + ...,
\eea
where the remaining terms have dimension $\Delta\geq 4$, and we will not have need of them. Similarly we will have no need of other types of multiplets here.   

\paragraph{Applications}

Like in the previous cases, we can consider the implications of the decompositions we found on the possible SUSY preserving relevant and marginal deformations. In this case there are no flavor symmetries, and all operators have integer dimensions. 

\subparagraph*{Relevant operators with dimension $\Delta=2$}

Ignoring multiplets containing higher spin currents, then only the $B_1 \bar{B}_1 [0;0]^{(0,2,0)}_{2}$ multiplet, corresponding to the energy-momentum tensor, can contain operators of type $L\bar{B}_1 [0;0]^{(\frac{4}{3})}_{2}$. This type of multiplts contains six of these, that transform in the $\bold{6}$ of the $SU(3)$, which is the commutant of the $\mathcal{N}=1$ $U(1)$ R-symmetry in the $\mathcal{N}=3$ R-symmetry. In Lagrangin theories, these correspond to quadratic invariants of the three adjoint chirals. As in any interacting $\mathcal{N}=4$ SCFT we expect to have one, and only one, energy-momentum tensor, these must contain precisely six dimension two SUSY preserving relevant deformations. 

\subparagraph*{Marginal operators}

We next consider the case of marginal operators. By going over the list, ignoring cases containing higher spin currents, we find a handful of types of possible $\mathcal{N}=1$ preserving marginal operators, differing by their representation under the $SU(3)$ symmetry. These are:

\begin{enumerate}

\item - $L\bar{B}_1 [0;0]^{(2),(3,0)}_{3}$ : These can only come from the $B_1\bar{B}_1 [0;0]^{(0,3,0)}_{3}$ type multiplets. These contain a dimension three Coulomb branch operator from the $\mathcal{N}=2$ viewpoint. In Lagrangian theories, these come from totally symmetric cubic invariants of the three adjoint chirals. This type of multiplets possesses a Kahler quotient, by itself, and leads to a two dimensional conformal manifold on a generic point of which the $SU(3)$ symmetry is broken. 

\item - $L\bar{B}_1 [0;0]^{(2),(1,1)}_{3}$ : This type of multiplets can only come from the $B_1\bar{B}_1 [0;0]^{(1,1,1)}_{3}$ type multiplets. In Lagrangian theories, these in principle can come from cubic invariants of the three adjoint chirals using the mixed symmetry state defined by the partition $[1,1]$. Nevertheless, it is interesting to note that this type of product is not gauge invariant for any simple Lie-group, and as a result, there is no known physical theory containing these types of multiplets, unless it contains decoupled parts. This type of multiplets possesses a Kahler quotient, by itself, and leads to a two dimensional conformal manifold on a generic point of which the $SU(3)$ symmetry is broken down to the Cartan subalgbra, assuming no additional symmetries.

\item - $L\bar{B}_1 [0;0]^{(2),(0,0)}_{3}$ : This type of multiplets can come from one of two sources. One is the $B_1\bar{B}_1 [0;0]^{(0,2,0)}_{2}$ multiplets, where it comes from the top component. As such it preserves the full $\mathcal{N}=4$ SUSY. In Lagrangian theories, this corresponds to the coupling constant. As it comes from the energy-momentum tensor multiplet, it is again expected that any interacting $\mathcal{N}=4$ SCFT has exactly one of these 

 The second type of mutiplets, containing such a deformation, is the $A_2\bar{B}_1 [0;0]^{(0,0,2)}_{3}$ type multiplets. These then would give an $\mathcal{N}=1$ preserving conformal manifold along which the $SU(3)$ symmetry is fully preserved. We are not aware of any $\mathcal{N}=4$ SCFT possessing such multiplets.

\end{enumerate} 

\subsection{Superconformal index}
\label{subsec:sindex}

In this section we summarize the contribution of the various superconformal multiplets to the superconformal index. For this we use the results of \cite{GRRYind}, which determined these relations for the case of $\mathcal{N}=1$ superconformal multiplets. Using the decomposition presented previously, it is straightforward to extend this to cases with extended supersymmetry. The results for the various contributions are:

\be \nonumber
L \bar{B}_1 [j;0]^{(r)}_{\frac{3}{2}r} \rightarrow (-1)^j\frac{t^{3r} \chi_j(y)}{(1-t^3 y)(1-\frac{t^3}{y})},
\ee

\be \nonumber
B_1 \bar{L} [0;\bar{j}]^{(r)}_{-\frac{3}{2}r} \rightarrow 0 ,
\ee

\be \nonumber
A_l \bar{A}_{\bar{l}} [j;\bar{j}]^{(\frac{1}{3}(j-\bar{j}))}_{2+\frac{1}{2}(j+\bar{j})} \rightarrow (-1)^{1+\bar{j}+j}\frac{t^{6+j+2\bar{j}} \chi_j(y)}{(1-t^3 y)(1-\frac{t^3}{y})},
\ee

\be \nonumber
L \bar{A}_{\bar{l}} [j;\bar{j}]^{(r)}_{2+\bar{j}+\frac{3}{2}r} \rightarrow (-1)^{1+\bar{j}+j}\frac{t^{3(2+\bar{j}+r)} \chi_j(y)}{(1-t^3 y)(1-\frac{t^3}{y})},
\ee

\be \nonumber
A_l \bar{L} [j;\bar{j}]^{(r)}_{2+j-\frac{3}{2}r} \rightarrow 0 ,
\ee

\be \nonumber
A_l \bar{B}_1 [j;0]^{(\frac{1}{3}(j+2))}_{1+\frac{j}{2}} \rightarrow (-1)^{j}\frac{(t^{2+j} \chi_j(y)-t^{5+j} \chi_{j-1}(y))}{(1-t^3 y)(1-\frac{t^3}{y})},
\ee

\be \nonumber
B_1 \bar{A}_{\bar{l}} [0;\bar{j}]^{(-\frac{1}{3}(\bar{j}+2))}_{1+\frac{\bar{j}}{2}} \rightarrow (-1)^{1+\bar{j}}\frac{t^{2(2+\bar{j})}}{(1-t^3 y)(1-\frac{t^3}{y})}.
\ee

Here we have deviated from our norm so far, and employed the notation of \cite{GRRYind}. It is related to the notaton employed in the rest of the article by $p=t^3 y$, $q=\frac{t^3}{y}$. We also note that $\chi_{j}(y) = \sum^{\frac{j}{2}}_{i=-\frac{j}{2}} y^{2i}$ for $j$ even, and $\chi_{j}(y) = \sum^{\frac{j-1}{2}}_{i=-\frac{j+1}{2}} y^{2i+1}$ for $j$ odd.


\begin{thebibliography}{40}

\bibitem{GER} 
  I.~G.~Etxebarria, D.~Regalado,
	JHEP {\bf 1603}, 083 (2016)
  [arXiv:1512.06434 [hep-th]].

\bibitem{ATsf} 
  O.~Aharony, Y.~Tachikawa,
	JHEP {\bf 1606}, 044 (2016)
  [arXiv:1602.08638 [hep-th]].

\bibitem{GER1} 
  I.~G.~Etxebarria, D.~Regalado,
	JHEP {\bf 1712}, 042 (2017)
  [arXiv:1611.05769 [hep-th]].

\bibitem{MN} 
  J.~A.~Minahan and D.~Nemeschansky,
  Nucl.\ Phys.\ B {\bf 482}, 142 (1996)
  [hep-th/9608047], 
  Nucl.\ Phys.\ B {\bf 489}, 24 (1997)
  [hep-th/9610076].

\bibitem{AW}
  P.~C.~Argyres, J.~R.~Wittig, 
  JHEP {\bf 0801}, 074 (2008)
  [arXiv:0712.2028 [hep-th]].

\bibitem{Gai} 
  D.~Gaiotto,
  JHEP {\bf 1208}, 034 (2012)
  [arXiv:0904.2715 [hep-th]].

\bibitem{CD}
  O.~Chacaltana, J.~Distler,
  JHEP {\bf 1011}, 099 (2010)
  [arXiv:1008.5203 [hep-th]].

\bibitem{CD1}
  O.~Chacaltana, J.~Distler,
  JHEP {\bf 1302}, 110 (2013)
  [arXiv:1106.5410 [hep-th]].

\bibitem{ArS}
  P.~C.~Argyres, N.~Seiberg, 
  JHEP {\bf 0712}, 088 (2007)
  [arXiv:0711.0054 [hep-th]]. 

\bibitem{ArD} 
  P.~C.~Argyres, M.~R.~Douglas,
	Nucl.\ Phys.\ B {\bf 488}, 93 (1995)
  [arXiv:9505062 [hep-th]].

\bibitem{APSW} 
  P.~C.~Argyres, M.~R.~Plesser, N.~Seiberg and E.~Witten,
	Nucl.\ Phys.\ B {\bf 461}, 71 (1996)
  [arXiv:9511154 [hep-th]].

\bibitem{ALLM} 
  P.~C.~Argyres, M.~Lotito, Y.~Lu, and M.~Martone,
	JHEP {\bf 1605}, 088 (2016)
  [arXiv:1602.02764 [hep-th]].

\bibitem{ZafTW} 
  G.~Zafrir,
	JHEP {\bf 1701}, 097 (2017)
  [arXiv:1605.08337 [hep-th]].

\bibitem{LS}
  R.~G.~Leigh, M.~J.~Strassler,
  Nucl.\ Phys.\ B {\bf 447}, 95 (1995)
  [arXiv:9505088 [hep-th]].

\bibitem{GKSTW}
  D.~Green, Z.~Komargodski, N.~Seiberg, Y.~Tachikawa and B.~Wecht,
  JHEP {\bf 1006}, 106 (2010)
  [arXiv:1005.3546 [hep-th]].

\bibitem{RSZcl} 
  S.~S.~Razamat, E.~Sabag, and G.~Zafrir,
	JHEP {\bf 2006}, 179 (2020)
  [arXiv:2004.07097 [hep-th]].

\bibitem{RZCM} 
  S.~S.~Razamat, G.~Zafrir,
	JHEP {\bf 1909}, 046 (2019)
  [arXiv:1906.05088 [hep-th]]. 

\bibitem{MS} 
  K.~Maruyoshi, J.~Song,
	Phys.\ Rev.\ Lett.\  118 (2017) no.15, 151602
  [arXiv:1606.05632 [hep-th]].

\bibitem{MS1} 
  K.~Maruyoshi, J.~Song,
	JHEP {\bf 1702}, 075 (2017)
  [arXiv:1607.04281 [hep-th]].

\bibitem{AMSad} 
  P.~Agarwal, K.~Maruyoshi and J.~Song,
	JHEP {\bf 1612}, 103 (2016)
  [arXiv:1610.05311 [hep-th]].	

\bibitem{ASS} 
  P.~Agarwal, A.~Sciarappa and J.~Song,
	JHEP {\bf 1710}, 211 (2017)
  [arXiv:1707.04751 [hep-th]].

\bibitem{BGad} 
  S.~Benvenuti, S.~Giacomelli,
	JHEP {\bf 1710}, 106 (2017)
  [arXiv:1707.05113 [hep-th]].

\bibitem{MNS} 
  K.~Maruyoshi, E.~Nardoni and J.~Song,
	Phys.\ Rev.\ Lett.\  122 (2019) no.12, 121601
  [arXiv:1806.08353 [hep-th]].	

\bibitem{GRW} 
  A.~Gadde, S.~S.~Razamat and B.~Willett,
	Phys.\ Rev.\ Lett.\  115 (2015) no.17, 171604
  [arXiv:1505.05834 [hep-th]].	
	
\bibitem{AMS} 
  P.~Agarwal, K.~Maruyoshi and J.~Song,
	JHEP {\bf 1805}, 193 (2018)
  [arXiv:1802.05268 [hep-th]].	

\bibitem{ZafE6} 
  G.~Zafrir,
  [arXiv:1912.09348 [hep-th]].

\bibitem{NT} 
  T.~Nishinaka, Y.~Tachikawa,
	JHEP {\bf 1609}, 116 (2016)
  [arXiv:1602.01503 [hep-th]].

\bibitem{ISin} 
  Y.~Imamura, S.~Yokoyama,
	J.\ Phys.\ A49, 43, 435401 (2016)
  [arXiv:1603.00851 [hep-th]].

\bibitem{AIin} 
  R.~Arai, Y.~Imamura,
	PTEP {\bf 2019}, no. 8, 083B04 (2019)
  [arXiv:1904.09776 [hep-th]].

\bibitem{Evt1} 
  M.~Evtikhiev,
	JHEP {\bf 1804}, 120 (2018)
  [arXiv:1708.08307 [hep-th]].

\bibitem{CC} 
  M.~Caorsi, S.~Cecotti,
	JHEP {\bf 1807}, 138 (2018)
  [arXiv:1801.04542 [hep-th]].

\bibitem{BMRl} 
  F.~Bonetti, C.~Meneghelli, and L.~Rastelli,
	JHEP {\bf 1905}, 155 (2019)
  [arXiv:1810.03612 [hep-th]].

\bibitem{TyZ} 
  Y.~Tachikawa, G.~Zafrir,
	JHEP {\bf 1912}, 176 (2019)
  [arXiv:1908.03346 [hep-th]].

\bibitem{ABMm1} 
  P.~C.~Argyres, A.~Bourget, and M.~Martone,
  [arXiv:1904.10969 [hep-th]].

\bibitem{ABMm2} 
  P.~C.~Argyres, A.~Bourget, and M.~Martone,
  [arXiv:1912.04926 [hep-th]].

\bibitem{CID} 
  C.~Cordova, T.~T.~Dumitrescu, and K.~Intriligator,
	JHEP {\bf 1903}, 163 (2019)
  [arXiv:1612.00809 [hep-th]].

\bibitem{LLMM} 
  M.~Lemos, P.~Liendo, C.~Meneghelli, and V.~Mitev,
	JHEP {\bf 1704}, 032 (2017)
  [arXiv:1612.01536 [hep-th]].

\bibitem{BPP} 
  T.~Bourton, A.~Pini, and E.~Pomoni,
	JHEP {\bf 1810}, 131 (2018)
  [arXiv:1804.05396 [hep-th]].

\bibitem{AoMe} 
  O.~Aharony, M.~Evtikhiev,
	JHEP {\bf 1604}, 040 (2016)
  [arXiv:1512.03524 [hep-th]].

\bibitem{ShTa} 
  A.~D.~Shapere, Y.~Tachikawa,
	JHEP {\bf 0809}, 109 (2008)
  [arXiv:0804.1957 [hep-th]].

\bibitem{AMdc} 
  P.~C.~Argyres, M.~Martone,
	JHEP {\bf 1703}, 145 (2017)
  [arXiv:1611.08602 [hep-th]].

\bibitem{Evt2} 
  M.~Evtikhiev,
	JHEP {\bf 2006}, 125 (2020)
  [arXiv:2004.03919 [hep-th]].

\bibitem{ALLM00} 
  P.~C.~Argyres, M.~Lotito, Y.~Lu, and M.~Martone,
	JHEP {\bf 1802}, 001 (2018)
  [arXiv:1505.04814 [hep-th]].

\bibitem{ALLM0} 
  P.~C.~Argyres, M.~Lotito, Y.~Lu, and M.~Martone,
	JHEP {\bf 1802}, 002 (2018)
  [arXiv:1601.00011 [hep-th]].

\bibitem{ALLM1} 
  P.~C.~Argyres, M.~Lotito, Y.~Lu, and M.~Martone,
	JHEP {\bf 1802}, 003 (2018)
  [arXiv:1609.04404 [hep-th]].

\bibitem{Amax} 
  K.~A.~Intriligator, B.~Wecht,
	Nucl.\ Phys.\ B667, 183-200 (2003)
  [arXiv:0304128 [hep-th]].

\bibitem{KPS} 
  D.~Kutasov, A.~Parnachev, and D.~A.~Sahakyan,
	JHEP {\bf 0311}, 013 (2003)
  [arXiv:0308071 [hep-th]].

\bibitem{Index} 
  J.~Kinney, J.~Maldacena, S.~Minwalla, and S.~Raju,
	Commun.\ Math.\ Phys.\ 275, 209-254 (2007)
  [arXiv:0510251 [hep-th]].

\bibitem{BG} 
  S.~Benvenuti, S.~Giacomelli,
	Phys.\ Rev.\ Lett.\  119 (2017) no.25, 251601
  [arXiv:1706.02225 [hep-th]].

\bibitem{DO} 
  F.~A.~Dolan, H.~Osborn,
	Nucl.\ Phys.\ B818, 137-178 (2009)
  [arXiv:0801.4947 [hep-th]].

\bibitem{RR} 
  L.~Rastelli, S.~S.~Razamat,
  [arXiv:1608.02965 [hep-th]].

\bibitem{GRRYpol}
  A.~Gadde, L.~Rastelli, S.~S.~Razamat, and W.~Yan,
	Commun.\ Math.\ Phys.\ 252:359-391 (2004)
  [arXiv:1110.3740 [hep-th]].

\bibitem{RZAS} 
  S.~S.~Razamat, G.~Zafrir,
	JHEP {\bf 2006}, 176 (2020)
  [arXiv:2003.01843 [hep-th]]. 

\bibitem{BLLPRR} 
  C.~Beem, M.~Lemos, P.~Liendo, W.~Peelaers, L.~Rastelli, and B.~C.~van Rees,
	Commun.\ Math.\ Phys.\ {\bf 336} no. 3, 1359-1433 (2015)
  [arXiv:1312.5344 [hep-th]].

\bibitem{ISS}
  K.~Intriligator, N.~Seiberg, and S.~H.~Shenker,
	Phys.\ Lett.\ B342, 152-154 (1995)
  [arXiv:9410203 [hep-th]].

\bibitem{BCI}
  J.~H.~Brodie, P.~L.~Cho, and K.~Intriligator,
	Phys.\ Lett.\ B429, 319-326 (1998)
  [arXiv:9410203 [hep-th]].

\bibitem{IntMA}
  K.~Intriligator,
	Nucl.\ Phys.\ B730, 239-251 (2005)
  [arXiv:050985 [hep-th]].

\bibitem{Vart}
  G.~S.~Vartanov,
	Phys.\ Lett.\ B696, 288-290 (2011)
  [arXiv:1009.2153 [hep-th]].

\bibitem{TachiRev}
  Y.~Tachikawa,
  [arXiv:1812.08946 [hep-th]].

\bibitem{SeiDul}
  N.~Seiberg, 
  Nucl.\ Phys.\ B435, 129-146 (1995)
  [arXiv:9411149 [hep-th]].

\bibitem{E6Index} 
  A.~Gadde, L.~Rastelli, S.~S.~Razamat and W.~Yan,
	JHEP {\bf 1008}, 107 (2010)
  [arXiv:1003.4244 [hep-th]].

\bibitem{RVZ}
  S.~S.~Razamat, C.~Vafa, and G.~Zafrir,
  JHEP {\bf 1704}, 064 (2017) 
  [arXiv:1610.09178 [hep-th]].

\bibitem{GKSW}
  D.~Gaiotto, A.~Kapustin, N.~Seiberg, and B.~Willett,
  JHEP {\bf 1502}, 172 (2015)
  [arXiv:1412.5148 [hep-th]].

\bibitem{GB} 
  C.~Beem, A.~Gadde,
	JHEP {\bf 1404}, 036 (2014)
  [arXiv:1212.1467 [hep-th]].

\bibitem{GRRY}
  A.~Gadde, L.~Rastelli, S.~S.~Razamat, and W.~Yan,
	Phys.\ Rev.\ Lett.\  106 (2011) no.24, 241602
  [arXiv:1104.3850 [hep-th]].

\bibitem{BTW} 
  F.~Benini, Y.~Tachikawa, and B.~Wecht,
	JHEP {\bf 1001}, 088 (2010)
  [arXiv:0909.1327 [hep-th]].

\bibitem{KuTh}
  S.~M.~Kuzenko, S.~Theisen,
	Class.\ Quant.\ Grav.\ {\bf 17}, 665696 (2000)
  [arXiv:9907107 [hep-th]].

\bibitem{TSZ} 
  Y.~Tachikawa, H.~Shimizu, and G.~Zafrir,
	JHEP {\bf 1712}, 127 (2017)
  [arXiv:1703.01013 [hep-th]].

\bibitem{CIDrmi} 
  C.~Cordova, T.~T.~Dumitrescu, and K.~Intriligator,
	JHEP {\bf 1611}, 135 (2016)
  [arXiv:1602.01217 [hep-th]].

\bibitem{BNP}
  M.~Buican, T.~Nishinaka, and C.~Papageorgakis,
	JHEP {\bf 1412}, 095 (2014)
  [arXiv:1407.2835 [hep-th]].

\bibitem{Man} 
  A.~Manenti,
	JHEP {\bf 2004}, 145 (2020)
  [arXiv:1910.12869 [hep-th]].

\bibitem{GRRYind}
  A.~Gadde, L.~Rastelli, S.~S.~Razamat, and W.~Yan,
	JHEP {\bf 1103}, 041 (2011)
  [arXiv:1011.5278 [hep-th]].









\end{thebibliography}
\end{document}